\documentclass[12pt]{article} 

\usepackage{geometry}
\pdfoutput=1

\usepackage{amsmath} % already in sashorthand but should be defined earlier here to use \numberwithin
\numberwithin{equation}{section}
\RequirePackage[pdftex,dvipsnames]{xcolor}
\definecolor{darkblue}{rgb}{0.1,0.1,.7}
\usepackage[colorlinks, linkcolor=darkblue, citecolor=darkblue, urlcolor=darkblue, linktocpage]{hyperref} 

\usepackage[square, comma, sort&compress,numbers]{natbib} % reference management

\usepackage[margin=10pt,font=small,labelfont=bf]{caption}

\usepackage{tikz}
\usetikzlibrary{external}

\usepackage[]{sashorthand} % Provides useful commands and packages. It has as three potions:
\renewcommand{\equref}[1]{\eqref{#1}} % This is actually defined in sashorthand, albeit with equ.~ in the front

\usetikzlibrary{trees}
\usetikzlibrary{decorations.pathmorphing}
\usetikzlibrary{decorations.markings}
\usetikzlibrary{shapes.misc}

\tikzset{
	threept/.style={
		circle,
		draw,
		inner sep=2pt,
	},
	twopt/.style={
		circle,
		draw,
		fill=black,
		inner sep=1pt,
		minimum size=1pt
	},
	cross/.style={
		cross out,
		draw=black, 
		minimum size=7pt, 
		inner sep=0pt,
		outer sep=0pt
	},
	scalar/.style={
		thick,
		dashed,
		postaction={
			decorate,
			decoration={
				markings,
				mark=at position 0.5 with {\arrow{>}}
			}
		}
	},
	spinning/.style={
		thick,
		postaction={
			decorate,
			decoration={
				markings,
				mark=at position 0.5 with {\arrow{>}}
			}
		}
	},
	spinning no arrow/.style={
		thick,
	},
	finite with arrow/.style={
		decoration={
			snake,
			amplitude=1pt,
			segment length=6pt,
			post length=2pt
		},
		decorate,
		thick,->
	},
	finite/.style={
		decoration={
			snake,
			amplitude=1pt,
			segment length=6pt,
		},
		decorate,
		thick
	}
}

\newcommand*{\uniq}{\raisebox{-0.7ex}{\scalebox{1.8}{$\cdot$}}}
\newcommand\sixj[6]{\ensuremath{\left\{\begin{array}{ccc} #1 & #2 & #6 \\ #3 & #4 & #5\end{array}\right\}}}
\newcommand\sixjBlock[6]{\ensuremath{\left(\begin{array}{ccc} #1 & #2 & #6 \\ #3 & #4 & #5\end{array}\right)}}

\newcommand{\sixjDecomp}[9]{
	\newcommand{\sixjDecompTemp}[4]{
		\ensuremath{
			\mathcal{J}^{#1}
			\footnotesize
			\left(\begin{array}{cccccc} #2 & #3 & #4 & #5 & #6 & #7 \\ #8 & #9 & ##1 & ##2 & ##3 & ##4\end{array}\right)
			\normalsize
		}
	}
	\sixjDecompTemp
}

\newcommand{\sixjDecompp}[9]{
	\newcommand{\sixjDecomppTemp}[5]{
		\ensuremath{
			\mathcal{J}^{#1}_{#2}
			\footnotesize
			\left(\begin{array}{cccccc} #3 & #4 & #5 & #6 & #7 & #8 \\ #9 & ##1 & ##2 & ##3 & ##4 & ##5\end{array}\right)
			\normalsize
		}
	}
	\sixjDecomppTemp
}

\newcommand{\opeFuncDecomp}[9]{
	\newcommand{\opeFuncDecompTemp}[5]{
		\ensuremath{
			\mathcal{K}^{#1}_{#2}
			\footnotesize
			\left(\begin{array}{cccccc} #3 & #4 & #5 & #6 & #7 & #8 \\ #9 & ##1 & ##2 & ##3 & ##4 & ##5\end{array}\right)
			\normalsize
		}
	}
	\opeFuncDecompTemp
}

\DeclareMathAlphabet{\mathpzc}{OT1}{pzc}{m}{it} % New font \mathpzc

\graphicspath{ {images/} }

\begin{document}
	\vspace*{-.6in} \thispagestyle{empty}
	\begin{flushright}
	\end{flushright}
	\vspace{.2in} {\Large
		\begin{center}
			{\bf The Inversion Formula and 6j Symbol for 3d Fermions \vspace{.1in}}
		\end{center}
	}
	\vspace{.2in}
	\begin{center}
		{\bf 
			Soner Albayrak$^{a}$, David Meltzer$^{b}$, David Poland$^{a}$}
		\\
		\vspace{.2in} 
		$^a$ {\it  Department of Physics, Yale University, New Haven, CT 06511}\\
		$^b$ {\it  Walter Burke Institute for Theoretical Physics, Caltech, Pasadena, CA 91125}
	\end{center}
	
	\vspace{.2in}
	
	\begin{abstract}
In this work we study the $6j$ symbol of the $3d$ conformal group for fermionic operators. In particular, we study 4-point functions containing two fermions and two scalars and also those with four fermions. By using weight-shifting operators and harmonic analysis for the Euclidean conformal group, we relate these spinning $6j$ symbols to the simpler $6j$ symbol for four scalar operators. As one application we use these techniques to compute $3d$ mean field theory (MFT) OPE coefficients for fermionic operators. We then compute corrections to the MFT spectrum and couplings due to the inversion of a single operator, such as the stress tensor or a low-dimension scalar. These results are valid at finite spin and extend the perturbative large spin analysis to include non-perturbative effects in spin.
	\end{abstract}
	
	\newpage
	
	\tableofcontents
	
	\newpage
	
% BODY
\section{Introduction}
Invariance under the Euclidean conformal group $\SO(d+1,1)$ is well-known to put strong constraints on the space and observables of quantum field theories \cite{Ferrara:1973yt,Polyakov:1974gs}. In conformal field theories (CFTs), one hope is that we can leverage this symmetry to map out the space of consistent theories. The modern conformal bootstrap, as reviewed in \cite{Poland:2018epd}, has made remarkable progress in this direction by thoroughly studying the constraints of crossing symmetry, unitarity, and the existence of a convergent operator product expansion (OPE). 

The fundamental ingredient in the bootstrap program, both for analytic and numerical studies, is the requirement that four-point functions can be consistently expanded in conformal blocks, $G_{\Delta,\ell}(x_i)$, in all channels. The numerical conformal bootstrap studies this crossing condition in the Euclidean regime to derive constraints on the low-dimension spectrum \cite{Rattazzi:2008pe}, while the analytic conformal bootstrap considers an intrinsically Lorentzian regime to derive the existence of multi-twist trajectories at large spin \cite{Fitzpatrick:2012yx,Komargodski:2012ek}.

However, as emphasized in a series of papers \cite{Costa:2012cb,Gadde:2017sjg,Caron-Huot:2017vep,Simmons-Duffin:2017nub,Kravchuk:2018htv,Liu:2018jhs}, we can also expand four-point functions in conformal partial waves $\Psi_{\Delta,\ell}(x_i)$, which are a complete, orthogonal basis for Euclidean correlators and can be expressed as a single-valued linear combination of two blocks. Orthogonality of Euclidean conformal partial waves (CPWs) gives us the Euclidean inversion formula and by Wick rotation one can derive the Lorentzian inversion formula \cite{Caron-Huot:2017vep,Simmons-Duffin:2017nub,Kravchuk:2018htv}. The Lorentzian inversion formula, when combined with bounds on the Regge growth of CFT correlators \cite{Maldacena:2015waa,Hartman:2015lfa}, makes manifest that non-perturbative CFTs are analytic in spin for $\ell>1$.

Furthermore, orthogonality and completeness imply that individual conformal partial waves are crossing symmetric. That is, a t-channel partial wave can be written in terms of s-channel partial waves in a dual channel. The inner product of an s- and t-channel partial wave in turn defines the $6j$ symbol of the conformal group. Although the $6j$ symbol can be defined in terms of Euclidean conformal integrals, in practice it has been calculated in $d=2$ and $4$ using the Lorentzian inversion formula and by first calculating the inversion of a single block \cite{Liu:2018jhs}.\footnote{In \cite{Liu:2018jhs} the $1d$ $6j$ symbol was calculated using the Euclidean definition directly.} 

The inversion of a single partial wave or block reveals new features which were not previously visible from the lightcone bootstrap directly. The lightcone bootstrap generates an asymptotic expansion in large spin by matching a sum over conformal blocks in one channel to a singular term in the crossed channel.\footnote{For a detailed analysis of tauberian theorems in CFTs see \cite{Qiao:2017xif}.} However, the inversion formula directly proves the existence of individual double-twist operators and that the OPE data is analytic in spin. Moreover, by inverting a single operator one can show there are corrections which are exponentially suppressed at large spin \cite{Liu:2018jhs}, which improve predictions from the analytic bootstrap for $3d$ CFTs such as the Ising or $O(2)$ model \cite{Albayrak:2019gnz} and also ensure predictions at large spin are consistent with bounds from causality \cite{Meltzer:2018tnm}.\footnote{For related work see \cite{Cardona:2018dov,Cardona:2018qrt,Li:2019dix,Chen:2019gka}.}

Thus far, most studies of the analytic bootstrap have focused on four external scalar operators. There are a few exceptions -- the analytic bootstrap for external, spinning operators has been studied using the lightcone bootstrap \cite{Li:2015itl,Hofman:2016awc,Li:2017lmh,Elkhidir:2017iov,Chowdhury:2018uyv,Albayrak:2019gnz}, Mellin space techniques \cite{Sleight:2018epi,Sleight:2018ryu}, and in mean field theory \cite{Karateev:2018oml}. In this work we further develop this program, building off of \cite{Albayrak:2019gnz} by studying the inversion formulas in $d=3$ for correlation functions containing Majorana fermions. In this way we can derive new results for the OPE coefficients and spectrum of double-twist operators containing fermions, including non-perturbative effects at finite spin.

Our strategy to study the inversion formula for spinning operators involves a combination of Euclidean and Lorentzian ingredients. Our starting point for relating the fermionic $6j$ symbol to the scalar one involves their Euclidean definition as an overlap of partial waves. Then we use weight-shifting operators \cite{Karateev:2017jgd}, which transform in a finite-dimensional representation of the conformal group, to expand the fermionic $6j$ symbol as a sum over scalar symbols. We can plug in the explicit form of the scalar $6j$ symbol as calculated via the Lorentzian inversion formula to obtain the fermionic $6j$ symbol in closed form.\footnote{The full $6j$ symbol is only known in $d=1, \ 2, \ 4$ but the poles and residues are computable in general dimensions.} Finally, since a partial wave for general external operators can be split as a sum over two blocks, there exists a similar split for the $6j$ symbol in terms of the inversion of two blocks. By splitting the scalar and fermionic $6j$ symbols, we then find the inversion of a single block when we have external fermions.

Finally, let us give a brief summary of the paper. In section \ref{sec:6jreview} we review the necessary ingredients to define the $6j$ symbols for external scalars, including conformal integrals and the partial wave decomposition. In section \ref{sec:review_int_pairings} we generalize the scalar analysis to general external operators in $3d$. In section \ref{sec:MFTOPE} we use the Euclidean inversion formula to calculate OPE coefficients for double-twist operators in mean field theory. In section \ref{sec:Inversion_Isolated_Ops} we introduce the weight-shifting operators and express fermionic $6j$ symbols in terms of $6j$ symbols for four external scalars. From the $6j$ symbol we then compute corrections to the double-twist spectrum and couplings from the inversion of isolated operators. We conclude in section \ref{sec:conclusion} with a discussion of future directions. Various technical details are given in the appendices. 

\section{Review: Scalar $6j$ Symbol and the Inversion Formula}
\label{sec:6jreview}
To start let us give a brief overview of the conformal partial wave expansion for external scalars, the $6j$ symbol, and its relation to the analytic bootstrap. This will allow us to describe the basic ideas and define quantities which will repeatedly appear in the following sections. This section will be based on previous work presented in \cite{Dobrev:1977qv,Liu:2018jhs,Karateev:2018oml}.

Following the notation of \cite{Karateev:2018oml}, we label an irreducible representation of $\SO(d+1,1)$ by its scaling dimension $\Delta$, and by $\rho$, an irreducible representation of $\SO(d)$. When we study conformally-invariant two- and three-point functions we will use $\cO$ as a shorthand for this representation.

Associated to every representation $\cO$ is a shadow representation $\widetilde{\cO}$ which has dimension $\widetilde{\Delta}=d-\Delta$ and the same $\SO(d)$ representation $\rho$.\footnote{In even dimensions we actually need the reflected representation $\rho^{R}$, but in odd dimensions the two are equivalent so to simplify the discussion we will ignore the distinction.} Then there exists a conformally-invariant pairing:
\be 
\left(\tl\cO^\dagger,\cO\right)=\int d^dx\tl\cO^\dagger_{\alpha_1\cdots\alpha_{2J}}(x) \cO^{\alpha_1\cdots\alpha_{2J}}(x),
\ee 
where $\tl \cO^{\dagger}$ has dimension $\widetilde{\Delta}$. In the above expression and in what follows we will implicitly contract indices when an operator and its shadow are being integrated over, and we will always use spinor indices. We will work with real operators, so we can drop the $\dagger$ in the expressions which follow.

As is well-known, in CFTs the two-point functions are fixed up to the normalization of the operator, and the three-point functions are fixed up to the OPE coefficients.  We will adopt the convention that physical, Euclidean correlation functions will be denoted as $\<\cdots\>_{\O}$, while two- and three-point functions without a subscript will denote a conformally-invariant tensor structure. In general we have several three-point tensor structures, $\<\cO_1\cO_2\cO_3\>^{a}$, labelled by the index $a$, but if $\cO_1$ and $\cO_2$ are scalars then there is a single structure.

We can then define the s-channel conformal partial wave (CPW) expansion:\footnote{Technically we need to include non-normalizable terms for blocks with $\Delta<\frac{d}{2}$, but we will ignore such terms here. For more details see \cite{Caron-Huot:2017vep,Simmons-Duffin:2017nub}.}
\begin{align}
\<\f_1\f_2\f_3\f_4\>&=\<\f_1\f_2\>\<\f_3\f_4\>+\sum\limits_{J=0}^{\infty}\int\limits_{\frac{d}{2}}^{\frac{d}{2}+i\infty} \frac{d\Delta}{2\pi i}\rho^{1234}(\Delta,J)\Psi^{1234}_{\Delta,J}(x_i),\label{eq:Scalar_CPW_Expansion}
\\
\Psi^{1234}_{\Delta,J}(x_i)&=\int d^{d}x_{5}\<\f_1(x_1)\f_2(x_2)\cO(x_5)\>\<\f_3(x_3)\f_4(x_4)\widetilde{\cO}(x_5)\>\label{eq:Scalar_CPW_Def1}
\\ &=S\big(\f_3\f_4[\widetilde{\Delta},J]\big)G^{1234}_{\Delta,J}(x_i)+S\left(\f_1\f_2[\Delta,J]\right)G^{1234}_{\widetilde{\Delta},J}(x_i). \label{eq:Scalar_CPW_Def2}
\end{align}

To define the factors of $S$ we need to introduce the shadow transform which maps a representation $\cO$ to $\widetilde{\cO}$:
\be 
\mathbf{S}[\cO](x)\equiv \int dy \cO(y)\<\tl\cO(y)\tl\cO(x)\>.
\label{eq: shadow definition}
\ee 
We will adopt the convention that the shadow transform always acts by multiplying the two-point function from the left. The distinction is immaterial for bosonic correlators, but when we introduce fermions the ordering does matter. 

The definition of the shadow coefficients are:
\begin{align}
\<\cO_1\cO_2\mathbf{S}[\cO_3]\>=S\left(\cO_1\cO_2[\cO_3]\right)\<\cO_1\cO_2\widetilde{\cO}_{3}\>.
\end{align}
To be clear, we will use $\widetilde{\cO}(x)$ to denote a representation which has scaling dimension $d-\Delta$ and representation $\rho$, while $\mathbf{S}[\cO](x)$ denotes the integral map $\mathbf{S}$ acting on $\cO(x)$.

The t- and u-channel expansions can be found by exchanging $1\leftrightarrow 3$ and $1\leftrightarrow 4$, respectively. The spectral integral in (\ref{eq:Scalar_CPW_Expansion}) runs over the Euclidean principal series:
\begin{align}
\Delta=\frac{d}{2}+i\nu, \quad \nu\geq0, \qquad J\in\mathbb{Z}.
\end{align}
In (\ref{eq:Scalar_CPW_Def1}) and (\ref{eq:Scalar_CPW_Def2}) we wrote down two equivalent definitions for the conformal partial wave, as a conformal integral over two three-point functions and as a linear combination of conformal blocks, each of which encodes the contribution of a irreducible representation to the four-point function. The exact definition of the shadow coefficients $S$ and the block are given in appendix \ref{app:Scalar_Conventions}.

The partial waves are orthogonal with respect to the following bilinear pairing:
\begin{align}
&\left(\widetilde{\Psi}^{1234}_{\frac{d}{2}+i\nu,J}\ ,\Psi^{1234}_{\frac{d}{2}+i\nu',J'}\right)=n_{\frac{d}{2}+i\nu,J}\delta_{JJ'}2\pi\delta(\nu-\nu') \qquad (\nu,\nu'\geq0),
\\
&\left(F,G\right)=\int \frac{d^{d}x_{1}...d^{d}x_{n}}{\vol(SO(d+1,1))}F(x_1,...,x_n)G(x_1,...,x_n),
\end{align}
where $n_{\Delta,J}$ is the normalization of the conformal partial wave, the bilinear pairing is defined for general $n$-point functions, and the shadowed partial wave is\footnote{In general we also need to conjugate the operators, but we will assume our operators are real.}
\begin{align}
\widetilde{\Psi}^{1234}_{\Delta,J}\equiv\Psi^{\tilde{1}\tilde{2}\tilde{3}\tilde{4}}_{\widetilde{\Delta},J}.
\end{align}

The identity contribution in (\ref{eq:Scalar_CPW_Expansion}) is orthogonal to all partial waves on the principal series, so by taking an inner-product with a shadowed partial wave we can write down the Euclidean inversion formula:
\begin{align}
\rho^{1234}(\Delta,J)=\frac{1}{n_{\Delta,J}}\left(\widetilde{\Psi}^{1234}_{\Delta,J},\<\f_1\f_2\f_3\f_4\>\right). \label{eq:scalar_E_Inversion}
\end{align}

The second definition (\ref{eq:Scalar_CPW_Def2}) allows us to relate the CPW expansion with the more physical conformal block expansion. To make the notation compact let us define
\begin{align}
\int\limits_{\mathcal{C}}d\cO=\sum\limits_{J=0}^{\infty}\int\limits_{\frac{d}{2}}^{\frac{d}{2}+i\infty}\frac{d\Delta}{2\pi i}, \qquad \qquad \int\limits_{\mathcal{C}'}d\cO=\sum\limits_{J=0}^{\infty}\int\limits_{\frac{d}{2}-i\infty}^{\frac{d}{2}+i\infty}\frac{d\Delta}{2\pi i}.
\end{align}
Then we can write
\begin{align}
\int\limits_{\mathcal{C}}d\cO\rho^{1234}(\Delta,J)\Psi^{1234}_{\Delta,J}(x_i)=\int\limits_{\mathcal{C}'}d\cO\rho^{1234}(\Delta,J)S(\f_3\f_4[\widetilde{\Delta},J])G^{1234}_{\Delta,J}(x_i),
\end{align}
where we used the following symmetry property, which can be derived from (\ref{eq:scalar_E_Inversion}):
\begin{align}
\rho^{1234}(\widetilde{\Delta},J)=\rho^{1234}(\Delta,J)\frac{S(\f_3\f_4[\widetilde{\Delta},J])}{S(\f_1\f_2[\widetilde{\Delta},J])}.
\end{align}
Using the conformal block decomposition
\begin{align}
\<\f_1\f_2\f_3\f_4\>=\sum\limits_\cO \lambda_{\f_1\f_2\cO}\lambda_{\f_3\f_4\cO}G_{\cO}^{1234}(x_i),
\end{align}
we find the following relation:
\begin{align}
\lambda_{\f_1\f_2\cO}\lambda_{\f_3\f_4\cO}=-\Res_{\Delta=\Delta_{\cO}}\rho^{1234}(\Delta,J)S(\f_3\f_4[\widetilde{\Delta},J]).
\end{align}
We can now ask the general question: what does the contribution of a single operator $\cO$ in the t-channel map to in the s-channel under crossing? As realized in \cite{Komargodski:2012ek,Fitzpatrick:2012yx}, by studying the lightcone limit, an isolated operator $\cO$ maps to double-twist operators in the crossed channel. To review this result in the current language, let us first introduce the $6j$ symbol of the conformal group.

The $6j$ symbol is defined as the overlap of a t- and s-channel partial wave, which for external scalars is
\begin{align}
\sixj{\f_1}{\f_2}{\f_3}{\f_4}{\cO_5}{\cO_6}&=\left(\widetilde{\Psi}^{1234}_{\Delta_{5},J_5},\Psi^{3214}_{\Delta_6,J_6}\right)
\nonumber \\ &=\int dx_1... dx_6 \<\widetilde{\f}_1\widetilde{\f}_2\widetilde{\cO}_5\>\<\widetilde{\f}_{3}\widetilde{\f}_{4}\cO_5\>\<\f_3\f_2\cO_6\>\<\f_1\f_4\widetilde{\cO}_{6}\>.\label{eq:6jDefScalar}
\end{align}
Using the $6j$ symbol it is possible to write a single t-channel partial wave as a spectral integral over s-channel partial waves:
\begin{align}
\Psi^{3214}_{\Delta_{6},J_6}(x_i)=\int\limits_{\mathcal{C}}d\cO_{5}\frac{1}{n_{\Delta_5,J_5}}\sixj{\f_1}{\f_2}{\f_3}{\f_4}{\cO_5}{\cO_6}\Psi^{1234}_{\Delta_5,J_5}(x_i).
\end{align}
In practice the $6j$ symbol (\ref{eq:6jDefScalar}) has been calculated using the Lorentzian inversion forumla in $d=2$ and $d=4$. The Lorentzian inversion formula gives another integral representation of the OPE function, but now with the correlator integrated over a causal diamond in Minkowski space:\footnote{Our convention is $\vol(\SO(n))=\vol(\SO(n-1))\vol(S^{n-1})$.}
\begin{align}
\rho^{1234}(\Delta,J)&=\alpha_{\Delta,J}\int\limits_{0}^{1}\int\limits_{0}^{1}dzd\bar{z}\mu(z,\bar{z})G^{1234}_{J+d-1,\Delta+1-d}(z,\bar{z})\frac{\<[\f_3,\f_2][\f_1,\f_4]\>}{T^{1234}}+(\text{u-channel}),
\\
\alpha_{\Delta,J}&=-\frac{\pi^{d-2}\Gamma(\frac{d-2}{2})\Gamma(J+d-2)}{2^{d+J+3}\text{vol}(\text{SO}(d-1))\Gamma(d-2)\Gamma(J+\frac{d-2}{2})}\frac{\Gamma(J+1)\Gamma(\Delta-\frac{d}{2})}{\Gamma(J+\frac{d}{2})\Gamma(\Delta-1)}
\nonumber \\ & \qquad \frac{\Gamma(\frac{\Delta_{12}+J+\Delta}{2})\Gamma(\frac{\Delta_{21}+J+\Delta}{2})\Gamma(\frac{\Delta_{34}+J+\widetilde{\Delta}}{2})\Gamma(\frac{\Delta_{43}+J+\widetilde{\Delta}}{2})}{\Gamma(J+\Delta)\Gamma(J+d-\Delta)}.
\end{align}
Here $T^{1234}$ is a kinematic, s-channel prefactor,
\begin{align} 
T^{1234}&=\frac{1}{|x_{12}|^{\Delta_1+\Delta_2}|x_{34}|^{\Delta_3+\Delta_4}}\left(\frac{|x_{24}|}{|x_{14}|}\right)^{\Delta_1-\Delta_2}\left(\frac{|x_{14}|}{|x_{13}|}\right)^{\Delta_{3}-\Delta_4}.
\end{align}
The u-channel term is the same but with $3\leftrightarrow 4$. Then by first inverting individual blocks we can find the $6j$ symbol:
\begin{align}
\sixj{\f_1}{\f_2}{\f_3}{\f_4}{\cO_5}{\cO_6}&=S(\f_3\f_4[\widetilde{\cO}_{6}])\sixjBlock{\f_1}{\f_2}{\f_3}{\f_4}{\cO_5}{\cO_{6}}+S(\f_1\f_2[\cO_{6}])\sixjBlock{\f_1}{\f_2}{\f_3}{\f_4}{\cO_5}{\widetilde{\cO}_{6}}, \label{eq:6j_symbol_split_Scalars}
\\
\sixjBlock{\f_1}{\f_2}{\f_3}{\f_4}{\cO_5}{\cO_{6}}&=\left(\Psi^{\tilde{1}\tilde{2}\tilde{3}\tilde{4}}_{\Delta_5,J_6},G^{3214}_{\Delta_6,J_6}\right)_{L}, \label{eq:scalar_Inv_Block}
\end{align}
where the subscript $L$ in (\ref{eq:scalar_Inv_Block}) is to emphasize that we use the Lorentzian inversion formula \cite{Liu:2018jhs}. We can drop the u-channel contribution because the u-channel double discontinuity of a t-channel block is zero.

We do not have a closed form expression for the $6j$ symbol in generic dimensions, but for $d=3$ it is straightforward to calculate its poles and residues by using dimensional reduction and the explicit $d=2$ expressions \cite{Hogervorst:2016hal,Albayrak:2019gnz}. Let us now focus on the problem of inverting a single operator. The contribution of a single block for $\cO_{6}$ exchange gives
\begin{align}
\<\f_1\f_2\f_3\f_4\>\supset \lambda_{\f_3\f_2\cO_{6}}\lambda_{\f_1\f_4\cO_6}\int\limits_{\mathcal{C}'}d\cO_{5}\frac{1}{n_{\cO_{5}}}\sixjBlock{\f_1}{\f_2}{\f_3}{\f_4}{\cO_5}{\cO_{6}}S(\f_3\f_4[\widetilde{\cO}_{5}])G^{1234}_{\cO_{5}}(x_i). \label{eq:ind_Block_Inv}
\end{align}
As a function of $\Delta_{5}$ the integrand of (\ref{eq:ind_Block_Inv}) has poles at the following locations:
\begin{eqnarray}
\Delta_{5}&=\Delta_{1}+\Delta_{2}+2n+J_5,
\\
\Delta_{5}&=\Delta_{3}+\Delta_{5}+2n+J_5.
\end{eqnarray}
These are the dimensions and spins of the double-twist operators $[\phi_1\phi_2]_{n,J}$ and $[\phi_3\phi_4]_{n,J}$ in mean field theory (MFT) \cite{Komargodski:2012ek,Fitzpatrick:2012yx}. A special case is when $\Delta_1+\Delta_2=\Delta_3+\Delta_4$ in which case we get single and double poles  corresponding to corrections to the MFT spectrum and OPE coefficients of $[\f_1\f_2]_{n,J}$. An important exception is when we are inverting the identity block, $\cO_{6}=\mathbb{1}$, in which case we get single poles and find the MFT OPE coefficients themselves. We will work out the MFT OPE coefficients for fermions in section \ref{sec:MFTOPE}.

\section{Inversion Formula and MFT for Fermions}
\label{sec:review_int_pairings}
In this section we will go into more detail on conformally-invariant integrals for fermions in $d=3$ and introduce our conventions in the process. We will also present some of our intermediate results, such as explicit expressions for shadow coefficients in $3d$ fermionic theories.

\subsection{Conformal Partial Waves and $6j$ Symbols}
\label{sec: conformal partial waves}

From here on we will set $d=3$ and to be general our operators will always be defined with spinor indices. To avoid possible sign ambiguities, we need to fix our conventions regarding pairing fermionic representations. We will assume an ascending index convention such that
\begin{align}
A\.B\equiv A_{\a_1\a_2\dots \a_{2l}}B^{\a_1\a_2\dots \a_{2l}}.
\end{align}
We will also normalize the two-point functions as \cite{Karateev:2017jgd}:
\be 
\<\cO_{\Delta,\ell}(X_1,S_1)\cO_{\Delta,\ell}(X_2,S_2)\>= i^{2l}\frac{\<S_1S_2\>^{2l}}{X_{12}^{2\De+2l}}, \label{eq:two_pt_convention}
\ee 
where $X$ and $S$ are position and polarization vectors in embedding space. Here we also introduced the shorthand that
\begin{align}
\cO_{\Delta,\ell}(x,s)=s_{\a_1}...s_{\a_{2\ell}}\cO_{\Delta,\ell}^{\alpha_1...\alpha_{2\ell}} \text{,}\qquad\cO_{\Delta,\ell}(X,S)=S_{I_1}...S_{I_{2\ell}}\cO_{\Delta,\ell}^{I_1...I_{2\ell}},
\end{align}
where $x$ and $s$ are the positions and polarizations in physical space.

We can expand a physical three-point function as
\be 
\<\cO_1\cO_2\cO_3\>_{\O}=\sum\limits_{a}\lambda^a_{\cO_1\cO_2\cO_3}\<\cO_1\cO_2\cO_3\>^a,
\ee 
where $\lambda^a$ are the OPE coefficients and $\<\cO_1\cO_2\cO_3\>^a$ are kinematic tensor structures. In particular, they are formal expressions and do not necessarily satisfy Fermi-statistics if operators are interchanged. Throughout this paper, we will refer to these functions as three-point structures.

When we calculate shadow transforms or partial waves, it will be important to keep track of signs that depend on the order of correlators in an integral, e.g.
\be 
\int dx  \<\cO_1 \cO(x) \cO_2 \>\<\cO_3 \tl \cO(x) \cO_4 \>  =(-1)^{2l_{\cO}}
\int dx  \<\cO_3 \tl \cO(x) \cO_4 \>\<\cO_1 \cO(x) \cO_2 \>\ .
\label{eq: pairing convention} 
\ee 
Details regarding our fermion conventions and the explicit choice of three-point structures are given in appendix~\ref{sec:conventions_fermions}.

Since there are in general multiple conformally-invariant three-point structures, the shadow transform of an operator now gives a matrix of shadow coefficients:
\begin{align}
\label{eq: shadow coefficients}
\<\cO_1\cO_2\mathbf{S}[\cO_3]\>^a=S^{a}_{b}(\cO_1\cO_2[\cO_3])\<\cO_1\cO_2\widetilde{\cO}_{3}\>^b.
\end{align}

\begin{figure}
	\centering
	\includegraphics[scale=1]{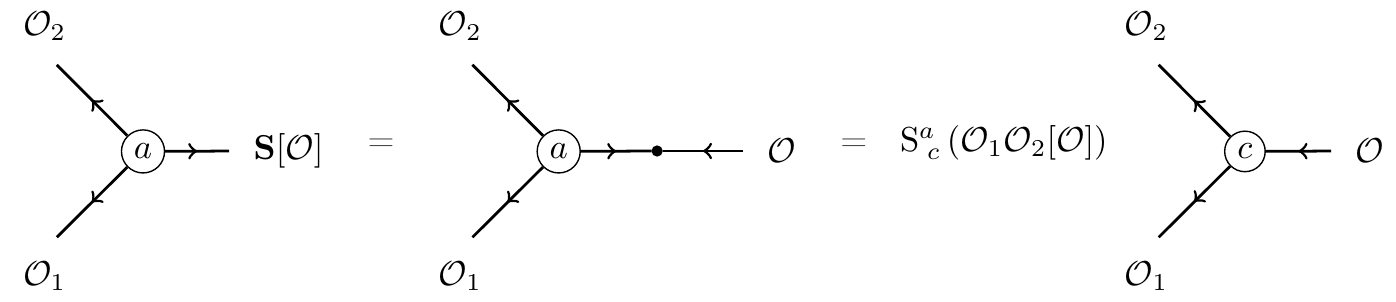}
	\caption{\label{fig:shadow} Diagrammatic definition of shadow coefficients. Note that $a$ and $c$ label the three-point structures, and arrows allow us to keep track of scaling dimensions. We use the standard convention where an operator $\cO_{\Delta,J}$ with an outgoing arrow from a three-point structure enters that structure as itself. On the contrary, changing the direction of arrow is equivalent to changing $\cO_{\Delta,J}$ to $\tl\cO_{\Delta,J}\equiv\cO_{\tl\Delta,J}\equiv \cO_{3-\Delta,J}$.}
\end{figure}

Following \cite{Karateev:2018oml}, we can also use the diagrammatic notation 
\be 
\includegraphics[scale=1.2]{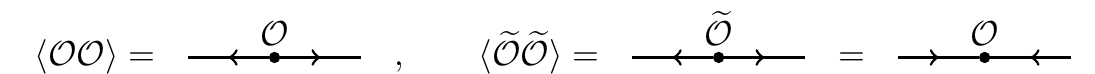},
\ee
where we see that taking the shadow is equivalent to changing the direction of arrow. In this language, pairing operators is gluing the arrows; for example the diagrammatic equation 
\be 
\includegraphics[scale=1.6]{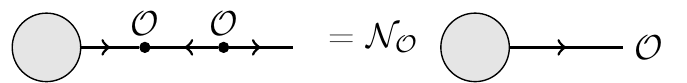}
\ee
stands for 
\be 
\<\cdots \mathbf{S}^{2}[\cO](x)\cdots\>=\int dx dy \<\cdots \cO(x)\cdots\>\<\tl\cO(x)\tl\cO(y)\>\<\cO(y)\cO(z)\>=\cN_{\cO}\<\cdots\cO(z)\cdots\>,
\ee 
or $\mathbf{S}^2[\cO]=\cN_{\cO}\cO $. This follows from the definition of the shadow transformation in \equref{eq: shadow definition} and the irreducibility of the representations. The factor $\cN$ in our conventions is
\be 
\cN_{\Delta,l} =\frac{\pi ^3 \tan (\pi  \left(\Delta+l\right) )}{(\Delta -\frac{3}{2}) (-\Delta +l+2) (\Delta +l-1)}.
\ee 

We can now study the partial wave expansion for a general four-point function:\footnote{To simplify the discussion we will again assume that there are no non-normalizable terms besides identity exchange.}
\begin{align}
\<\cO_1\cO_2\cO_3\cO_4\>&=\<\cO_1\cO_2\>\<\cO_3\cO_4\>+\int\limits_{\mathcal{C}}d\cO \rho^{(s)}_{ab}(\cO)\Psi^{(s),ab}_{\cO}(x_i), \label{eq:CPW_General}
\\
\Psi^{(s),ab}_{\cO_5}(x_i)&=\int d^{d}x_{5}\<\cO_1\cO_2\cO_5\>^a\<\cO_3\cO_4\widetilde{\cO}_{5}\>^{b},
\end{align}
where $a,b$ label different three-point function tensor structures. We have suppressed Lorentz indices, but the operators $\cO_i$ can be bosonic or fermionic. 

To be precise, in the measure $d\cO$ we now have a sum over either integer or half-integer spin, depending on the four-point function.\footnote{In general dimensions $d$ we have to sum over all allowed $\SO(d)$ representations.} In odd dimensions we can also have the discrete series of dimensions, but they will be cancelled by poles in $\rho^{(s)}_{ab}(\mathcal{O})$, so we will not include them explicitly.

The inner product between two s-channel partial waves is now a matrix:
\be[eq: definition of eta] 
\left(\widetilde{\Psi}^{(s),ab}_{\cO'},\Psi^{(s)cd}_{\cO}\right) =\eta^{(ac)(bd)}_\cO 2\pi\delta(\nu-\nu')\delta_{JJ'}, \label{eq:overlap_Fermion}
\ee
and we can also define the inverse matrix, $\eta_{(ac)(bd)}^\cO \eta^{(ce)(df)}_\cO=\delta^{e}_{a}\delta^{f}_{b}$.

We will use the same convention when writing inverses for three-point structures and their pairings, e.g.
\be 
\frac{\left(\<ABC\>^a,\<\tl A\tl B\tl C\>^b\right)}{\left(\<ABC\>^c,\<\tl A\tl B\tl C\>^a\right)}=
\frac{\left(\<ABC\>^b,\<\tl A\tl B\tl C\>^a\right)}{\left(\<ABC\>^a,\<\tl A\tl B\tl C\>^c\right)}=
\delta_{bc}.
\ee 
This notation is schematic and should be understood as inverting a matrix of pairings. Then the Euclidean inversion formula for this correlator is
\begin{align}
\rho^{(s)}_{ab}(\cO)=\eta^{\cO}_{(ac)(bd)}\left(\widetilde{\Psi}^{(s),cd}_{\cO_{5}},\<\cO_1\cO_2\cO_3\cO_4\>\right). \label{eq:GenEucInvSpin}
\end{align}
Given the general partial wave decomposition (\ref{eq:CPW_General}) we still have to find the relation between the residues of the OPE function and the actual OPE coefficients. The relation between the s-channel partial waves and the conformal blocks is given by
\begin{align}
\Psi^{(s);ab}_{\cO_5}(x_i)=(-1)^{\Sigma_{55}}\left[S^{b}_{c}(\cO_3\cO_4[\widetilde{\cO}_{5}])G^{(s);ac}_{\cO_5}(x_i)+S^{a}_{c}(\cO_1\cO_2[\cO_5])G^{(s);cb}_{\cO_5}(x_i)\right].
\end{align}
Here $\Sigma_{ij}$ is 1 if both the operators $\cO_i$ and $\cO_j$ are fermions, and is 0 otherwise. Then the corresponding map for the OPE coefficients is
\begin{align}
P_{ab}^{(s)}(\cO_5) \equiv \lambda_{125,a}\lambda_{345,b}=(-1)^{\Sigma_{55}+1}\Res_{\Delta=\Delta_{5}}\rho^{(s)}_{ac}(\Delta,J)S^c_b(\cO_{3}\cO_{4}[\widetilde{\cO}])\bigg|_{J=J_{5}}.\label{eqn:OPEfunc_To_Coefs_schannel}
\end{align}

Next we need to write down the same expansion, but for the $\cO_3\cO_2\rightarrow \cO_1\cO_4$ channel, or in the t-channel. We first rewrite the correlator as
\begin{align}
\<\cO_1\cO_2\cO_3\cO_4\>=(-1)^{\Sigma_{23}+\Sigma_{12}+\Sigma_{13}}\<\cO_3\cO_2\cO_1\cO_4\> \label{eqn:crossing_Gen_Spin}.
\end{align} 
Then we define the t-channel CPW as
\begin{align}
\<\cO_3\cO_2\cO_1\cO_4\>&=\<\cO_3\cO_2\>\<\cO_1\cO_4\>+\int d\cO \rho^{(t)}_{ab}(\cO)\Psi^{(t),ab}_{\cO}(x_i),
\\
\Psi^{(t),ab}_{\cO_5}(x_i)&=\int d^{d}x_{5}\<\cO_3\cO_2\cO_5\>\<\cO_1\cO_4\widetilde{\cO}_{5}\>.
\end{align}
The relation between the OPE coefficients and the OPE function in the t-channel is now the same as before, with $1\leftrightarrow 3$:
\begin{align}
P_{ab}^{(t)}(\cO_6) \equiv \lambda_{326,a}\lambda_{146,b}=(-1)^{\Sigma_{66}+1}\Res_{\Delta=\Delta_{6}}\rho^{(t)}_{ac}(\Delta,J)S^{c}_{b}(\cO_1\cO_4[\widetilde{\cO}])\bigg|_{J=J_{6}}.
\label{eqn:OPEfunc_To_Coefs_tchannel}
\end{align}

We will then define the $6j$ symbol for a general four-point function as
\begin{align}
\hspace{-.25in}\sixj{\cO_1}{\cO_2}{\cO_3}{\cO_4}{\cO_5}{\cO_6}^{abcd}=\left(\widetilde{\Psi}^{(s),ab}_{\cO_5},\Psi^{(t),cd}_{\cO_6}\right)=\int d^{d}x_{1}...d^{d}x_{6}\<\widetilde{\cO}_1\widetilde{\cO}_2\widetilde{\cO}_{5}\>\<\widetilde{\cO}_3\widetilde{\cO}_4\cO_5\>\<\cO_3\cO_2\cO_6\>\<\cO_1\cO_4\widetilde{\cO}_{6}\>. \label{eqn:Gen6jSymbol}
\end{align}

As in the scalar case we can split this into two pieces, corresponding to the inversion of individual blocks:
\begin{align}
\sixj{\cO_1}{\cO_2}{\cO_3}{\cO_4}{\cO_5}{\cO_6}^{abcd}&=(-1)^{\Sigma_{66}}\bigg[S^{d}_{e}\left(\cO_1\cO_4[\widetilde{\cO}_{6}]\right)\sixjBlock{\cO_1}{\cO_2}{\cO_3}{\cO_4}{\cO_5}{\cO_6}^{abce} \nonumber
\\ &\hspace{.75in}+S^{c}_{e}\left(\cO_3\cO_2[\cO_{6}]\right)\sixjBlock{\cO_1}{\cO_2}{\cO_3}{\cO_4}{\cO_5}{\widetilde{\cO}_6}^{abed}\bigg],\label{eqn:GenSplit6j}
\\
\sixjBlock{\cO_1}{\cO_2}{\cO_3}{\cO_4}{\cO_5}{\cO_6}^{abcd}&=\left(\widetilde{\Psi}^{(s),ab}_{\cO_{5}},G^{(t),cd}_{\cO_{6}}\right)_{L}.&\label{eqn:InvSingBlockGenSpin}
\end{align}

Then using (\ref{eq:GenEucInvSpin}), (\ref{eqn:OPEfunc_To_Coefs_schannel}), and  (\ref{eqn:crossing_Gen_Spin}), we find that inverting a single block $G^{(t),fg}_{\cO_6}$ yields the correction:
\begin{multline}
\lambda_{125,a}\lambda_{345,b}\bigg|_{G^{(t),fg}_{\cO_6}}=(-1)^{1+\Sigma_{55}+\Sigma_{12}+\Sigma_{13}+\Sigma_{23}}\lambda_{326,f}\lambda_{146,g}\\\x\Res_{\Delta=\Delta_{5}}\eta_{(ad)(ce)}^{\De,J}\sixjBlock{\cO_1}{\cO_2}{\cO_3}{\cO_4}{\cO_{\De,J}}{\cO_6}^{defg}S^c_b(\cO_{3}\cO_{4}[\widetilde{\cO}])\bigg|_{J=J_{5}}. \label{eqn:InvSingBlock_to_OPEdata}
\end{multline}

The problem now is how to reduce the full $6j$ symbol, (\ref{eqn:Gen6jSymbol}), to a sum of scalar $6j$ symbols, (\ref{eq:6jDefScalar}). To do this we will need to use weight-shifting operators, which allow us to reduce general conformal integrals involving fermions to those involving scalars. We will introduce these operators in the next subsection to calculate shadow coefficients. 

\subsection{Shadow Matrices and Weight-Shifting Operators}
\label{sec: shadow matrices}
The work \cite{Karateev:2017jgd} introduced differential operators, $\mathcal{D}^{A}$, which transform in a finite-dimensional representation of the conformal group $\SO(d+1,1)$ given by the index $A$. By acting with weight-shifting operators we can transform a conformally-invariant tensor structure involving an operator $\cO$ to a conformally covariant structure involving a new operator $\cO'$. Here, we will use weight-shifting operators which change the spin by half-integers.

In $d=3$ the double cover of the conformal group $\SO(3,2)$ is $\Sp(4,\mathbb{R})$, so we will use weight-shifting operators which transform in the fundamental representation of $\Sp(4,\mathbb{R})$. We will use the notation of  \cite{Karateev:2017jgd} and write the four possible operators as
\begin{align}
\mathcal{D}^{\pm\pm}: [\Delta,\ell]\rightarrow [\Delta\pm \frac{1}{2},\ell\pm\frac{1}{2}],
\end{align}
which in embedding space are given by
\bea[eq: differential operators in embedding space]
\left(\cD^{-+}\right)_a=& S_a,\\
\left(\cD^{--}\right)_a=& X_{ab}\pdr{}{S_b},\\
\left(\cD^{++}\right)_a=& -2(\Delta-1)S_b(\partial_X)^b_{\;\;a}
-S_aS_b(\partial_X)^b_{\;\;c}\pdr{}{S_c},
\\
\left(\cD^{+-}\right)_a=&
4(1+l-\De)(\Delta-1)\Omega_{ab}\pdr{}{S_b}
+2(1+l-\De)X_{ab}(\partial_X)^b_{\;c}\pdr{}{S_c}
\nonumber
\\
&-S_a\pdr{}{S_b}X_{bc}(\partial_X)^c_{\;d}\pdr{}{S_d}.
\eea
In the rest of the paper, we will suppress spinor index of the weight-shifting operators.

We will use these operators to relate conformal integrals involving fermionic tensor structures to known integrals involving bosonic structures. To start we will follow \cite{Karateev:2018oml} and use the weight-shifting operators to compute the shadow coefficient matrix. Before getting to this point we need to define the adjoint of a weight-shifting operator with respect to our bilinear pairing:
\begin{align}
&\left(\mathcal{D}\cO,\widetilde{\cO}'\right)=\left(\cO,\mathcal{D}^{*}\widetilde{\cO}'\right),
\\
&\left(\cO,\widetilde{\cO}\right)=\int d^dx\cO(x)\widetilde{\cO}(x).
\end{align}
We should stress that here $\cO$ is shorthand for some representation of the conformal group and does not need to obey the spin-statistics theorem. We will define the adjoint by moving from the left to the right, where we recall there are suppressed spinor indices.

It is not hard to see that
\begin{align}
\label{eq: adjoints of weight-shifting operators}
\left(\mathcal{D}^{pq}\right)^{*}\bigg|_{\Delta,\ell}=\zeta^{pq}_{\ell}\mathcal{D}^{p,-q}\bigg|_{\widetilde{\Delta}\mp \frac{p}{2},\ell\pm \frac{q}{2}},
\end{align}
where $p,q=\pm1$ and we have emphasized that the adjoint of a weight-shifting operator depends explicitly on the representation it acts on, although the coefficient $\zeta$ only depends on the spin. 

To calculate the adjoints in practice we go to the Poincar\'e section, or work in physical space. The result is summarized as:
\begin{eqnarray}
&\zeta^{--}_{\ell}=-2\ell, \qquad &\zeta^{-+}_{\ell}=\frac{1}{2\ell+1},
\\
&\zeta^{+-}_{\ell}=2\ell, \qquad  &\zeta^{++}_{\ell}=-\frac{1}{2\ell+1}.
\end{eqnarray}
 
The next ingredient we need is the crossing equation for covariant two-point functions:\footnote{To avoid clutter, we will use $\cD_n^{\pm\pm}$ to denote a weight-shifting operator acting on the $n^{\text{th}}$ operator in the correlator that follows the weight-shifting operator. For example, $\cD_2^{ab}\<\cO_{\De_1,l_1}(x_1)\cO_{\De_2,l_2}(x_2)\cO_{\De_3,l_3}(x_3)\>$ stands for $\cD^{ab}(X_2,S_2)\<\cO_{\De_1,l_1}(X_1,S_1)\cO_{\De_2,l_2}(X_2,S_2)\cO_{\De_3,l_3}(X_3,S_3)\>$ in the embedding space.}
\begin{align}
\mathcal{D}_{1}^{pq}\<\cO_{\Delta,\ell}(x_1)\cO_{\Delta +\frac{p}{2},\ell+\frac{q}{2}}(x_2)\>=\alpha^{pq}_{\Delta,\ell}\mathcal{D}_{2}^{-p,-q}\<\cO_{\Delta,\ell}(x_1)\cO_{\Delta+\frac{p}{2},\ell+\frac{q}{2}}(x_2)\>,
\end{align}
and we find
\bea[eq: Definition of alpha]
\a^{--}_{\De,l}=&{} \frac{i l}{(2 \Delta -3) (\Delta +l-1)},\\
\a^{-+}_{\De,l}=&{} -\frac{i}{2 (2 \Delta -3) (2 l+1)(-\Delta +l+2)},\\
\a^{+-}_{\De,l}=&{} -8 i (\Delta -1) l (-\Delta +l+1),\\
\a^{++}_{\De,l}=&{} \frac{4 i (\Delta -1) (\Delta +l)}{2 l+1}.
\eea 

We can now find the shadow transform of weight-shifted operators:
\begin{align}
\mathbf{S}[\mathcal{D}^{pq}\cO_{\Delta,\ell}](x)&=\int d^{d}y\left(\mathcal{D}^{pq}\cO_{\Delta,\ell}(y)\right)\<\widetilde{\cO}_{\Delta+\frac{p}{2},\ell+\frac{q}{2}}(y)\widetilde{\cO}_{\Delta+\frac{p}{2},\ell+\frac{q}{2}}(x)\>
\nonumber \\ &=\int d^{d}y \zeta^{pq}_{\ell}\alpha^{p,-q}_{\widetilde{\Delta}-\frac{p}{2},\ell+\frac{q}{2}}\cO_{\Delta,\ell}(y)\mathcal{D}^{-a,b}_{2}\<\widetilde{\cO}_{\Delta,\ell}(y)\widetilde{\cO}_{\Delta,\ell}(x)\>
\nonumber \\ &= \zeta^{pq}_{\ell}\alpha^{p,-q}_{\widetilde{\Delta}-\frac{p}{2},\ell+\frac{q}{2}}\mathcal{D}^{-pq}\mathbf{S}[\cO_{\Delta,\ell}](x),
\end{align}
or
\begin{align}
\mathbf{S}[\mathcal{D}^{pq}\cO_{\Delta,\ell}](x)&=\chi^{pq}_{\Delta,\ell}\mathcal{D}^{-p,q}\mathbf{S}[\cO_{\Delta,\ell}],\label{eq:SDtoDS}
\\
\chi^{pq}_{\Delta,\ell}&=\zeta^{pq}_{\ell}\alpha^{p,-q}_{\widetilde{\Delta}-\frac{p}{2},\ell+\frac{q}{2}}. \label{eq:SDtoDS_Coef}
\end{align}

This gives a way to push the shadow transform past the weight-shifting operators. Then to calculate the shadow transform of a three-point tensor structure we will relate the fermionic structures to the bosonic ones. For simplicity we focus on three-point function structures involving one fermion and one scalar:
\begin{align}
\<\psi_{\Delta_1}\phi_{\Delta_2}\cO_{\Delta_3,\ell_3}\>,
\end{align}
where $\cO$ has half-integer spin. We want to use weight-shifting operators to write such a three-point function in terms of a structure involving two scalars:
\begin{align}
\<\phi_{\Delta_{1}'}\phi_{\Delta_{2}}\cO_{\Delta_{3}',\ell_{3}'}\>,
\end{align}
where the third operator $\cO$ now has integer spin. There are two equivalent ways to do this:
\bea
\<\psi_{\Delta_1}\phi_{\Delta_2}\cO_{\Delta_3,\ell_3}\>^a=\sum\limits_{p}\kappa^{a}_{1,p}(\psi_{\Delta_1}\phi_{\Delta_2}\cO_{\Delta_3,\ell_3})\mathcal{D}_{1}^{-p,+}\mathcal{D}_{3}^{-+}\<\phi_{\Delta_1+\frac{p}{2}}\phi_{\Delta_2}\cO_{\Delta_{3}+\frac{1}{2},\ell-\frac{1}{2}}\>,\label{eq: fsO decomposition 1}
\\
\<\psi_{\Delta_1}\phi_{\Delta_2}\cO_{\Delta_3,\ell_3}\>^a=\sum\limits_{p}\kappa^{a}_{2,p}(\psi_{\Delta_1}\phi_{\Delta_2}\cO_{\Delta_3,\ell_3})\mathcal{D}_{1}^{-p,+}\mathcal{D}_{3}^{++}\<\phi_{\Delta_1+\frac{p}{2}}\phi_{\Delta_2}\cO_{\Delta_{3}-\frac{1}{2},\ell-\frac{1}{2}}\>, \label{eq: fsO decomposition 2}
\eea
where each matrix, $\kappa_{1,2}$, is invertible.

The simplest example is calculating the matrix $S^{a}_{b}(\psi_1 [\phi_2]\cO_3)$:
\begin{align}
\<\psi_{\Delta_1}&\mathbf{S}[\phi_{\Delta_2}]\cO_{\Delta_3,\ell_3}\>^{a}\\
&=\mathbf{S}_{2}\sum\limits_{p}\kappa^{a}_{1,p}(\psi_{\Delta_1}\phi_{\Delta_2}\cO_{\Delta_3,\ell_3})\mathcal{D}_{1}^{-p,+}\mathcal{D}_{3}^{-+}\<\phi_{\Delta_1+\frac{p}{2}}\phi_{\Delta_2}\cO_{\Delta_{3}+\frac{1}{2},\ell-\frac{1}{2}}\>,
\nonumber \\ &= \sum\limits_{p}\kappa^{a}_{1,p}(\psi_{\Delta_1}\phi_{\Delta_2}\cO_{\Delta_3,\ell_3})S(\phi_{\Delta_1+\frac{p}{2}}[\phi_{\Delta_2}]\cO_{\Delta_{3}+\frac{1}{2},\ell-\frac{1}{2}})\mathcal{D}_{1}^{-p,+}\mathcal{D}_{3}^{-+}\<\phi_{\Delta_1+\frac{p}{2}}\phi_{\widetilde{\Delta}_2}\cO_{\Delta_{3}+\frac{1}{2},\ell-\frac{1}{2}}\>,
\nonumber \\ &= \sum\limits_{p,b}\kappa^{a}_{1,p}(\psi_{\Delta_1}\phi_{\Delta_2}\cO_{\Delta_3,\ell_3})S(\phi_{\Delta_1+\frac{p}{2}}[\phi_{\Delta_2}]\cO_{\Delta_{3}+\frac{1}{2},\ell-\frac{1}{2}})\left(\kappa^{-1}_{1}(\psi_{\Delta_1}\phi_{\widetilde{\Delta}_{2}}\cO_{\Delta_3,\ell_3})\right)^{p}_{b}\<\psi_{\Delta_1}\phi_{\widetilde{\Delta}_{2}}\cO_{\Delta_{3},\ell_{3}}\>^b.
\end{align}
Here we first wrote the fermionic structure in terms of the bosonic one, then acted with the shadow transform on the simple three point function involving two scalars, and then finally acted with the weight-shifting operators. After acting with the weight-shifting operators we expressed the answer in the original basis.  Therefore, the relevant shadow matrix is
\begin{align}
S^{a}_{b}(\psi_{\Delta_1}[\phi_{\Delta_2}]\cO_{\Delta_3,\ell_3})=\sum\limits_{p}\kappa^{a}_{1,p}(\psi_{\Delta_1}\phi_{\Delta_2}\cO_{\Delta_3,\ell_3})S(\phi_{\Delta_1+\frac{p}{2}}[\phi_{\Delta_2}]\cO_{\Delta_{3}+\frac{1}{2},\ell-\frac{1}{2}})\left(\kappa^{-1}_{1}(\psi_{\Delta_1}\phi_{\widetilde{\Delta}_{2}}\cO_{\Delta_3,\ell_3})\right)^{p}_{b}.
\end{align}
We will follow this strategy for other correlators, i.e. use weight shifting operators and integration by parts to reduce the shadow transform of a fermionic structure to the shadow transform of a simpler bosonic structure. Then we can always re-express the final answer in our original, fermionic basis of three-point functions.

To find the shadow transform of $\psi_{\Delta_1}$, we now have to pass the shadow transform past the weight-shifting operators using (\ref{eq:SDtoDS}) and (\ref{eq:SDtoDS_Coef}):
\begin{align}
S^{a}_{b}(&[\psi_{\Delta_1}]\phi_{\Delta_2}\cO_{\Delta_3,\ell_3}) \\
&=\sum\limits_{p}\kappa^{a}_{1,p}(\psi_{\Delta_1}\phi_2\cO_{\Delta_3,\ell_3})\chi^{-p,+}_{\Delta+\frac{p}{2},0}S\left([\phi_{\Delta_1+\frac{p}{2}}]\phi_{\Delta_2}\cO_{\Delta_3+\frac{1}{2},\ell-\frac{1}{2}}\right)\left(\kappa^{-1}_{1}(\psi_{\widetilde{\Delta}_{1}}\phi_{\Delta_2}\cO_{\Delta_3,\ell_3})\right)^{-p}_{b}. \nonumber
\end{align}
Similarly the shadow transform of the $\cO_{\Delta_3,\ell_3}$ is given by
\begin{align}
S^{a}_{b}(\psi_{\Delta_1}\phi_{\Delta_2}[\cO_{\Delta_3,\ell_3}])=\sum\limits_{p}\kappa^{a}_{1,p}(\psi_{\Delta_1}\phi_2\cO_{\Delta_3,\ell_3})\chi^{-,+}_{\Delta_3+\frac{1}{2},\ell_3-\frac{1}{2}}\left(\kappa^{-1}_{2}(\psi_{\Delta_1}\phi_{\Delta_2}\cO_{\widetilde{\Delta}_3,\ell_3})\right)^{p}_{b}.
\end{align}

Now we turn to correlators containing two fermions. To calculate shadow transforms of $\<\psi_{\Delta_1}\psi_{\Delta_2}\cO_{\Delta_3,\ell_3}\>$ we will utilize the expansion
\begin{align}
\label{eq: expansion of psipsiO}
\<\psi_{\Delta_1}\psi_{\Delta_2}\cO_{\Delta_3,\ell_3}\>^a=\sum\limits_{p,q}\kappa^{a}_{3,pq}(\psi_{\Delta_1}\psi_{\Delta_2}\cO_{\Delta_3,\ell_3})\mathcal{D}_{1}^{-p,+}\mathcal{D}_{2}^{-q,+}\<\phi_{\Delta_1+\frac{p}{2}}\phi_{\Delta_2+\frac{q}{2}}\cO_{\Delta_{3}}\>.
\end{align}
As a reminder, for $\<\psi_1\psi_2\cO_{3}\>^a$ there are four tensor structures and it is possible to invert $\kappa_3$.

In this case the simplest example would be to take the shadow transform of $\cO_{\Delta_3,\ell_3}$, since the differential operators are unaffected by the shadow transform:
\begin{align}
S^{a}_{b}(\psi_1\psi_2[\cO_{\Delta_3,\ell_3}])=\sum\limits_{p,q}\kappa^{a}_{3,pq}(\psi_{\Delta_1}\psi_{\Delta_2}\cO_{\Delta_3,\ell_{3}})S(\phi_{\Delta_1+\frac{p}{2}}\phi_{\Delta_2+\frac{q}{2}}[\cO_{\Delta_3,\ell_3}])\left(\kappa^{-1}_{3}(\psi_{\Delta_1}\psi_{\Delta_2}\cO_{\widetilde{\Delta}_3,\ell_3})\right)^{pq}_{b}.
\end{align}
Finally, the shadow transform of one of the $\psi$ operators yields
\begin{align}
S^{a}_{b}(&[\psi_1]\psi_2\cO_3]) \\
&=\sum\limits_{p,q}\kappa^{a}_{3,pq}(\psi_{\Delta_1}\psi_{\Delta_2}\cO_{\Delta_3,\ell_{3}})S([\phi_{\Delta_1+\frac{p}{2}}]\phi_{\Delta_2+\frac{q}{2}}\cO_{\Delta_3,\ell_3})\chi^{-p,+}_{\Delta_1+\frac{p}{2},0}\left(\kappa^{-1}_{3}(\psi_{\widetilde{\Delta}_1}\psi_{\Delta_2}\cO_{\Delta_3,\ell_3})\right)^{-p,q}_{b},\nonumber
\end{align}
with a similar formula for the shadow transform of $\psi_2$. In the next section we will use these $3d$ shadow coefficients to compute OPE coefficients in mean field theory.

\section{Mean Field Theory OPE Coefficients}
\label{sec:MFTOPE}
In mean field theory we have the factorized correlator
\be 
\<\cO_1\cO_2\cO_3\cO_4\>=\<\cO_1\cO_2\>\<\cO_3\cO_4\>
+(-1)^{\Sigma_{23}}\<\cO_1\cO_3\>\<\cO_2\cO_4\>+\<\cO_1\cO_4\>\<\cO_2\cO_3\>.\ee 
The two-point function $\<\cO_i(x_i)\cO_j(x_j)\>$ is only non-zero when $\cO_i=\cO_j$, but we will leave this restriction implicit in this section.

We will now expand the MFT four-point function in s-channel partial waves and extract the partial wave expansion coefficient. The first term is automatically separated in the s-channel partial wave expansion, so we can focus on the latter two Wick contractions.

The $\<\cO_1\cO_4\>\<\cO_2\cO_3\>$ contraction gives
\be 
\rho_{ab}^{(s)}(\cO)\bigg|_{\<14\>\<23\>} &= \eta^{\cO_5}_{(ac)(bd)}\left(\widetilde{\Psi}^{(s),cd}_{\cO_5},\<\cO_1\cO_4\>\<\cO_2\cO_3\>\right)
\\ &= \eta^{\cO_5}_{(ac)(bd)}\int d^{d}x_{1}...d^{d}x_{5}\<\widetilde{\cO}_{1}\widetilde{\cO}_{2}\widetilde{\cO}_{5}\>^c\<\widetilde{\cO}_3\widetilde{\cO}_4\cO_5\>^d\<\cO_1\cO_4\>\<\cO_2\cO_3\>
\\ &= \eta^{\cO_5}_{(ac)(bd)} (-1)^{\Sigma_{11}+\Sigma_{22}}S^{d}_{e}(\widetilde{\cO}_2[\widetilde{\cO}_1]\cO_5)S^{e}_{f}([\widetilde{\cO}_2]\cO_{1}\cO_5)\left(\<\widetilde{\cO}_{1}\widetilde{\cO}_{2}\widetilde{\cO}_{5}\>^c,\<\cO_2\cO_1\cO_5\>^f\right),
\ee 
where compared with (\ref{eq:GenEucInvSpin}) we have made the replacements $\cO_{3,4}\leftrightarrow \cO_{2,1}$, respectively.  

The other contraction gives a similar result:
\be
\rho_{ab}^{(s)}(\cO)\bigg|_{\<13\>\<24\>}&=(-1)^{\Sigma_{23}} \eta^{\cO_5}_{(ac)(bd)}\left(\widetilde{\Psi}^{(s),cd}_{\cO_5},\<\cO_1\cO_3\>\<\cO_2\cO_4\>\right)
\\
&=(-1)^{\Sigma_{23}+\Sigma_{11}+\Sigma_{22}}\eta^{\cO_5}_{(ac)(bd)}S^{d}_{e}([\widetilde{\cO}_{1}]\widetilde{\cO}_{2}\cO_5)S^{e}_{f}(\cO_1[\widetilde{\cO}_{2}]\cO_5)\left(\<\widetilde{\cO}_{1}\widetilde{\cO}_{2}\widetilde{\cO}_{5}\>^{c},\<\cO_1\cO_2\cO_5\>^f\right).
\ee

The full result in MFT is then
\be
\rho^{(s),\text{MFT}}_{ab}(\cO)=\hat{\delta}_{\cO_1\cO_4}\hat{\delta}_{\cO_2\cO_3}\rho_{ab}^{(s)}(\cO)\bigg|_{\<14\>\<23\>}+\hat{\delta}_{\cO_2\cO_4}\hat{\delta}_{\cO_1\cO_3}\rho_{ab}^{(s)}(\cO)\bigg|_{\<13\>\<24\>}, \label{eq: generic MFT result}
\ee
where $\hat{\delta}_{\cO_1\cO_2}=\delta_{\Delta_1\Delta_2}\delta_{J_1J_2}$ and should not be confused with the delta function on the principal series. Next we will apply this explicitly to various correlators containing fermions in order to calculate their MFT coefficients.

\subsection{$\<\psi\phi\phi\psi\>$}
For the correlator $\<\psi\phi\phi\psi\>$, the identity operator only appears in the t-channel, hence \equref{eq: generic MFT result} becomes
\be 
\label{eq: 2f2s MFT rho}
\rho_{ab}^{(s)}(\cO)=
& (-1)\eta^{\cO}_{(ac)(bd)}S^{d}_{e}(\widetilde{\phi}[\widetilde{\psi}]\cO)S^{e}_{f}([\widetilde{\phi}]\psi\cO)\left(\<\widetilde{\psi}\widetilde{\phi}\widetilde{\cO}\>^c,\<\phi\psi\cO\>^f\right).
\ee 
By explicit calculation, we find
\begin{multline}
\rho^{(s)}_{ac}(\cO)
S^{c}_{b}(\phi\psi[\tl\cO])
=
\frac{\pi ^{5/2} (2 \Delta -3) (2 J+1) 4^{\Delta _{\phi }-2} \Gamma \left(\Delta -\frac{3}{2}\right) (-\Delta +J+2)
	(\Delta +J-1)
}{\Gamma (\Delta -1) \Gamma \left(J+\frac{3}{2}\right)
	\Gamma \left(\Delta _{\psi }-1\right) \Gamma \left(\Delta _{\psi }+\frac{1}{2}\right) }
\\
\x \frac{ \Gamma (J+1)\csc \left(\pi  \Delta _{\psi }\right) \sin \left(\pi  \Delta _{\phi }\right)\csc \left(2 \pi  \Delta _{\phi }\right) \csc (\pi  (J-\Delta ))}{\Gamma \left(2 \Delta _{\phi
	}-1\right) \Gamma (-J+\Delta -1) \Gamma (J+\Delta )}
\left(
\begin{array}{cc}
c_1 & 0 \\
0 & c_2 \\
\end{array}
\right),
\end{multline}
with coefficients
\scriptsize
\bea 
c_1=& -\frac{\Gamma \left(\frac{1}{4} \left(2 J+2 \Delta _{\mathcal{O}\phi \psi }-1\right)\right) \Gamma \left(\frac{1}{4}
	\left(2 J+2 \Delta _{\mathcal{O}\psi \phi }+1\right)\right) \Gamma \left(\frac{1}{4} \left(2 J+2 \Delta _{\phi \psi
		\mathcal{O}}-1\right)\right) \Gamma \left(\frac{1}{4} \left(2 J+2 \Delta +2 \Delta _{\phi }+2 \Delta _{\psi
	}-5\right)\right)}{\Gamma \left(\frac{1}{4} \left(2 J-2 \Delta _{\mathcal{O}\phi \psi }+5\right)\right) \Gamma
	\left(\frac{1}{4} \left(2 J-2 \Delta _{\mathcal{O}\psi \phi }+7\right)\right) \Gamma \left(\frac{1}{4} \left(2 J-2
	\Delta _{\phi \psi \mathcal{O}}+5\right)\right) \Gamma \left(\frac{1}{4} \left(2 J-2 \Delta -2 \Delta _{\phi }-2 \Delta
	_{\psi }+13\right)\right)},
\\
c_2=& \frac{\Gamma \left(\frac{1}{4} \left(2 J+2 \Delta _{\mathcal{O}\phi \psi }+1\right)\right) \Gamma \left(\frac{1}{4}
	\left(2 J+2 \Delta _{\mathcal{O}\psi \phi }-1\right)\right) \Gamma \left(\frac{1}{4} \left(2 J+2 \Delta _{\phi \psi
		\mathcal{O}}+1\right)\right) \Gamma \left(\frac{1}{4} \left(2 J+2 \Delta +2 \Delta _{\phi }+2 \Delta _{\psi
	}-7\right)\right)}{\Gamma \left(\frac{1}{4} \left(2 J-2 \Delta _{\mathcal{O}\phi \psi }+7\right)\right) \Gamma
	\left(\frac{1}{4} \left(2 J-2 \Delta _{\mathcal{O}\psi \phi }+5\right)\right) \Gamma \left(\frac{1}{4} \left(2 J-2
	\Delta _{\phi \psi \mathcal{O}}+7\right)\right) \Gamma \left(\frac{1}{4} \left(2 J-2 \Delta -2 \Delta _{\phi }-2 \Delta
	_{\psi }+11\right)\right)},
\eea 
\normalsize
where $\De_{abc}\equiv\De_a+\De_b-\De_c$.

We see that the first component has residues at $\Delta=\Delta_\psi+\Delta_\phi+J-\half+2n$, which corresponds to the double twist families $[\phi\psi_\a]_{n,J}$. In contrast, the last component has residues at $\Delta=\Delta_\psi+\Delta_\phi+J+\half+2n$, which corresponds to the double twist families $[\phi\partial_{\a\b}\psi^\b]_{n,J}$.

By taking their respective residues, we can find the OPE coefficients. For the leading ($n=0$) tower, they read as
\footnotesize
\bea 
P_{11}^{(s)}\left([\phi\psi]_{0,J}\right)=&
-\frac{\Gamma \left(J+\Delta _{\psi }\right) \Gamma \left(J+\Delta _{\phi }-\frac{1}{2}\right) \Gamma
	\left(J+\Delta _{\phi }+\Delta _{\psi }-1\right)}{\Gamma \left(J+\frac{1}{2}\right) \Gamma \left(\Delta _{\psi
	}+\frac{1}{2}\right) \Gamma \left(\Delta _{\phi }\right) \Gamma \left(2 J+\Delta _{\phi }+\Delta _{\psi
	}-\frac{3}{2}\right)},
\\
P_{22}^{(s)}\left([\phi\partial\psi]_{0,J}\right)=&
-\frac{ (2 J+1) 2^{3-2 \Delta _{\phi }} \Gamma (J+1) \Gamma \left(\Delta _{\psi }\right) \Gamma
	\left(J+\Delta _{\psi }\right) \left(\Delta _{\psi }+\Delta _{\phi }+J-1\right)}{\sqrt{\pi } \left(2 \Delta _{\psi
	}-1\right) \Gamma (2 J+3) \Gamma \left(2 \Delta _{\psi }-2\right) \Gamma \left(\Delta _{\phi }\right) \Gamma
	\left(J+\Delta _{\phi }+\Delta _{\psi }-\frac{1}{2}\right) }
\nn\\&\x\frac{\Gamma \left(J+\Delta _{\phi
	}+\frac{1}{2}\right) \Gamma \left(2 J+2 \Delta _{\phi }+2 \Delta _{\psi }-3\right)}{\Gamma \left(2 J+\Delta _{\phi }+\Delta _{\psi
	}-\frac{1}{2}\right)}.
\eea 
\normalsize
Results for higher-twist towers are given in appendix~\ref{app:MFT_higher_twist}.

At large $J$, the leading behavior is
\bea[eq: MFT coefficients at large spin for two fermion and two scalars]
P_{11}^{(s)}\left([\phi\psi]_{0,J}\right)\sim&
-\frac{\sqrt{\pi } 2^{-\Delta_ \psi -\Delta_ \phi -2 J+\frac{5}{2}} J^{\Delta_ \psi +\Delta_ \phi
		-1}}{\Gamma \left(\Delta_ \psi +\frac{1}{2}\right) \Gamma (\Delta _\phi )},
\\
P_{22}^{(s)}\left([\phi\partial\psi]_{0,J}\right)\sim&
-\frac{\sqrt{\pi }  \left(\Delta _{\psi }-1\right) 2^{-\Delta _{\psi }-\Delta _{\phi }-2 J+\frac{3}{2}}
	J^{\Delta _{\psi }+\Delta _{\phi }-2}}{\Gamma \left(\Delta _{\psi }+\frac{1}{2}\right) \Gamma \left(\Delta _{\phi
	}\right)}.
\eea 

As a reminder, in our partial wave convention we have defined $P^{(s)}_{ab}=\lambda_{\psi\phi\cO,a}\lambda_{\phi\psi\cO,b}$. We are working with three-point structures such that $\lambda_{\phi\psi\cO,a}=(-1)^{J-\half}\lambda_{\psi\phi\cO,a}$, so we see that
\bea 
\left(\lambda_{\psi\phi\cO,1}\right)^2\sim& -(-1)^{J-\half}
\frac{\sqrt{\pi }2^{-\Delta_ \psi -\Delta_ \phi -2 J+\frac{5}{2}} J^{\Delta_ \psi +\Delta_ \phi
		-1}}{\Gamma \left(\Delta_ \psi +\frac{1}{2}\right) \Gamma (\Delta _\phi )},
\label{eq: OPE coefficient squared for fermion scalar parity even double twist family}
\\
\left(\lambda_{\psi\phi\cO,2}\right)^2\sim& -(-1)^{J-\half}
\frac{\sqrt{\pi } \left(\Delta _{\psi }-1\right) 2^{-\Delta _{\psi }-\Delta _{\phi }-2 J+\frac{3}{2}}
	J^{\Delta _{\psi }+\Delta _{\phi }-2}}{\Gamma \left(\Delta _{\psi }+\frac{1}{2}\right) \Gamma \left(\Delta _{\phi
	}\right)}.
\eea 

In the free theory limit with $\{\De_\phi,\De_\psi\}\rightarrow\{\half,1\}$, these become
\be 
\left(\lambda_{\psi\phi\cO,1}\right)^2\sim -(-1)^{J-\half}4^{1-J}\sqrt{\frac{J}{\pi}}\;,\quad \left(\lambda_{\psi\phi\cO,2}\right)^2\sim 0.
\ee 
We see that this matches the results of \cite{Iliesiu:2015akf} once we take the normalizations into account. 
Specifically, they normalize their operators as
\begin{align} 
\<\cO_{\Delta,\ell}(X_1,S_1)\cO_{\Delta,\ell}(X_2,S_2)\> &= c^{\text{there}}_{\cO}i^{2l}\frac{\<S_1S_2\>^{2l}}{X_{12}^{2\De+2l}}, 
\\
c^{\text{there}}_{\cO} &=\frac{4^{\Delta -1} (-1)^{\frac{1}{2}-J} \Gamma \left(J+\frac{3}{2}\right)}{\left(\frac{1}{2}\right)_{J+\frac{1}{2}}},
\end{align}
while we take $c^{\text{here}}_{\cO}=1$.
Thus we have
\be 
c_\cO^{\text{there}}\left(\lambda_{\psi\phi\cO,1}\right)^2\sim -4 J,
\ee 
which can be compared with Eq. (4.4) of \cite{Iliesiu:2015akf}.

\subsection{$\<\psi_1\psi_2\psi_2\psi_1\>$}
Now we turn to correlators containing four fermions. For a correlator of the form $\<\psi_1\psi_2\psi_2\psi_1\>$, the identity operator only appears in the $14\rightarrow23$ channel, hence \equref{eq: generic MFT result} becomes
\be 
\rho_{ab}^{(s)}(\cO)
=
\eta^{\cO}_{(ac)(bd)} S^{d}_{e}(\widetilde{\psi}_2[\widetilde{\psi}_1]\cO)S^{e}_{f}([\widetilde{\psi}_2]\psi_{1}\cO)\left(\<\widetilde{\psi}_{1}\widetilde{\psi}_{2}\widetilde{\cO}\>^c,\<\psi_2\psi_1\cO\>^f\right).
\ee 

When we calculate $\rho$, which we will not reproduce here, we see that it is block-diagonal, with the upper $2\x2$ block having poles at $\Delta=\Delta_{\psi_1}+\Delta_{\psi_2}+J-1+2n$, and with the lower $2\x2$ block having poles at $\Delta=\Delta_{\psi_1}+\Delta_{\psi_2}+J+2n$. These correspond to the double-twist families $[\psi_{1\a}\psi_{2\b}]_{n,J}$ and $[\psi_{1\a}\psi_2^\a]_{n,J}$ respectively. We then can read off the explicit results for OPE coefficients. 

For the leading tower ($n=0$) we obtain:
\small
\be 
{}&P^{(s)}_{[\psi_{1\a}\psi_{2\b}]_{0,J}}
=
-\frac{\Gamma \left(J+\Delta _1-\frac{1}{2}\right) \Gamma \left(J+\Delta _2-\frac{1}{2}\right) \Gamma \left(J+\Delta
	_1+\Delta _2-1\right)}{\Gamma \left(\Delta _1+\frac{1}{2}\right) \Gamma \left(\Delta _2+\frac{1}{2}\right) \Gamma (J)
	\Gamma \left(2 J+\Delta _1+\Delta _2-2\right)}\begin{pmatrix}
	0&0\\0&1
\end{pmatrix},
\\
&P^{(s)}_{[\psi_{1\a}\psi_2^\a]_{0,J}}
=
-\frac{\left(\Delta _1+\Delta _2+2 J-1\right) \Gamma \left(J+\Delta _1-\frac{1}{2}\right) \Gamma \left(J+\Delta
	_2-\frac{1}{2}\right) \Gamma \left(J+\Delta _1+\Delta _2-2\right)}{4 \Gamma \left(\Delta _1+\frac{1}{2}\right) \Gamma
	\left(\Delta _2+\frac{1}{2}\right) \Gamma (J+2) \Gamma \left(2 J+\Delta _1+\Delta _2\right)}
\\
&\x \begin{pmatrix}
	\left(\Delta _1+\Delta _2-2\right) \left(\Delta _1+J-\frac{1}{2}\right) \left(\Delta _2+J-\frac{1}{2}\right) +c_1& \left(\Delta _1-\Delta _2\right) J (J+1) \left(\Delta _1+\Delta _2+J-\frac{3}{2}\right)\\
	-\left(\Delta _1-\Delta _2\right) J (J+1) \left(\Delta _1+\Delta _2+J-\frac{3}{2}\right)& \left(\Delta _1+\Delta _2-2\right) \left(\Delta _1+J-\frac{1}{2}\right) \left(\Delta _2+J-\frac{1}{2}\right)-c_1
\end{pmatrix},
\ee 
\normalsize
with
\begin{multline}
c_1= \frac{1}{4} \left(2 \Delta _1-1\right) \left(\Delta _1+\Delta _2-2\right) \left(2 \Delta _2-1\right)\\
+\left(\Delta
_1+\Delta _2-2\right) J^3+\left(\Delta _1^2+\left(4 \Delta _2-\frac{9}{2}\right) \Delta _1+\Delta _2^2-\frac{9 \Delta
	_2}{2}+3\right) J^2\\+\left(\Delta _1+\Delta _2-1\right) \left(\Delta _1 \left(2 \Delta _2-1\right)-\Delta _2\right) J.
\end{multline}
Results for higher-twist towers are given in appendix~\ref{app:MFT_higher_twist}.

\subsection{$\<\psi\psi\psi\psi\>$}
For identical fermions $\<\psi\psi\psi\psi\>$, both the t- and u-channels contribute, hence we have
\be 
\rho_{ab}^{(s)}(\cO)
=\, &\eta^{\cO}_{(ac)(bd)} S^{d}_{e}(\widetilde{\psi}[\widetilde{\psi}]\cO)S^{e}_{f}([\widetilde{\psi}]\psi\cO)\left(\<\widetilde{\psi}\widetilde{\psi}\widetilde{\cO}\>^c,\<\psi\psi\cO\>^f\right)
\nonumber 
\\
&-\eta^{\cO}_{(ac)(bd)}S^{d}_{e}([\widetilde{\psi}]\widetilde{\psi}\cO)S^{e}_{f}(\psi[\widetilde{\psi}]\cO)\left(\<\widetilde{\psi}\widetilde{\psi}\widetilde{\cO}\>^{c},\<\psi\psi\cO\>^f\right).
\ee 

For the leading ($n=0$) tower, the explicit expressions for the OPE coefficients are as follows:
\small
\be
\label{eq: MFT coefficients for four fermions}
P^{(s)}_{[\psi_\a\psi_\b]_{n,J}}
=& -\frac{2 (-1)^J \left(\Delta _{\psi }+J-1\right) \Gamma \left(J+\Delta _{\psi
	}-\frac{1}{2}\right){}^2 \Gamma \left(J+2 \Delta _{\psi }-1\right)}{\Gamma (J) \Gamma \left(\Delta _{\psi
	}+\frac{1}{2}\right){}^2 \Gamma \left(2 J+2 \Delta _{\psi }-1\right)}\begin{pmatrix}
	0&0\\0&1+(-1)^J
\end{pmatrix},
\\
P^{(s)}_{[\psi_\a\psi^\a]_{n,J}}
=&-
\frac{(-1)^{J} 2^{-2 J} J \Gamma \left(\Delta _{\psi }\right) \Gamma \left(J+\Delta _{\psi }+\frac{1}{2}\right) \Gamma
	\left(J+2 \Delta _{\psi }\right)}{\left(1-2 \Delta _{\psi }\right){}^2 \Gamma (J+1) \Gamma \left(\Delta _{\psi
	}-\frac{1}{2}\right) \Gamma \left(2 \Delta _{\psi }-2\right) \Gamma \left(J+\Delta _{\psi }\right)}\left(
\begin{array}{cc}
	\frac{1+(-1)^J}{J \left(J+2 \Delta _{\psi }-2\right)} & 0 \\
	0 & \frac{1-(-1)^{J}}{(J+1) \left(J+2 \Delta _{\psi }-1\right)} \\
\end{array}
\right).
\ee
\normalsize
After an appropriate change of basis, these results match perfectly to those calculated using the lightcone bootstrap at large $J$~\cite{Albayrak:2019gnz}.

\section{Analytic Bootstrap for Fermions}
\label{sec:Inversion_Isolated_Ops}

Now we will return to the problem of inverting a single partial wave (or block) with external spinning operators. This will allow us to compute corrections to the anomalous dimensions of double twist operators. As a reminder, the general form of the $6j$ symbol is given by
\begin{align}
\sixj{\cO_1}{\cO_2}{\cO_3}{\cO_4}{\cO_5}{\cO_6}^{abcd} &=\left(\widetilde{\Psi}^{(s),ab}_{\cO_5},\Psi^{(t),cd}_{\cO_6}\right) \nonumber\\
&=\int d^{d}x_{1}...d^{d}x_{6}\<\widetilde{\cO}_1\widetilde{\cO}_2\widetilde{\cO}_{5}\>\<\widetilde{\cO}_3\widetilde{\cO}_4\cO_5\>\<\cO_3\cO_2\cO_6\>\<\cO_1\cO_4\widetilde{\cO}_{6}\>. \label{eqn:Gen6jSymbol2}
\end{align}
Our strategy will in a way be the reverse of the scalar case \cite{Liu:2018jhs}. Instead of using the Lorentzian inversion formula, we will follow the strategy outlined in \cite{Karateev:2017jgd} and use weight-shifting operators to calculate the $6j$ symbol for external fermions in terms of the $6j$ symbol for external scalars. We then use the expression (\ref{eq:6j_symbol_split_Scalars}), which splits the scalar $6j$ symbol into two pieces from inverting the physical block and its shadow to find the corresponding split for the fermionic $6j$ symbol.

To start, we use the results of \cite{Karateev:2017jgd} to write the t-channel spinning partial wave as a differential operator acting on the partial wave for external scalars: 
\begin{align}
\Psi^{(t),ab}_{\cO}(x_i)=\mathfrak{D}^{ab}_{t}\Psi^{(t),\text{scalar}}_{\cO'}(x_i),
\end{align}
where $\Psi^{(t),\text{scalar}}_{\cO'}(x_i)$ is a partial wave for four external scalars, $\<\f_3\f_2\f_1\f_4\>$. We are being very schematic here and it should be understood that $\mathfrak{D}^{ab}_{t}$ is a sum of multiple weight-shifting operators. For each term in the sum we may have to choose a different shifted operator $\cO'$, with scaling dimension and spin shifted from $\cO$, as well as different external scaling dimensions $\Delta_i$ of the scalar partial wave.

Given this expression we can simplify the spinning $6j$ symbol by taking the adjoint of this operator:
\begin{align}
\left(\widetilde{\Psi}^{(s),ab}_{\cO_5},\Psi^{(t),ab}_{\cO_{6}}\right)&=\left(\mathfrak{D}^{ab}_{s}\widetilde{\Psi}^{(s),\text{scalar}}_{\cO'_{5}},\mathfrak{D}^{ab}_{t}\Psi^{(t),\text{scalar}}_{\cO'_{6}}\right) \nonumber
\\ &=\left(\widetilde{\Psi}^{(s),\text{scalar}}_{\cO'_{5}},\mathfrak{D}^{*,ab}_{s}\mathfrak{D}^{ab}_{t}\Psi^{(t),\text{scalar}}_{\cO'_{6}}\right).
\end{align}
The two weight-shifting operators acting on the t-channel conformal partial wave then give a linear combination of undifferentiated t-channel conformal partial waves, at the price of more shifts for the internal and external labels. In the end we are left with an equality of the form
\begin{align}
\sixj{\cO_1}{\cO_2}{\cO_3}{\cO_4}{\cO_5}{\cO_6}^{abcd}=\sum\limits_{\f_i,\cO'_{5,6}}
\sixjDecomp{abcd}{\cO_1}{\cO_2}{\cO_3}{\cO_4}{\cO_5}{\cO_6}{\f_1}{\f_2}{\f_3}{\f_4}{\cO'_{5}}{\cO'_{6}}
\sixj{\f_1}{\f_2}{\f_3}{\f_4}{\cO'_{5}}{\cO'_{6}}, \label{eq:6jDecompSpintoScalars}
\end{align}
where the sum runs over some set of scaling dimensions for the fictitious external scalars $\f_i$, whose dimensions are related to $\cO_i$ by some (half-)integer shift, and over both scaling dimensions and spins for $\cO'_{5,6}$, which are again related to the $\cO_{5,6}$ by (half-)integer shifts in both labels. We now turn to how to compute these coefficients.

\subsection{Spinning Down the $6j$ Symbol}

The general strategy to compute the decomposition factors $J^{abcd}$ in \equref{eq:6jDecompSpintoScalars} can be systematized by the following procedure:
\begin{itemize}
	\item Write a three-point structure in terms of weight-shifting operators acting on three-point structures of lower spins.
	\item Use integration by parts and crossing symmetry of covariant three-point structures to move the weight-shifting operators such that they act on the same operator.
	\item By using irreducibility of the representations, the weight-shifting operators become multiples of the identity.
	\item Repeat until all three-point structures are of the form $\<\f\f\cO\>$, i.e.~we are left with three-point functions involving two scalars.
\end{itemize}
Below, we will unpack this procedure further by detailing each step in the explicit decomposition of the $6j$ symbol of two external scalars and two external fermions.

In our conventions, the $6j$-symbol for the correlator $\<\phi\phi\psi\psi\>$ reads as
\be 
\label{eq: fermion-scalar 6j symbol}
\left\{
\begin{matrix}
	\f_1 & \f_2& \cO_6 \\
	\psi_3 & \psi_4 & \cO_5
\end{matrix}
\right\}^{\uniq s_2t_1t_2}
=\int d^dx_1...d^dx_6\<\tl\f_1\tl\f_2\tl\cO_5\>\.\< \tl\psi_3\tl\psi_4\cO_5\>^{s_2}\.\<\psi_3\phi_2\cO_6\>^{t_1}\.\< \f_1\psi_4\tl\cO_6\>^{t_2},
\ee 
where the first three-point structure is already in the appropriate form for a $6j$ symbol of four external scalars, so all we need to do is to massage the remaining structures. As the first step, we use \equref{eq: expansion of psipsiO} to rewrite $\<\tl\psi_3\tl\psi_4\cO_5\>^{s_2}$:\footnote{In this section, we use the following shorthand notation for brevity: $\f^a_i\equiv\f_{\De_i+a}$, $\tl\f^a_i\equiv\f_{3-\De_i+a}$, \mbox{$\cO_i^{a,b}\equiv\cO_{\De_i+a,l_i+b}$}, and $\tl\cO_i^{a,b}\equiv\cO_{3-\De_i+a,l_i+b}$.}
\begin{multline}
\label{eq: decomposition of fermionic 6j symbol, step 0}
\int d^dx_3d^dx_4d^dx_6\<\tl\psi_3\tl\psi_4\cO_5\>^{s_2}\.\<\psi_3\f_2\cO_6\>^{t_1}\.\< \f_1\psi_4\tl\cO_6\>^{t_2}\\=\int d^dx_3d^dx_4d^dx_6\sum\limits_{a,b}\kappa^{s_2}_{3,ab}(\tl\psi_3\tl\psi_4\cO_5)
\cD_1^{-a,+}\cD_2^{-b,+}\<\tl\f_3^a\tl\f_4^b\cO_5 \>
\<\psi_3\f_2\cO_6\>^{t_1}\.
\<\f_1\psi_4\tl\cO_6\>^{t_2}.
\end{multline}
After that, we integrate by parts to obtain
\begin{multline}
\label{eq: decomposition of fermionic 6j symbol, step 1}
\int d^dx_3d^dx_4d^dx_6\<\tl\psi_3\tl\psi_4\cO_5\>^{s_2}\.\<\psi_3\f_2\cO_6\>^{t_1}\.\< \f_1\psi_4\tl\cO_6\>^{t_2}\\=\int d^dx_3d^dx_4d^dx_6\sum\limits_{a,b}\kappa^{s_2}_{3,ab}(\tl\psi_3\tl\psi_4\cO_5)
\<\tl\f_3^a\tl\f_4^b\cO_5 \>
\.\left[\left(\cD_1^{-a,+}\right)^*_{A}\<\psi_3\f_2\cO_6\>^{t_1}\right]\.
\left[\left(\cD_2^{-b,+}\right)^{A*}
\<\f_1\psi_4\tl\cO_6\>^{t_2}\right],
\end{multline}
where we are showing the spinor indices of weight-shifting operators explicitly. With \equref{eq: adjoints of weight-shifting operators}, the equation further reduces to
\begin{align}
\label{eq: decomposition of fermionic 6j symbol, step 2}
\int d^dx_3&d^dx_4d^dx_6\<\tl\psi_3\tl\psi_4\cO_5\>^{s_2}\.\<\psi_3\f_2\cO_6\>^{t_1}\.\< \f_1\psi_4\tl\cO_6\>^{t_2}
\nonumber
\\=&\int d^dx_3d^dx_4d^dx_6\sum\limits_{a,b}\kappa^{s_2}_{3,ab}(\tl\psi_3\tl\psi_4\cO_5)
\zeta_{0}^{-a,+}\zeta_{0}^{-b,+}
\<\tl\f_3^a\tl\f_4^b\cO_5 \>
\nonumber
\\ &\.\left[\left(\cD_1^{-a,-}\right)_{A}\<\psi_3\f_2\cO_6\>^{t_1}\right]\.
\left[E_{\phi\psi\cO\rightarrow \cO\phi\psi}^{t_2t_2'}\left(\cD_3^{-b,-}\right)^{A}
\<\tl\cO_6\f_1\psi_4\>^{t_2'}\right],
\end{align}
where we defined the exchange matrix $E$
\be 
\<\cO_2\cO_1\cO\>^a
=E^{ab}_{21\cO\to 12\cO}\<\cO_1\cO_2\cO\>^b
\ee 
to reorder the last tensor structure so that the weight-shifting operator acts on the third operator.\footnote{We do this because we will use the convention of \cite{Karateev:2017jgd} for finite-dimensional representations, where the weight-shifting operator acts on the first (third) operator in the $s$ ($t$) channel.} We can then use the crossing for conformally covariant three-point structures as derived in \cite{Karateev:2017jgd}, which reads as
\be \label{eq:3ptcrossing}
\left(\cD^{-a,-b}_3\right)^A\<\cO_1\cO_2\cO_3^{a,b}\>^{m}=\sum_{c,d,n}
\left\{
\begin{matrix}
	\cO_1 & \cO_2 & \cO_1^{c,d}\\
	\cO_3 & S & \cO_3^{a,b}
\end{matrix}
\right\}^{mn}
\left(\cD^{-c,-d}_1\right)^A\<\cO_1^{c,d}\cO_2\cO_3\>^{n}
\ee 
for the fermionic representation of weight-shifting operators in $3d$. Thus we obtain
\begin{align}
\label{eq: decomposition of fermionic 6j symbol, step 3}
\int d^dx_3&d^dx_4d^dx_6\<\tl\psi_3\tl\psi_4\cO_5\>^{s_2}\.\<\psi_3\f_2\cO_6\>^{t_1}\.\< \f_1\psi_4\tl\cO_6\>^{t_2}
\nonumber \\=& \int d^dx_3d^dx_4d^dx_6\sum\limits_{a,b}\kappa^{s_2}_{3,ab}(\tl\psi_3\tl\psi_4\cO_5)
\zeta_{0}^{-a,+}\zeta_{0}^{-b,+}
\<\tl\f_3^a\tl\f_4^b\cO_5 \>
\nonumber \\\. &\left[\left(\cD_1^{-a,-}\right)_{A}\<\psi_3\f_2\cO_6\>^{t_1}\right]\.
\left[E_{\phi\psi\cO\rightarrow \cO\phi\psi}^{t_2t_2'}\sum_{c,d}
\left\{
\begin{matrix}
\tl\cO_6 & \f_1 & \tl\cO_6^{c,d}\\
\f_4^{-b} &S& \psi_4 
\end{matrix}
\right\}^{t_2'\uniq}
\left(\cD^{-c,-d}_1\right)^A
\<\tl\cO_6\f_1\phi_4^{-b}\>\right].
\end{align}

We can integrate by parts and use the crossing again to get both weight-shifting operators to act on the same operator:
\begin{align}
\int & d^dx_3d^dx_4d^dx_6\<\tl\psi_3\tl\psi_4\cO_5\>^{s_2}\.\<\psi_3\f_2\cO_6\>^{t_1}\.\< \f_1\psi_4\tl\cO_6\>^{t_2}
\nonumber \\
=&\int d^dx_3d^dx_4d^dx_6\sum\limits_{a,b,c,d}\kappa^{s_2}_{3,ab}(\tl\psi_3\tl\psi_4\cO_5)
\zeta_{0}^{-a,+}\zeta_{0}^{-b,+}\zeta_{l_t+d}^{-c,-d}\left\{
\begin{matrix}
\tl\cO_6 & \f_1 & \tl\cO_6^{c,d}\\
\f_4^{-b} &S& \psi_4 
\end{matrix}
\right\}^{t_2'\uniq} \x E_{\phi\psi\cO\rightarrow \cO\phi\psi}^{t_2t_2'}
\nonumber \\ & \<\tl\f_3^a\tl\f_4^b\cO_5 \>\.\bigg[
\sum_{e,f,n}
\left\{
\begin{matrix}
\psi_3 & \f_2 & \cO_3^{e,f}\\
\cO_6^{-c,d} & S & \cO_6
\end{matrix}
\right\}^{t_1n}
\left(\cD_1^{-a,-}\right)_{A}
\left(\cD^{-e,-f}_1\right)^A\<\cO_3^{e,f}\f_2\cO_6^{-c,d}\>^{n}
\bigg]\.
\<\tl\cO_6\f_1\phi_4^{-b}\>. \label{eq: decomposition of fermionic 6j symbol, step 4}
\end{align}

By the irreducibility of the representations, we have $\left(\cD^{-a,-b}\right)_A\left(\cD^{c,d}\right)^A\cO_{\De,l}= \delta^{ac}\delta^{bd}\beta^{\De,l}_{ab}\cO_{\De,l}$  with which we finally obtain\footnote{One can derive $\beta$ by acting on the two point function with weight shifting operators in embedding space. In our conventions, we have 
\be 
\beta_{a,b}^{\De,l}\equiv b (a+2 \Delta -3) (b+2 l+1) (2 a b \Delta +a (2 b+1) (a+b-2)+2 l)\quad\text{ for } a,b=\pm 1 \.
\ee 
}
\begin{align}
\int & d^dx_3d^dx_4d^dx_6\<\tl\psi_3\tl\psi_4\cO_5\>^{s_2}\.\<\psi_3\f_2\cO_6\>^{t_1}\.\< \f_1\psi_4\tl\cO_6\>^{t_2}
\nonumber \\ &=d^dx_3d^dx_4d^dx_6\sum\limits_{a,b,c,d}\beta_{a,+}^{\Delta_3-a,0}\kappa^{s_2}_{3,ab}(\tl\psi_3\tl\psi_4\cO_5)
\zeta_{0}^{-a,+}\zeta_{0}^{-b,+}\zeta_{l_t+d}^{-c,-d}E_{\phi\psi\cO\rightarrow \cO\phi\psi}^{t_2t_2'}
\nonumber \\ &\x  \left\{
\begin{matrix}
\tl\cO_6 & \f_1 & \tl\cO_6^{c,d}\\
\f_4^{-b} &S& \psi_4 
\end{matrix}
\right\}^{t_2'\uniq}\left\{
\begin{matrix}
\psi_3 & \f_2 & \phi_3^{-a}\\
\cO_6^{-c,d} & S & \cO_6
\end{matrix}
\right\}^{t_1\uniq} \<\tl\f_3^a\tl\f_4^b\cO_5 \>\.
\<\phi_3^{-a}\f_2\cO_6^{-c,d}\>^{n}
\.
\<\f_1\phi_4^{-b}\tl\cO_6^{c,d}\>. \label{eq: decomposition of fermionic 6j symbol, step 5}
\end{align}
Inserting this into \equref{eq: fermion-scalar 6j symbol}, we get
\be
\label{eq: 6j decomposition of 2f2s}
\left\{
\begin{matrix}
	\f_1 & \f_2& \cO_6 \\
	\psi_3 & \psi_4 & \cO_5
\end{matrix}
\right\}^{\uniq s_2t_1t_2}
=
\sum\limits_{a,b,c,d}
\sixjDecomp{\uniq s_2t_1t_2}{\f_1}{\f_2}{\psi_3}{\psi_4}{\cO_5}{\cO_6}{\f_1}{\f_2}{\f_3^{-a}}{\f_4^{-b}}{\cO_5}{\cO_6^{-c,d}}
\left\{
\begin{matrix}
	\f_1 &\f_2& \cO_6^{-c,d} \\
	\f_3^{-a} & \f_4^{-b} & \cO_5
\end{matrix}
\right\},
\ee
where
\begin{multline}
\label{eq: J for 2f2s}
\sixjDecomp{\uniq s_2t_1t_2}{\f_1}{\f_2}{\psi_3}{\psi_4}{\cO_5}{\cO_6}{\f_1}{\f_2}{\f_3^{-a}}{\f_4^{-b}}{\cO_5}{\cO_6^{-c,d}}=\beta_{a,+}^{\Delta_3-a,0}\kappa^{s_2}_{3,ab}(\tl\psi_3\tl\psi_4\cO_5)
\zeta_{0}^{-a,+}\zeta_{0}^{-b,+}\zeta_{l_t+d}^{-c,-d}\\\x E_{\phi\psi\cO\rightarrow \cO\phi\psi}^{t_2t_2'}
  \left\{
\begin{matrix}
\tl\cO_6 & \f_1 & \tl\cO_6^{c,d}\\
\f_4^{-b} &S& \psi_4 
\end{matrix}
\right\}^{t_2'\uniq}\left\{
\begin{matrix}
\psi_3 & \f_2 & \phi_3^{-a}\\
\cO_6^{-c,d} & S & \cO_6
\end{matrix}
\right\}^{t_1\uniq}.
\end{multline}

Note that if $l_5=0$, there are only 2 independent structures for $\<\psi_3\psi_4\phi_5\>^{s_2}$, hence we need to take a $2\times 2$ invertible submatrix of $\kappa_3$. We can do this by restricting to the structures $\<\psi_3\psi_4\phi_5\>^{1,3}$ in \equref{eq: 3pt structures}, and fixing $b=\half$ in \equref{eq: 6j decomposition of 2f2s} instead of summing over $b=\pm\half$.

\begin{figure}
	\centering
	$\begin{aligned}
	\includegraphics[scale=.55]{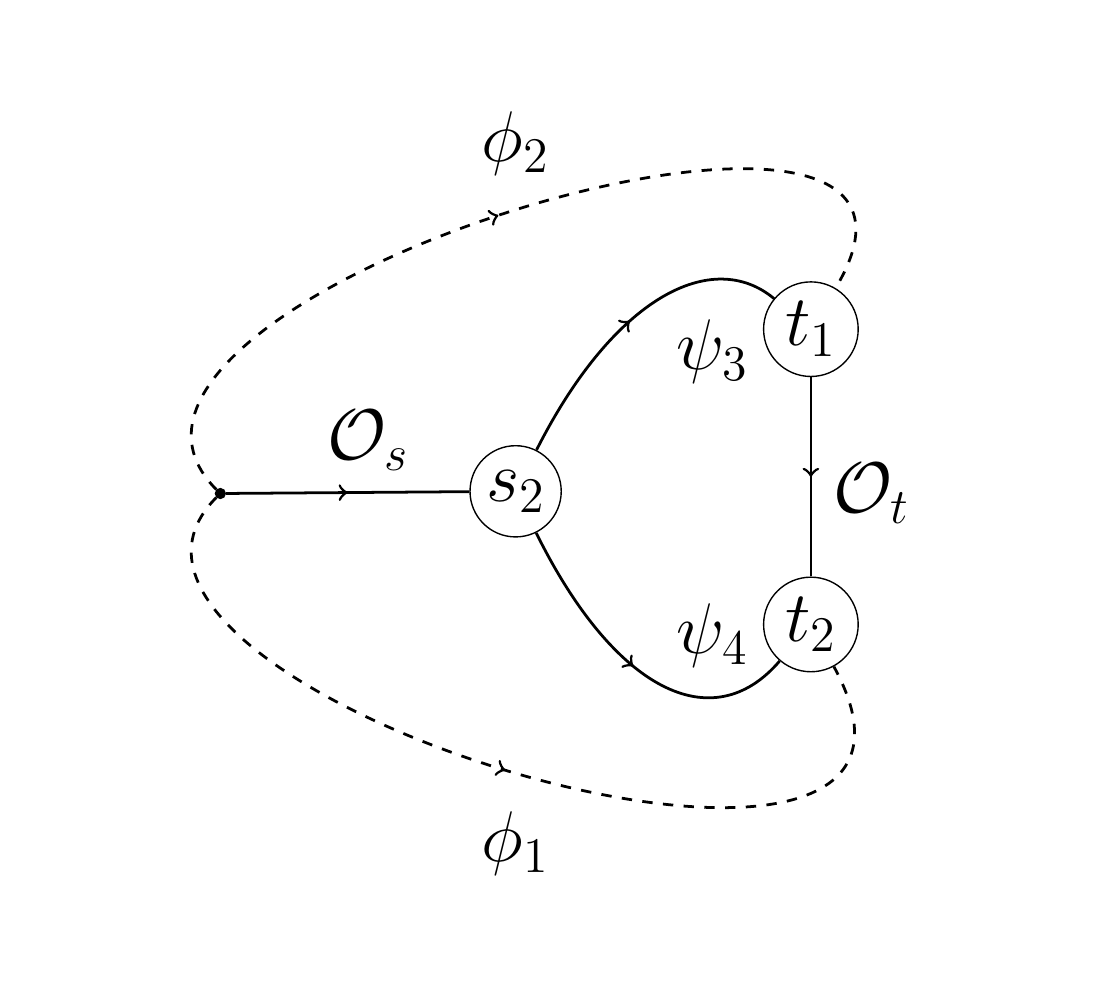}
	\end{aligned}$ $\Rightarrow$ $\begin{aligned}
	\includegraphics[scale=.55]{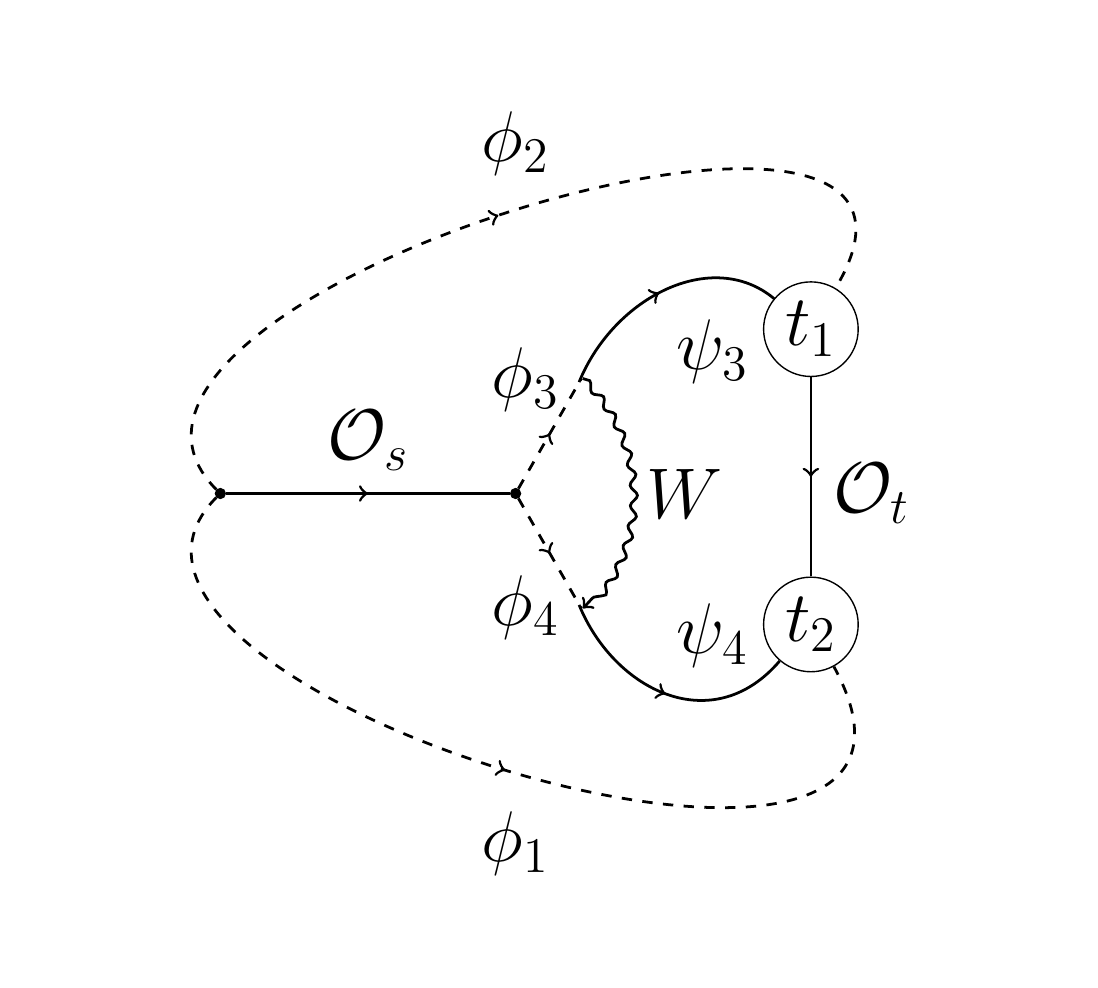}
	\end{aligned}$ $\Rightarrow$ $\begin{aligned}
	\includegraphics[scale=.55]{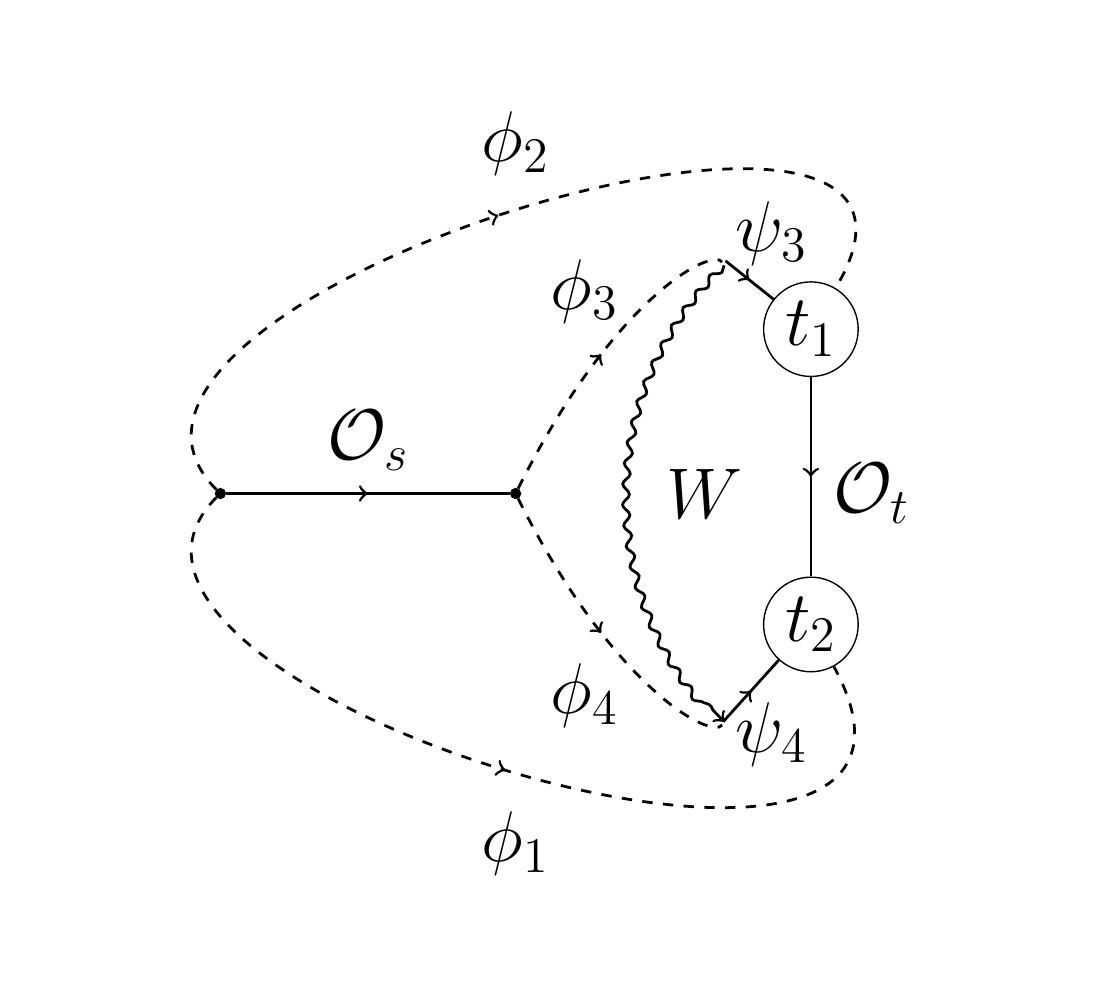}
	\end{aligned}$ $\Rightarrow$ $\begin{aligned}
	\includegraphics[scale=.55]{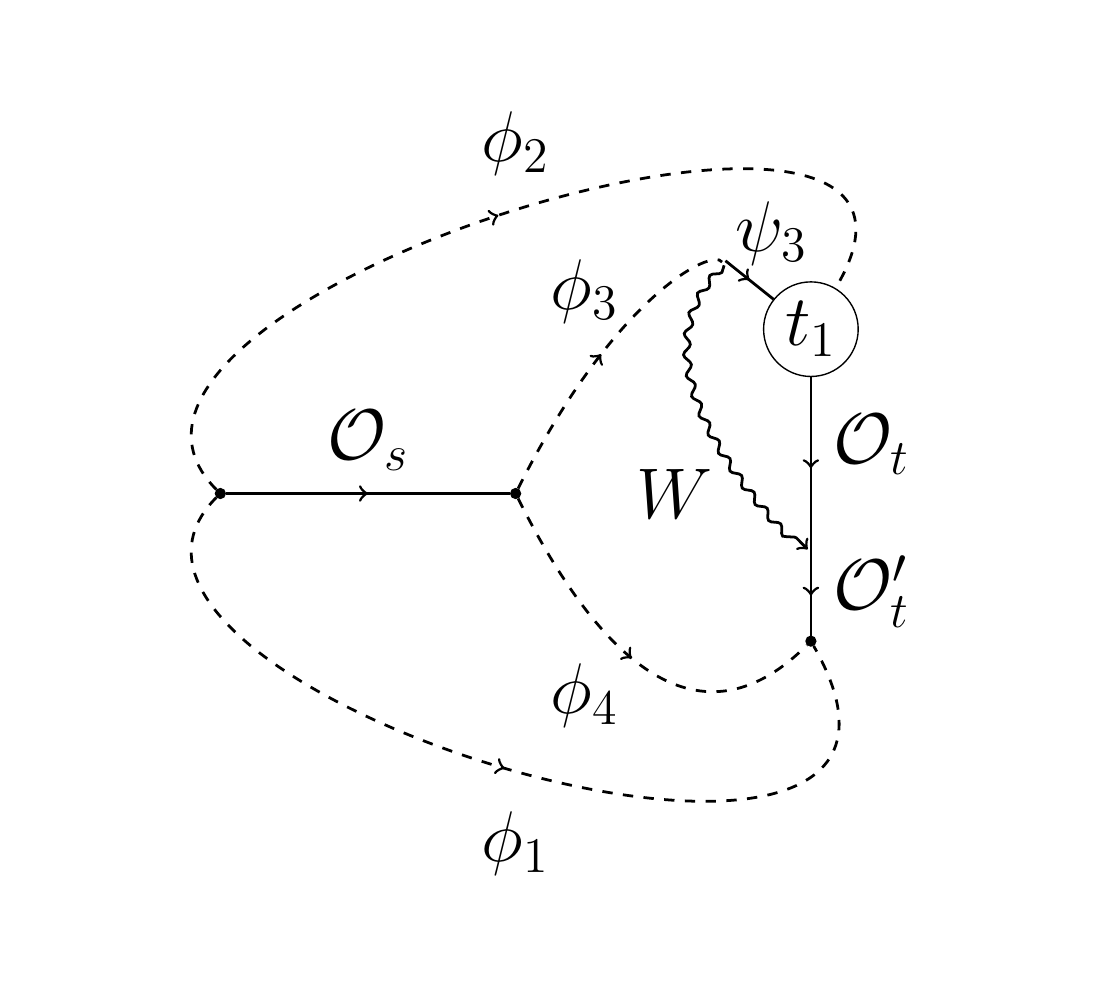}
	\end{aligned}$ $\Rightarrow$ $\begin{aligned}
	\includegraphics[scale=.55]{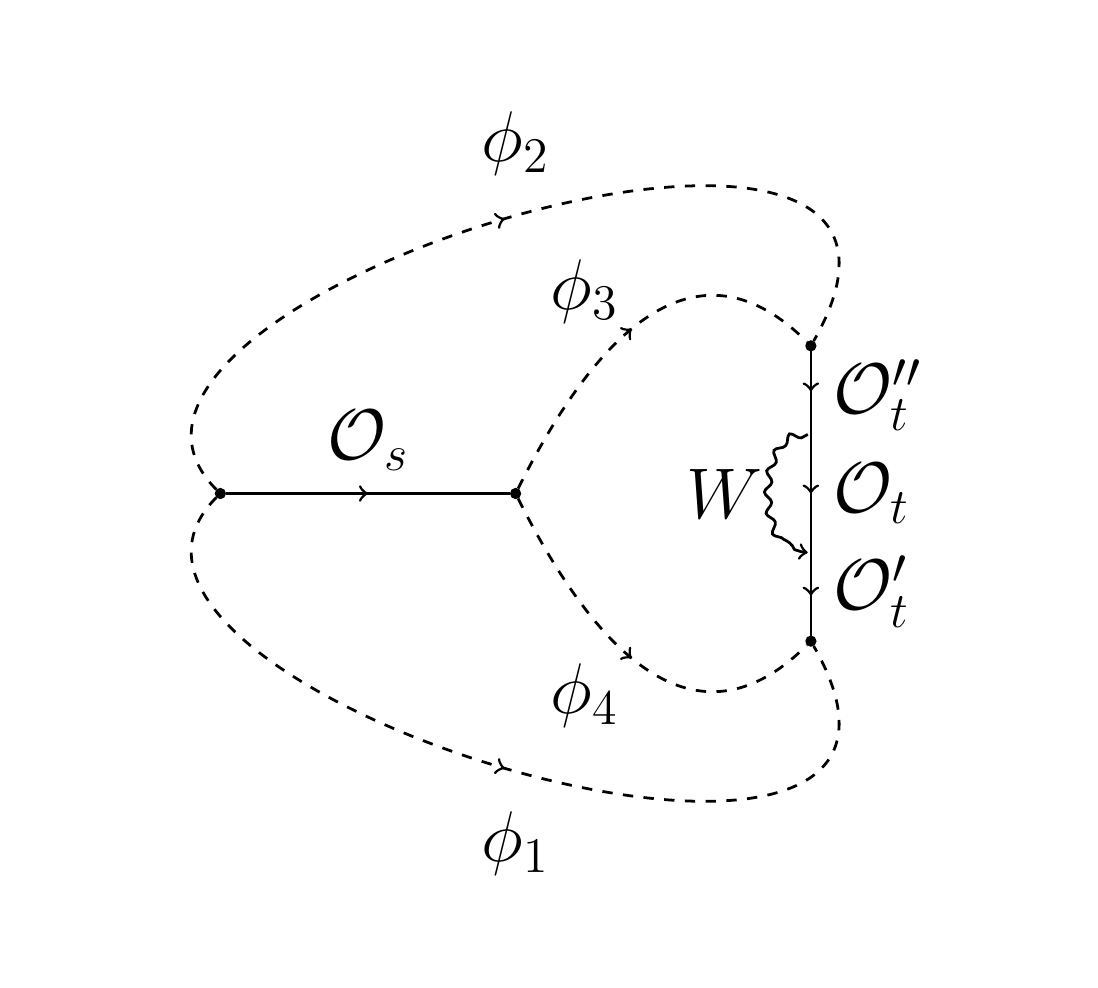}
	\end{aligned}$ $\Rightarrow$ $\begin{aligned}
	\includegraphics[scale=.55]{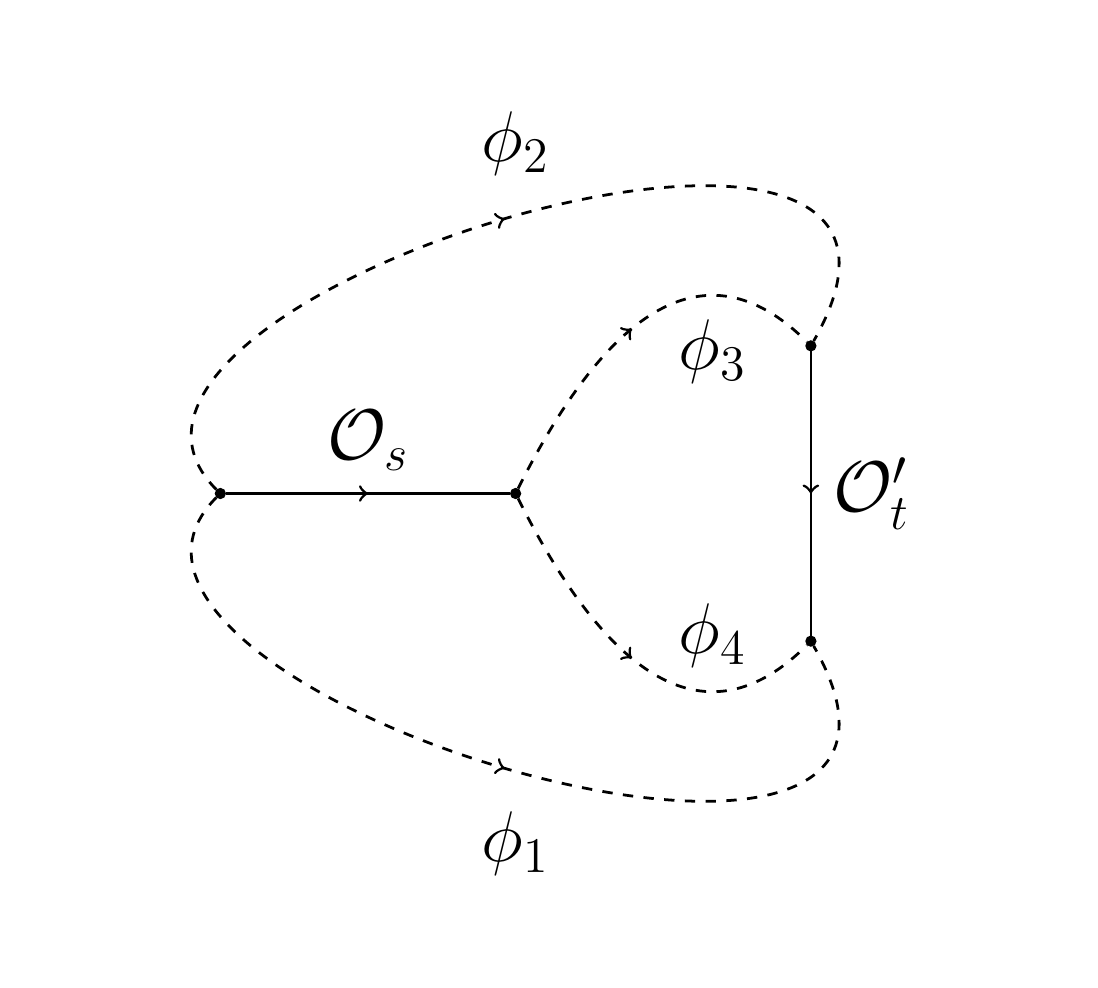}
	\end{aligned}$
	\caption{\label{fig: ssff decomposition}
	Step by step diagrammatic illustration for the decomposition of the fermionic $6j$ symbol relevant for $\<\phi \psi \psi \phi\>$ into the scalar $6j$ symbol. The idea is as follows: one re-expresses a fermionic three-point structure, $\<\psi\psi\cO\>$, in terms of weight-shifting operators acting on a scalar three-point structure, $\<\phi\phi\cO\>$. The weight-shifting operators are then moved inside the diagrammatic loop until they act on the same leg. By irreducibility of the representations, i.e. $\left(\cD^{a,b}\right)_A\left(\cD^{-c,-d}\right)^A\sim \delta^{ac}\delta^{bd}$, the diagram reduces to that of a scalar $6j$ symbol. To be able to move around the weight-shifting operators, one either integrates by parts or uses the crossing relation for covariant three-point structures as explained in the main text. The diagrams above correspond to the equations \equref{eq: fermion-scalar 6j symbol}, \equref{eq: decomposition of fermionic 6j symbol, step 0}, \equref{eq: decomposition of fermionic 6j symbol, step 2}, \equref{eq: decomposition of fermionic 6j symbol, step 3}, \equref{eq: decomposition of fermionic 6j symbol, step 4}, and \equref{eq: decomposition of fermionic 6j symbol, step 5} respectively.
}
\end{figure}

Despite the complicated and lengthy expressions, the procedure is actually quite straightforward and most easily tractable in the diagrammatic notation, see figure \ref{fig: ssff decomposition} for a summary of the decomposition above. However, one should only use the diagrammatic expressions as a guide, as there are ambiguities in their meaning, most notably sign ambiguities as they do not carry the information of the order of operators in the equations. 

For spinning correlators beyond $\<\f\f\psi\psi\>$ we can repeat the procedure above and recursively spin-down the $6j$ symbol. For this, we first define a generalized form of \equref{eq:6jDecompSpintoScalars}:
\be
\sixj{\cO_1}{\cO_2}{\cO_3}{\cO_4}{\cO_5}{\cO_6}^{abcd}=\sum\limits_{\f_i,\cO'_{5,6}}
\sixjDecompp{abcd}{efgh}{\cO_1}{\cO_2}{\cO_3}{\cO_4}{\cO_5}{\cO_6}{\cO'_{1}}{\cO'_{2}}{\cO'_{3}}{\cO'_{4}}{\cO'_{5}}{\cO'_{6}}
\sixj{\cO'_{1}}{\cO'_{2}}{\cO'_{3}}{\cO'_{4}}{\cO'_{5}}{\cO'_{6}}^{efgh},
\ee
where we obtain the ultimate result $J^{abcd}$ of \equref{eq:6jDecompSpintoScalars} by summing over these intermediate factors $J^{abcd}_{efgh}$.

For four external fermions, we only need to repeat this process twice, where in the first step we reduce from four external fermions to two external fermions and two external scalars, and then in the second step we reduce from two external fermions and two external scalars to four external scalars. The second step is already what we derived above, so the only new ingredient is the first step:
\begin{multline}
\sixj{\psi_1}{\psi_2}{\psi_3}{\psi_4}{\cO_5}{\cO_6}^{s_1s_2t_1t_2}=\sum\limits_{a,b,c,d,t_1',t_2'}
\sixjDecompp{s_1s_2t_1t_2}{\uniq s_2t_1't_2'}{\psi_1}{\psi_2}{\psi_3}{\psi_4}{\cO_5}{\cO_6}{\f_1^{-a}}{\f_2^{-b}}{\psi_3}{\psi_4}{\cO_5}{\cO_6^{-c,d}}\\\x
\sixj{\f_1^{-a}}{\f_2^{-b}}{\psi_3}{\psi_4}{\cO_5}{\cO_6^{-c,d}}^{\uniq s_2t_1't_2'}.
\end{multline}
Combining this with \equref{eq: 6j decomposition of 2f2s}, we obtain the final result
\begin{multline}
\label{eq: 6j decomposition of 4f}
\sixj{\psi_1}{\psi_2}{\psi_3}{\psi_4}{\cO_5}{\cO_6}^{s_1s_2t_1t_2}=\sum\limits_{a,\dots,h}
\sixjDecomp{s_1s_2t_1t_2}{\psi_1}{\psi_2}{\psi_3}{\psi_4}{\cO_5}{\cO_6}{\f_1^{-a}}{\f_2^{-b}}{\f_3^{-e}}{\f_4^{-f}}{\cO_5}{\cO_6^{-c-g,d+h}}\\\x
\sixj{\f_1^{-a}}{\f_2^{-b}}{\f_3^{-e}}{\f_4^{-f}}{\cO_5}{\cO_6^{-c-g,d+h}},
\end{multline}
where
\begin{multline}
\sixjDecomp{s_1s_2t_1t_2}{\psi_1}{\psi_2}{\psi_3}{\psi_4}{\cO_5}{\cO_6}{\f_1^{-a}}{\f_2^{-b}}{\f_3^{-e}}{\f_4^{-f}}{\cO_5}{\cO_6^{-c-g,d+h}} =\\ \sum\limits_{t_1',t_2'} 
\sixjDecompp{s_1s_2t_1t_2}{\uniq s_2t_1't_2'}{\psi_1}{\psi_2}{\psi_3}{\psi_4}{\cO_5}{\cO_6}{\f_1^{-a}}{\f_2^{-b}}{\psi_3}{\psi_4}{\cO_5}{\cO_6^{-c,d}}
\sixjDecomp{\uniq s_2t_1't_2'}{\f_1^{-a}}{\f_2^{-b}}{\psi_3}{\psi_4}{\cO_5}{\cO_6^{-c,d}}{\f_1^{-a}}{\f_2^{-b}}{\f_3^{-e}}{\f_4^{-f}}{\cO_5}{\cO_6^{-c-g,d+h}}.
\end{multline}

\begin{figure}
	\centering
	$\begin{aligned}
	\includegraphics[scale=.55]{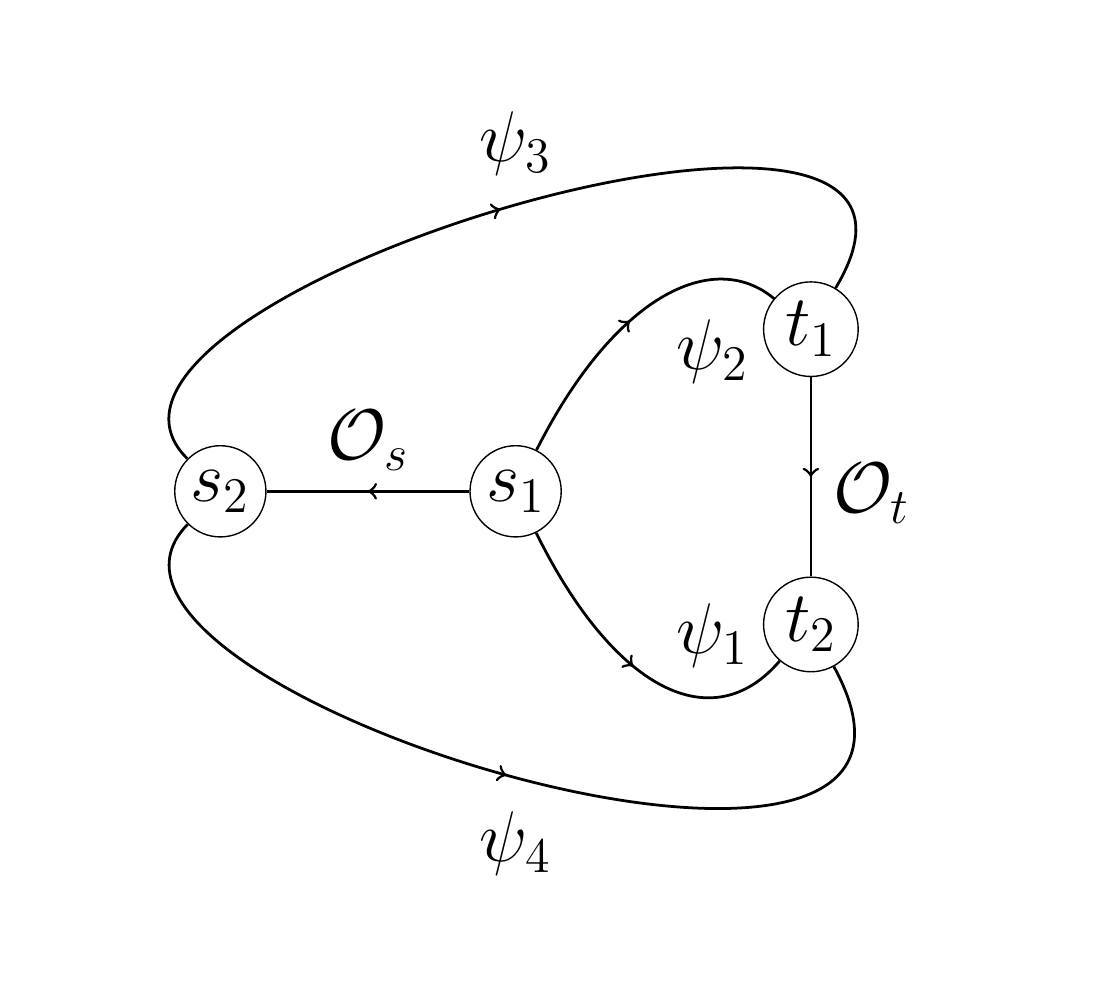}
	\end{aligned}$ $\Rightarrow$ $\begin{aligned}
	\includegraphics[scale=.55]{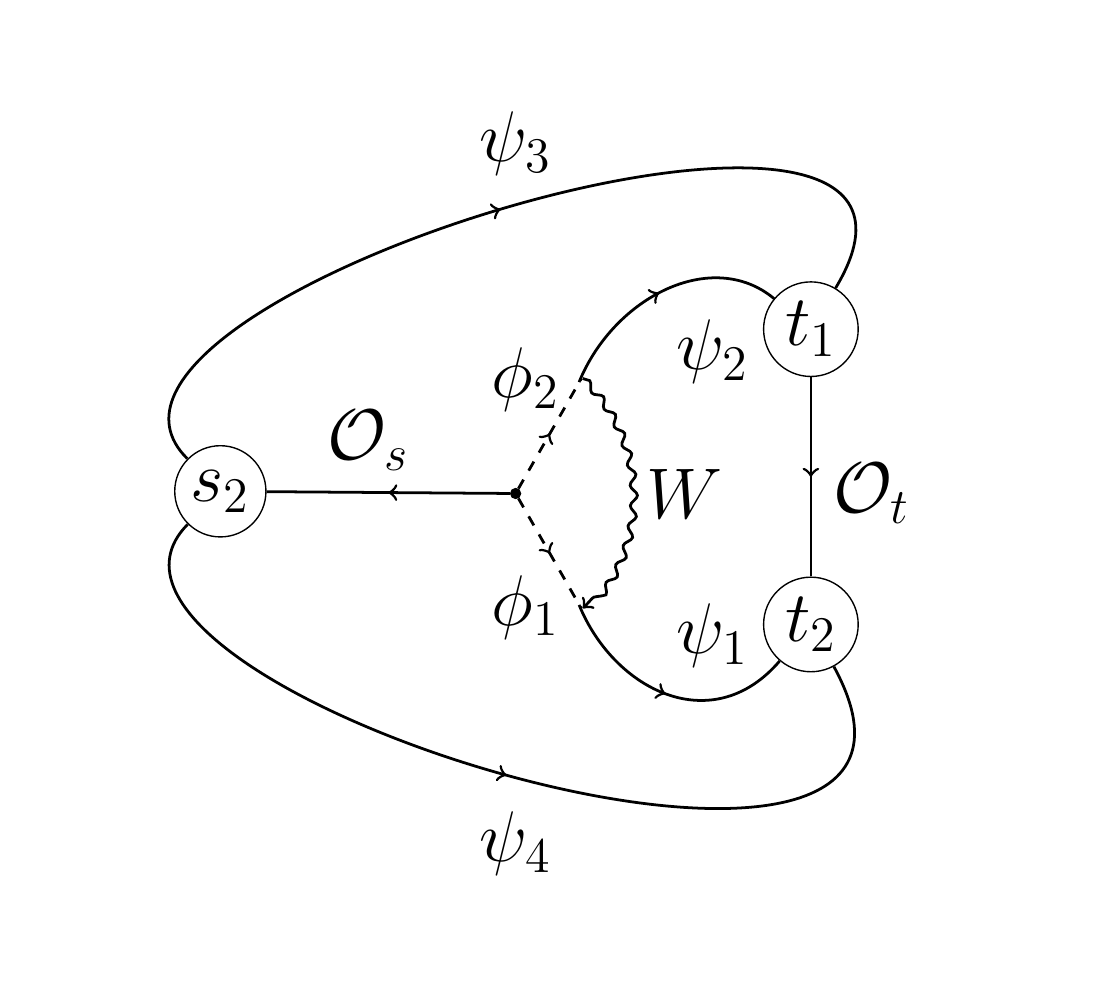}
	\end{aligned}$ $\Rightarrow$ $\begin{aligned}
	\includegraphics[scale=.55]{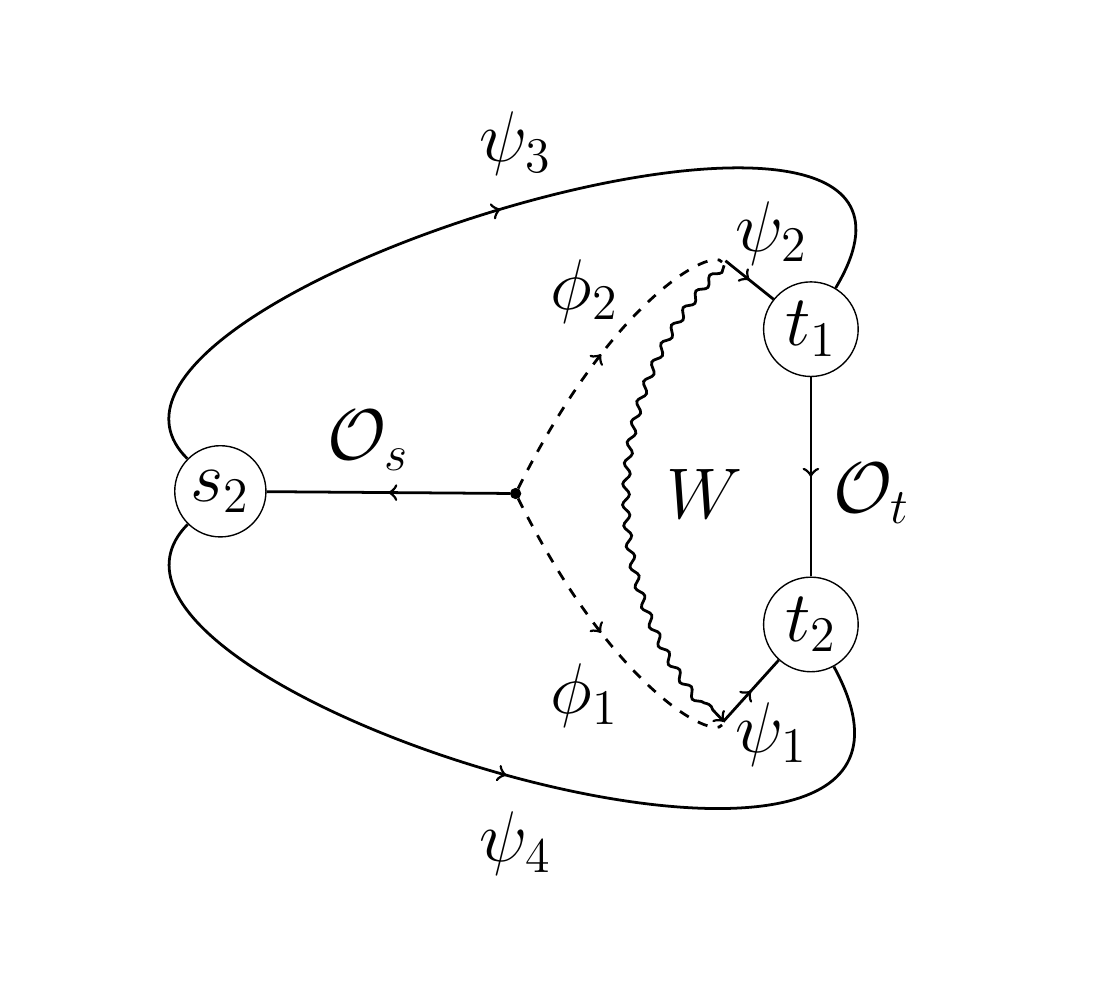}
	\end{aligned}$ $\Rightarrow$ $\begin{aligned}
	\includegraphics[scale=.55]{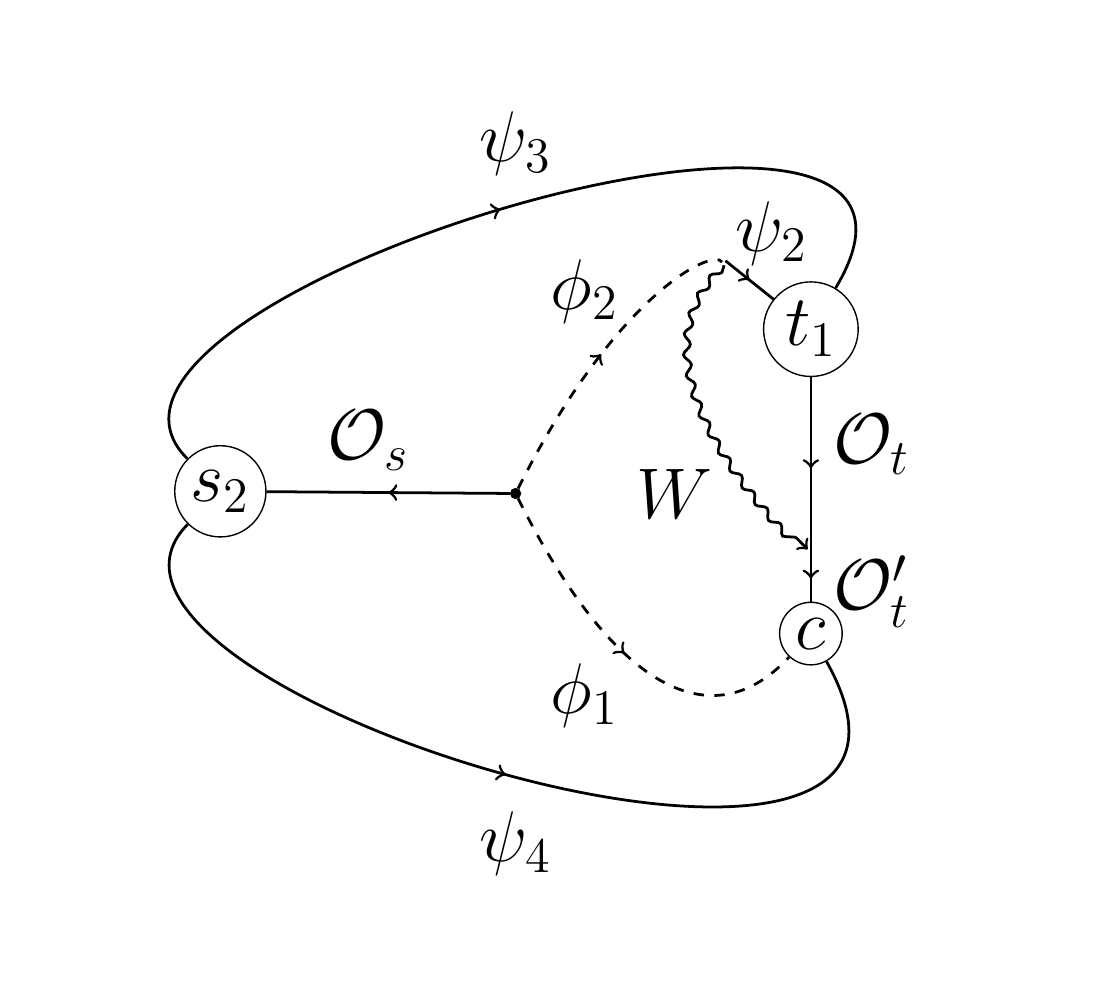}
	\end{aligned}$ $\Rightarrow$ $\begin{aligned}
	\includegraphics[scale=.55]{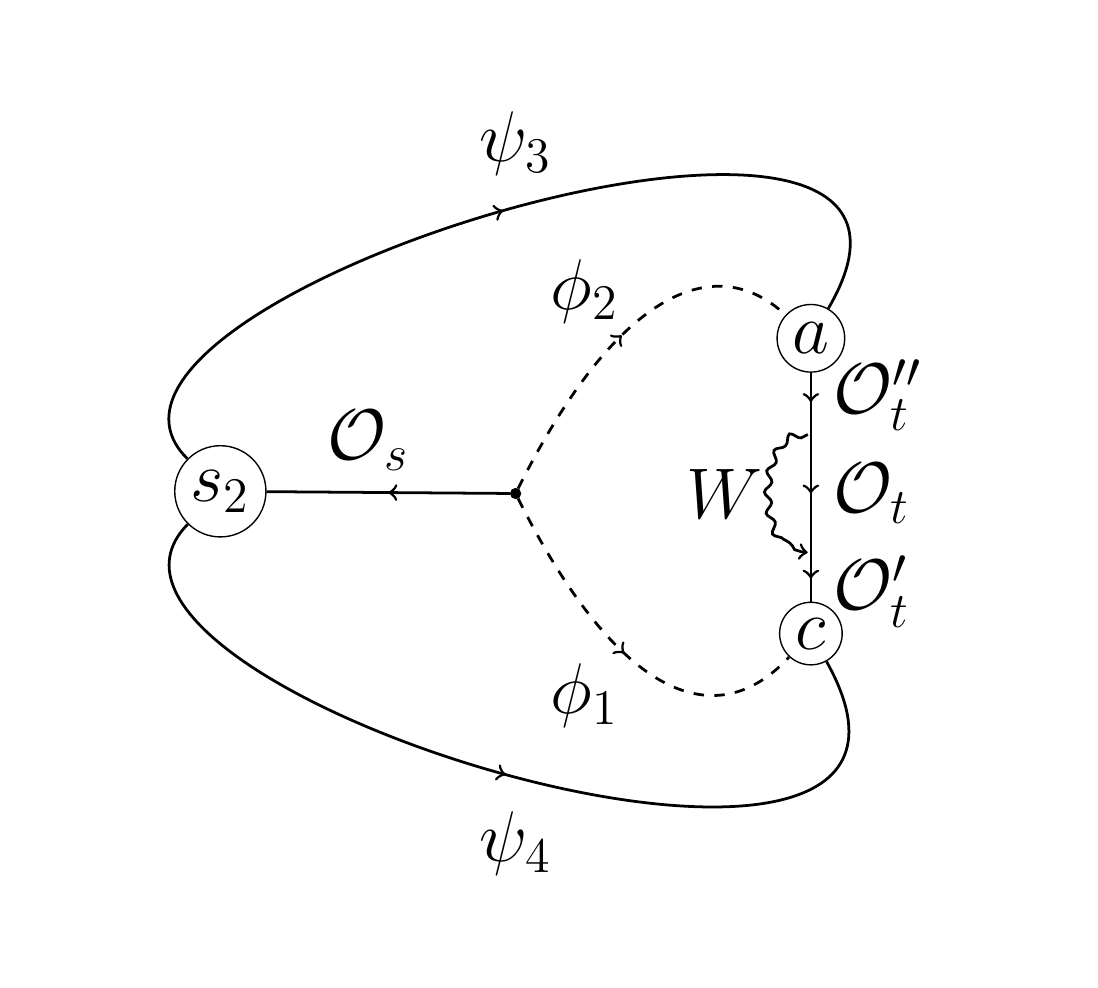}
	\end{aligned}$ $\Rightarrow$ $\begin{aligned}
	\includegraphics[scale=.55]{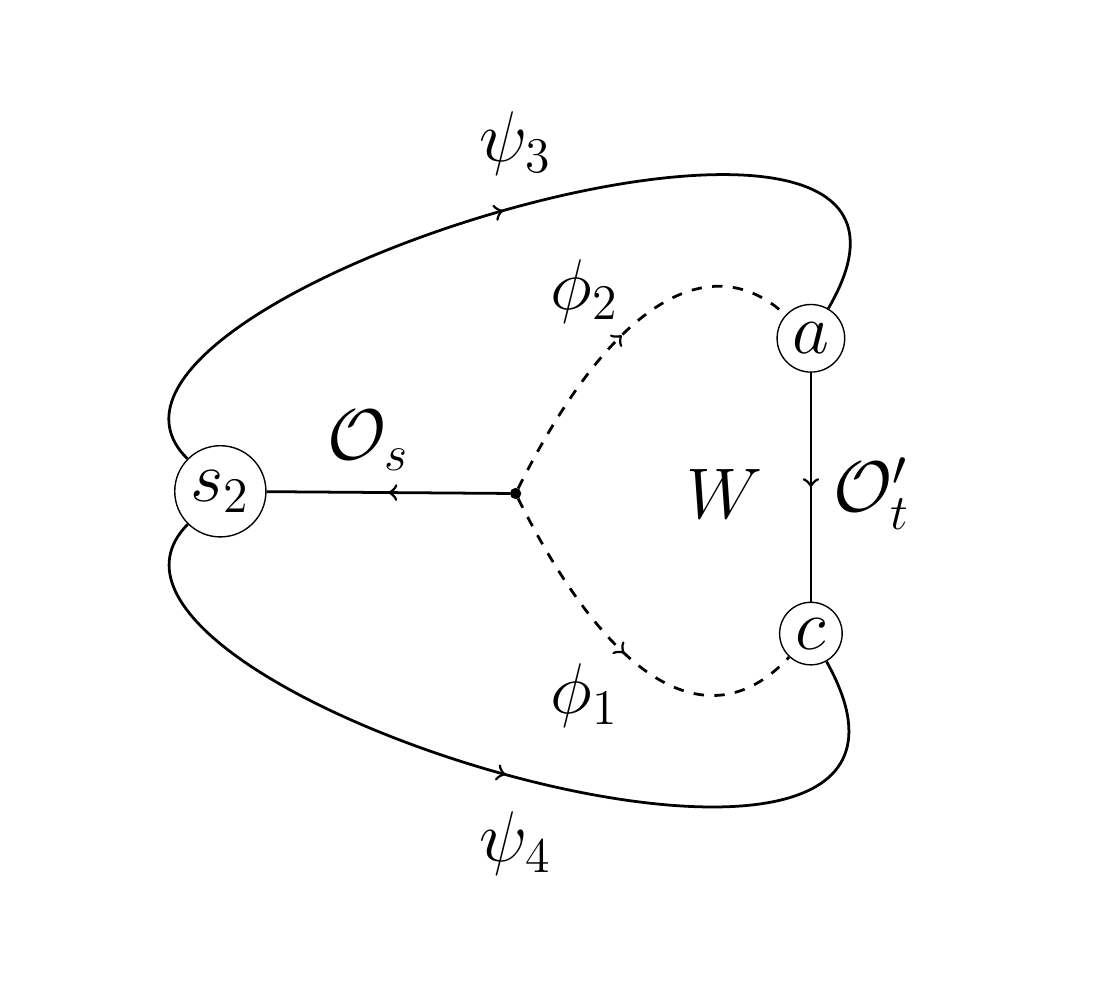}
	\end{aligned}$
	\caption{\label{fig: ffff decomposition}
		Step by step diagrammatic illustration for the decomposition of $6j$ symbol of four external fermions, $\<\psi \psi \psi \psi\>$, in terms of $6j$ symbols of two external fermions and two external scalars.
	}
\end{figure}

We can derive $J^{s_1s_2t_1t_2}_{\uniq s_2t_1't_2'}$ in a similar manner to how we derived $J^{\uniq s_2t_1t_2}$. For brevity we skip the intermediate steps, illustrated diagrammatically in figure \ref{fig: ffff decomposition}, and only present the final result here:
\begin{multline}
\sixjDecompp{s_1s_2t_1t_2}{\uniq s_2t_1't_2'}{\psi_1}{\psi_2}{\psi_3}{\psi_4}{\cO_5}{\cO_6}{\f_1^{-a}}{\f_2^{-b}}{\psi_3}{\psi_4}{\cO_5}{\cO_6^{-c,d}}=\sum\limits_{u_1,u_2,u_1',u_2'}
\kappa^{s_1}_{3,ab}(\tl\psi_1\tl\psi_2\tl\cO_5)
\zeta_{0}^{-b,+}\zeta_{0}^{-a,+}\zeta_{l_t+d}^{-c,-d}\beta_{b,+}^{\Delta_2-b,0}
\\\times
E_{\psi_1\psi_2\cO\rightarrow \psi_2\psi_1\cO}^{t_1u_1}
\left\{
\begin{matrix}
\psi_2 & \psi_3 & \f_2^{-b}\\
\cO_6^{-c,d} & S & \cO_6
\end{matrix}
\right\}^{u_1u_1'}
E_{\phi\psi\cO\rightarrow \psi\phi\cO}^{u_1't_1'}
E_{\psi_1\psi_2\cO\rightarrow \cO\psi_2\psi_1}^{t_2u_2}
\left\{
\begin{matrix}
\tl\cO_6 & \psi_4 & \tl\cO_6^{c,d}\\
\f_1^{-a} & S & \psi_1
\end{matrix}
\right\}^{u_2u_2'} 
E_{\cO\psi\phi\rightarrow \phi\psi\cO}^{u_2't_2'}.
\end{multline} 

\subsection{OPE Function and its Decomposition}
In \secref{\ref{sec: conformal partial waves}} we discussed how $6j$ symbols are related to OPE coefficients. We reproduce \equref{eqn:InvSingBlock_to_OPEdata} for convenience:
\begin{align}
\lambda_{125,a}\lambda_{345,b}\bigg|_{G^{(t),fg}_{\cO_6}}&=-\Res_{\Delta=\Delta_{5}}\rho_{ac}^{(s)}(\cO)S^c_b(\cO_3\cO_4[\tl\cO])\bigg|_{G^{(t),fg}_{\cO_6}}
\nonumber \\ &=
(-1)^{1+\Sigma_{55}+\Sigma_{12}+\Sigma_{13}+\Sigma_{23}}\lambda_{326,f}\lambda_{146,g}
\nonumber \\
& \quad \x\Res_{\Delta=\Delta_{5}}\eta_{(ad)(ce)}^{\De,J}\sixjBlock{\cO_1}{\cO_2}{\cO_3}{\cO_4}{\cO_{\De,J}}{\cO_6}^{defg}S^c_b(\cO_{3}\cO_{4}[\widetilde{\cO}])\bigg|_{J=J_{5}}.
\end{align}

We aim to relate the inversion of a single block for external fermions to the inversion of a single block for a scalar four-point function. For that, by comparing (\ref{eq:6j_symbol_split_Scalars}), (\ref{eqn:GenSplit6j}), and (\ref{eq:6jDecompSpintoScalars}) we write down
\begin{multline}
\sixjBlock{\cO_1}{\cO_2}{\cO_3}{\cO_4}{\cO_5}{\cO_6}^{abcd}=(-1)^{\Sigma_{66}}\left(S^{-1}(\cO_1\cO_4[\widetilde{\cO_6}])\right)^{d}_{e}\\\x\sum\limits_{\f_i,\cO'_{5,6}}
\sixjDecomp{abce}{\cO_1}{\cO_2}{\cO_3}{\cO_4}{\cO_5}{\cO_6}{\f_1}{\f_2}{\f_3}{\f_4}{\cO'_{5}}{\cO'_{6}} K^{\f_1\f_4}_{\widetilde{\cO}'_{6}}\sixjBlock{\f_1}{\f_2}{\f_3}{\f_4}{\cO'_{5}}{\cO'_{6}}.
\end{multline}
We then find the following expression for the OPE function:
\begin{align}
\label{eq: OPE function in terms of scalar 6j symbol}
\rho_{ac}^{(s)}(\cO)S^c_b(\cO_3\cO_4[\tl\cO])\evaluated_{G^{(t),fg}_{\cO_6}}=\lambda_{326,f}\lambda_{146,g}\sum\limits_{\f_i,\cO',\cO'_{6}}
&\opeFuncDecomp{fg}{ab}{\cO_1}{\cO_2}{\cO_3}{\cO_4}{\cO}{\cO_6}{\f_1}{\f_2}{\f_3}{\f_4}{\cO'}{\cO'_{6}}
 \nonumber \\ & \qquad\frac{S(\phi_3\phi_4[\widetilde{\cO'}])}{\eta^{(s)}_{\cO'}}\sixjBlock{\f_1}{\f_2}{\f_3}{\f_4}{\cO'}{\cO'_{6}},
\end{align}
where $\eta^{(s)}_{\cO_5}$ is the normalization of the scalar partial wave for $\<\f_1\f_2\f_3\f_4\>$ and we have defined
\begin{multline}
\label{eq: K coefficients}
\opeFuncDecomp{fg}{ab}{\cO_1}{\cO_2}{\cO_3}{\cO_4}{\cO}{\cO_6}{\f_1}{\f_2}{\f_3}{\f_4}{\cO'}{\cO'_{6}}\equiv (-1)^{\Sigma_{55}+\Sigma_{66}+\Sigma_{12}+\Sigma_{13}+\Sigma_{23}} K^{\f_1\f_4}_{\widetilde{\cO}'_{6}}
\frac{S^c_b(\cO_{3}\cO_{4}[\widetilde{\cO}])}{S(\f_3\f_4[\tl \cO'])}\\\x \left(S^{-1}(\cO_1\cO_4[\widetilde{\cO_6}])\right)^{g}_{h} \eta_{\cO'}^{(s)}\eta_{(ad)(ce)}^{(s)\cO}\sixjDecomp{defh}{\cO_1}{\cO_2}{\cO_3}{\cO_4}{\cO}{\cO_6}{\f_1}{\f_2}{\f_3}{\f_4}{\cO'}{\cO'_{6}}. 
\end{multline}

We expect the physical poles for inverting a fermionic block to come from the physical poles from inverting a scalar block. We then have for example
\begin{multline}
\lambda_{125,a}\lambda_{345,b}\evaluated_{G^{(t),fg}_{\cO_6}}=-\lambda_{326,f}\lambda_{146,g}\sum\limits_{\f_i,\cO',\cO'_{6}}\opeFuncDecomp{fg}{ab}{\cO_1}{\cO_2}{\cO_3}{\cO_4}{\cO_5}{\cO_6}{\f_1}{\f_2}{\f_3}{\f_4}{\cO'_{5}}{\cO'_{6}}
\\\x \Res_{\Delta=\Delta_{5}}\frac{S(\phi_3\phi_4[\widetilde{\cO'_{\De,J}}])}{\eta^{(s)}_{\cO'_{\De,J}}}\sixjBlock{\f_1}{\f_2}{\f_3}{\f_4}{\cO_{\De,J}'}{\cO'_{6}}\evaluated_{J=J_5}. \label{eq: OPE data in terms of scalar 6j symbol} 
\end{multline}

Here we have assumed the inversion of a single block just has single poles. In general when studying a correlator $\<\cO_1\cO_2\cO_3\cO_4\>$ with $\Delta_1+\Delta_2=\Delta_3+\Delta_4$ we find both single and double poles. The double poles give the OPE coefficients times the anomalous dimensions while the single poles gives the OPE coefficients themselves \cite{Caron-Huot:2017vep}.

Equation \equref{eq: OPE function in terms of scalar 6j symbol} is the main result of the paper: by using weight-shifting operators successively, we can express CFT data of spinning operators in terms of $6j$ symbols of external scalars and the decomposition coefficients $\cK$. The former can be calculated efficiently using the Lorentzian inversion formula whereas the latter is given in terms of partial wave normalization factors, shadow matrices, and $6j$ symbol decomposition coefficients $\cJ$, each of which we have computed explicitly. 

We presented the most general form in \equref{eq: OPE function in terms of scalar 6j symbol}, however one can in fact choose either $\cO_5$ or $\cO_6$ to stay the same by moving the weight-shifting operators through the other leg only. Indeed, we kept $\cO_5$ the same in the calculation of the $\cJ$ coefficients both for $\<\f\f\psi\psi\>$ and for $\<\psi\psi\psi\psi\>$, as we can observe in \equref{eq: 6j decomposition of 2f2s} and \equref{eq: 6j decomposition of 4f}. One can similarly compute $\cJ$ while keeping $\cO_6$ constant, though a separate calculation is not necessary: there are several identities between various $\cJ$ coefficients, which follow from the symmetries of the $6j$ symbols that we have summarized in appendix \ref{sec: symmetries of 6j symbols}. In particular, via \equref{eq: 6j symmetry regarding exchange of 5 and 6}, we have
\begin{multline}
\label{eq: relation between J coefficients for 5 and 6 exchange}
\sixjDecompp{abcd}{efgh}{\cO_1}{\cO_2}{\cO_3}{\cO_4}{\cO_5}{\cO_6}{\cO_1^{\delta_1,\epsilon_1}}{\cO_2^{\delta_2,\epsilon_2}}{\cO_3^{\delta_3,\epsilon_3}}{\cO_4^{\delta_4,\epsilon_4}}{\cO_5^{\delta_5,\epsilon_5}}{\cO_6^{\delta_6,\epsilon_6}}=(-1)^{\Sigma_{55}+\Sigma_{5'5'}+\Sigma_{66}+\Sigma_{6'6'}}
\\\x
\sixjDecompp{dcba}{hgfe}{\tl\cO_1}{\tl\cO_4}{\tl\cO_3}{\tl\cO_2}{\cO_6}{\cO_5}{\tl\cO_1^{-\delta_1,\epsilon_1}}{\tl\cO_4^{-\delta_4,\epsilon_4}}{\tl\cO_3^{-\delta_3,\epsilon_3}}{\tl\cO_2^{-\delta_2,\epsilon_2}}{\cO_6^{\delta_6,\epsilon_6}}{\cO_5^{\delta_5,\epsilon_5}}.
\end{multline}

Let us now turn to the explicit results for the $\cK^{fg}_{ab}$ coefficients. Despite the complicated intermediate steps, the final form they take is quite simple as they are relatively short meromorphic functions in scaling dimensions and spins. For example, the only nonzero $\cK$ coefficients for $\<\psi\f\f\psi\>$ arising from the exchange of a scalar $\f_6$ in the t-channel are
\small
\bea[eq: K coefficients for 2f2s with an exchange of a scalar t channel]
\opeFuncDecomp{\uniq 1}{11}{\psi_1}{\f_2}{\f_2}{\psi_1}{\cO_5}{\f_6}{\f_1^{\half}}{\f_2}{\f_2}{\f_1^{\half}}{\cO_{5}^{\half,-\half}}{\f_6} ={}&{} i, \\
\opeFuncDecomp{\uniq 1}{11}{\psi_1}{\f_2}{\f_2}{\psi_1}{\cO_5}{\f_6}{\f_1^{\half}}{\f_2}{\f_2}{\f_1^{\half}}{\cO_{5}^{-\half,\half}}{\f_6} ={}&{}  \frac{i \left(\Delta _5-\frac{3}{2}\right) \left(l_5+\frac{1}{2}\right) \left(l_5-\De_{25}^1+\frac{5}{2}\right)^2}{4\left(\Delta _5-2\right) \left(l_5+1\right) \left(l_5-\Delta _5+2\right) \left(l_5-\Delta _5+3\right)},
 \\
\opeFuncDecomp{\uniq 1}{22}{\psi_1}{\f_2}{\f_2}{\psi_1}{\cO_5}{\f_6}{\f_1^{\half}}{\f_2}{\f_2}{\f_1^{\half}}{\cO_{5}^{\half,\half}}{\f_6} ={}&{} -\frac{i \left(l_5+\frac{1}{2}\right)}{l_5+1}, \\
\opeFuncDecomp{\uniq 1}{22}{\psi_1}{\f_2}{\f_2}{\psi_1}{\cO_5}{\f_6}{\f_1^{\half}}{\f_2}{\f_2}{\f_1^{\half}}{\cO_{5}^{-\half,-\half}}{\f_6} ={}&{} -\frac{i \left(\Delta _5-\frac{3}{2}\right) \left(\De_{25}^1+l_5-\frac{3}{2}\right)^2}{4 \left(\Delta _5-2\right)
	\left(\Delta _5+l_5-2\right) \left(\Delta _5+l_5-1\right)},
\\
\opeFuncDecomp{\uniq 3}{12}{\psi_1}{\f_2}{\f_2}{\psi_1}{\cO_5}{\f_6}{\f_1^{-\half}}{\f_2}{\f_2}{\f_1^{\half}}{\cO_{5}^{\half,\half}}{\f_6} ={}&{} \frac{i \left(l_5+\frac{1}{2}\right) \left(l_5-\De_{12}^5+\frac{5}{2}\right)}{\left(\Delta _6-1\right)\left(l_5+1\right)},
\\
\opeFuncDecomp{\uniq 3}{12}{\psi_1}{\f_2}{\f_2}{\psi_1}{\cO_5}{\f_6}{\f_1^{-\half}}{\f_2}{\f_2}{\f_1^{\half}}{\cO_{5}^{-\half,-\half}}{\f_6} ={}&{}  \frac{i \left(\Delta_5-\frac{3}{2}\right) \left(\De_{15}^2+l_5-\frac{3}{2}\right) \left(\De_{25}^1+l_5-\frac{3}{2}\right) \left(\De_{125}+l_5-\frac{9}{2}\right)}{4 \left(\Delta _5-2\right) \left(\Delta
	_6-1\right) \left(\Delta _5+l_5-2\right) \left(\Delta _5+l_5-1\right)},
\\
\opeFuncDecomp{\uniq 3}{21}{\psi_1}{\f_2}{\f_2}{\psi_1}{\cO_5}{\f_6}{\f_1^{-\half}}{\f_2}{\f_2}{\f_1^{\half}}{\cO_{5}^{\half,\half}}{\f_6} ={}&{}  \frac{i\left(\De_{12}^5+l_5-\frac{3}{2}\right)}{\Delta _6-1},
\\
\opeFuncDecomp{\uniq 3}{21}{\psi_1}{\f_2}{\f_2}{\psi_1}{\cO_5}{\f_6}{\f_1^{-\half}}{\f_2}{\f_2}{\f_1^{\half}}{\cO_{5}^{-\half,\half}}{\f_6} ={}&{} \frac{i \left(\Delta _5-\frac{3}{2}\right) \left(l_5+\frac{1}{2}\right) \left(l_5-\De_{15}^2+\frac{5}{2}\right)}{4 \left(\Delta _5-2\right)
	\left(\Delta _6-1\right) \left(l_5+1\right) \left(l_5-\Delta _5+2\right)}
\nn\\
&\qquad \x \frac{\left(l_5-\De_{25}^1+\frac{5}{2}\right) \left(l_5-\De_{125}+\frac{11}{2}\right)}{\left(-\Delta _5+l_5+3\right)},
\eea 
\normalsize
where we defined $\De_{ijk}\equiv\De_i+\De_j+\De_k$ and $\De_{ij}^k\equiv \De_i+\De_j-\De_k$.

We would like to emphasize two points. Firstly, as there is not a unique way to write \equref{eq: OPE data in terms of scalar 6j symbol}, the statement that the $\cK$ in \equref{eq: K coefficients for 2f2s with an exchange of a scalar t channel} are the only nonzero coefficients for $\<\psi\f\f\psi\>$ with an exchange of a scalar $\f_6$ in the t-channel should be understood for a particularly chosen decomposition in \equref{eq: OPE data in terms of scalar 6j symbol}. One can of course change the decomposition, which would then require a new set of $\cK$ coefficients. For example, we used a set of $\cK$ coefficients with $\cO_6$ held constant in \equref{eq: K coefficients for 2f2s with an exchange of a scalar t channel}; another set with $\cO_5$ held constant instead can be immediately obtained via \equref{eq: relation between J coefficients for 5 and 6 exchange}.\footnote{It should be noted that not all different decompositions are related to each other via symmetries. For example, $\cO_4'=\f_1^{\half}$ whereas $\cO_1'=\f_1^{\pm\half}$ in \equref{eq: K coefficients for 2f2s with an exchange of a scalar t channel}: this follows from fixing $b=\half$ in \equref{eq: J for 2f2s} as we noted after the equation. If we were to fix $b=-\half$ instead, we would then have a set of $\cK$ coefficients with $\cO_4'=\f_1^{-\half}$ and $\cO_1'=\f_1^{\pm\half}$, and these new coefficients are not related to \equref{eq: K coefficients for 2f2s with an exchange of a scalar t channel} in any manifestly symmetric way.} Secondly, we note that the absence of $\cK^{\uniq 1}_{12}$, $\cK^{\uniq 1}_{21}$, $\cK^{\uniq 3}_{11}$, and $\cK^{\uniq 3}_{22}$ is not coincidental: they are forbidden by the parity symmetry as we work in a parity-definite basis.\footnote{\label{footnote: parity of three-point structures}We would like to caution the reader that this statement follows from the symmetries of three-point structures $\<\cO_1\cO_2\cO_3\>^a$ under the transformation $X\rightarrow -X$ in embedding space, hence it is true whether the relevant physical theory has parity symmetry or not, i.e. $\<\cO_1\cO_2\cO_3\>^a$ are merely formal entities and should not be thought of as physical three-point structures.}

We list the $\cK$ coefficients for all the remaining cases for $\<\psi\f\f\psi\>$ explicitly in appendix~\ref{sec: K coefficients}, and the coefficients for $\<\psi_1 \psi_2 \psi_2 \psi_1\>$ and $\<\psi\psi\psi\psi\>$ are given in an attached \texttt{Mathematica} file.

\subsection{Applications and Examples}
\label{sec:applications}
In this section, we will use the techniques we developed in previous section to calculate the anomalous dimensions of the double-twist operators exchanged in the $\<\psi\f\f\psi\>$ and $\<\psi\psi\psi\psi\>$ correlators. For this, we will primarily focus on using the coefficient of the double poles in \equref{eq: OPE function in terms of scalar 6j symbol} to obtain $\delta h P$, and we will divide it by $P^{\text{MFT}}$ to obtain $\delta h=\gamma/2$.\footnote{One can further improve these results by considering the residues of \equref{eq: OPE function in terms of scalar 6j symbol} to obtain $\delta P$ as in (\ref{eq: OPE data in terms of scalar 6j symbol}) or the example in~\ref{app:OPE}, with which one schematically has $\delta h=\frac{\delta h P}{P^{\text{MFT}}+\delta P}$ up to possible mixing between different twist towers. We have provided all the $\cK$ coefficients needed for these computations.} We also give an example of calculating OPE coefficient corrections in appendix~\ref{app:OPE}.

As we reviewed in section~\ref{sec:6jreview}, $6j$ symbols develop double poles when certain relations are satisfied; in the case of \equref{eq: OPE function in terms of scalar 6j symbol}, we have double poles in $\De'$ at \mbox{$\De'=\De_1+\De_2+J'$} if $\Delta_1+\Delta_2=\Delta_3+\Delta_4$. Here these correspond to double-twist operators $[\f_1\f_2]$. We will stick to decomposition coefficients $\cK$ with $\cO_6'=\cO_6$, so taking the coefficient of double poles of \equref{eq: OPE function in terms of scalar 6j symbol} roughly translates into relating $\delta h P_{[\cO_1\cO_2]}\evaluated_{\cO_6}$ to $\delta h P_{[\f_1\f_2]}\evaluated_{\cO_6}$, where here $\f_i$ are some fictitious scalar operators. For later convenience, we define
\small
\bea 
\mathfrak{dp}_1^{J,n}(\f_1,\f_2,\cO_6)\equiv &
\lim\limits_{\De\rightarrow\De_1+\De_2+J+2n} \left(\De -\De_1-\De_2-J-2n\right)^2 \frac{S(\f_3 \f_4 [\widetilde{\cO}_{5}])}{\eta^{(s)}_{\cO_{5}}} \sixjBlock{\f_1}{\f_2}{\f_2}{\f_1}{\cO_{\De,J}}{\cO_{6}},
\\
\mathfrak{dp}_2^{J,n}(\f_1,\f_2,\cO_6)\equiv &
\lim\limits_{\De\rightarrow\De_1+\De_2+J+2n} \left(\De -\De_1-\De_2-J-2n\right)^2  \frac{S(\f_3 \f_4 [\widetilde{\cO}_{5}])}{\eta^{(s)}_{\cO_{5}}}  \sixjBlock{\f_1}{\f_2}{\f_1}{\f_2}{\cO_{\De,J}}{\cO_{6}},
\eea
\normalsize
which describe the corrections $\delta h P$ with external scalars after the OPE coefficients are stripped off:
\be 
\delta h P_{[\f_1\f_2]}\evaluated_{\cO_6\in \phi_1\x \phi_1}=&\lambda_{\f_1\f_1\cO_6}\lambda_{\f_2\f_2\cO_6}\mathfrak{dp}_{1}^{J,n}(\f_1,\f_2,\cO_6),
\\
\delta h P_{[\f_1\f_2]}\evaluated_{\cO_6\in \phi_1\x \phi_2}=&\lambda_{\f_1\f_2\cO_6}^2\mathfrak{dp}_{2}^{J,n}(\f_1,\f_2,\cO_6)\.
\ee 
As a reminder, in these formulas $\eta$ is the normalization for the scalar partial wave.

The particular decomposition we will be using in \equref{eq: OPE function in terms of scalar 6j symbol} will  be the one with $\cO_6$ kept constant, hence by taking the coefficients of double poles on both sides, we obtain
\begin{multline}
\label{eq: general form for deltahP}
(\delta h P)_{ab}([\cO_1\cO_2]_{n,J})\evaluated_{G^{(t),fg}_{\cO_6}}=
\\
-\lambda_{326,f}\lambda_{146,g}\sum\limits_{\f_i,J',n'}
\opeFuncDecomp{fg}{ab}{\cO_1}{\cO_2}{\cO_3}{\cO_4}{[\cO_1\cO_2]_{n,J}}{\cO_6}{\f_1}{\f_2}{\f_2}{\f_1}{[\f_1\f_2]_{n',J'}}{\cO_{6}}
\mathfrak{dp}_1^{J',n'}(\f_1,\f_2,\cO_6)
\\
-\lambda_{326,f}\lambda_{146,g}\sum\limits_{\f_i,J',n'}
\opeFuncDecomp{fg}{ab}{\cO_1}{\cO_2}{\cO_3}{\cO_4}{[\cO_1\cO_2]_{n,J}}{\cO_6}{\f_1}{\f_2}{\f_1}{\f_2}{[\f_1\f_2]_{n',J'}}{\cO_{6}}
\mathfrak{dp}_2^{J',n'}(\f_1,\f_2,\cO_6).
\end{multline}

For example, for the leading parity-even tower in the s-channel of $\<\psi\f\f\psi\>$, we find  
\begin{multline}
\label{eq: parity even tower of two fermions and two scalars}
(\delta h P)_{11}([\f\psi]^+_{0,J})\evaluated_{G^{(t)}_{\cO_6}}=-i \lambda_{\f\f\cO_6}\lambda_{\psi\psi\cO_6}^1\mathfrak{dp}_1^{J-\half,0}(\psi^{\half},\f,\cO_6)
\\
-i\lambda_{\f\f\cO_6}\lambda_{\psi\psi\cO_6}^2(1-\delta_{0,l_6})\bigg(\frac{\left(J+\frac{1}{2}\right) (J+1)}{\left(\Delta _6-1\right) l_6}
\mathfrak{dp}_1^{J+\half,0}(\psi^{-\half},\f,\cO_6)\\
+\frac{2 \left(\Delta _{\psi }+J-1\right){}^2 \left(\Delta _{\psi }+\Delta _{\phi
	}+J-2\right) \left(2 \Delta _{\psi }+2 \Delta _{\phi }+2 J-5\right)}{\left(\Delta
	_6-1\right) l_6 \left(2 \Delta _{\psi }+2 \Delta _{\phi }+4 J-5\right) \left(2 \Delta
	_{\psi }+2 \Delta _{\phi }+4 J-3\right)}\mathfrak{dp}_1^{J-\half,0}(\psi^{-\half},\f,\cO_6)
\\
+\frac{\left(-2 \Delta _{\psi }-\Delta _6+l_6+4\right) \left(2 \Delta _{\psi }-\Delta
	_6+l_6-1\right)}{4 \left(\Delta _6-1\right) l_6} \mathfrak{dp}_1^{J-\half,0}(\psi^{\half},\f,\cO_6)
\bigg).
\end{multline}
For the parity-odd tower, we instead have
\small
\begin{multline}
\label{eq: parity odd tower of two fermions and two scalars}
(\delta h P)_{22}\left([\f\psi]^-_{0,J}\right)\evaluated_{G^{(t)}_{\cO_6}}=i \lambda_{\f\f\cO_6}\lambda_{\psi\psi\cO_6}^1\bigg(
\frac{J+\frac{1}{2}}{J+1}
\mathfrak{dp}_1^{J+\half,0}(\psi^{\half},\f,\cO_6)\\
+\frac{2 \left(2 \Delta _{\phi }+2 J-1\right)^2 \left(\Delta _{\psi }+\Delta _{\phi
	}+J-1\right)}{\left(2 \Delta _{\psi }+2 \Delta _{\phi }+2 J-3\right) \left(2 \Delta
	_{\psi }+2 \Delta _{\phi }+4 J-3\right) \left(2 \Delta _{\psi }+2 \Delta _{\phi }+4
	J-1\right)}\mathfrak{dp}_1^{J-\half,0}(\psi^{\half},\f,\cO_6)\bigg)
\\
+i \lambda_{\f\f\cO_6}\lambda_{\psi\psi\cO_6}^2(1-\delta_{0,l_6})\bigg(
\frac{1}{\left(\Delta _6-1\right) l_6}\mathfrak{dp}_1^{J-\half,1}(\psi^{-\half},\f,\cO_6)
\\
+\frac{(2 J+1) \left(-2 \Delta _{\psi }-\Delta _6+l_6+4\right) \left(2 \Delta _{\psi
	}-\Delta _6+l_6-1\right)}{8 \left(\Delta _6-1\right) (J+1) l_6}\mathfrak{dp}_1^{J+\half,0}(\psi^{\half},\f,\cO_6)
\\
+\frac{\left(J+\frac{1}{2}\right) \left(\Delta _{\psi }-1\right){}^2 \left(\Delta _{\psi
	}+\Delta _{\phi }-\frac{5}{2}\right) \left(\Delta _{\psi }+\Delta _{\phi
	}+J-1\right)}{\left(\Delta _6-1\right) (J+1) l_6 \left(\Delta _{\psi }+\Delta _{\phi
	}-\frac{3}{2}\right) \left(\Delta _{\psi }+\Delta _{\phi }+J-\frac{3}{2}\right)}\mathfrak{dp}_1^{J-\half,0}(\psi^{\half},\f,\cO_6)
\\
+
\frac{\left(2 \Delta _{\phi }+2 J-1\right){}^2 \left(-2 \Delta _{\psi }-\Delta
	_6+l_6+4\right) \left(2 \Delta _{\psi }-\Delta _6+l_6-1\right) \left(\Delta _{\psi
	}+\Delta _{\phi }+J-1\right)}{2 \left(\Delta _6-1\right) l_6 \left(2 \Delta _{\psi }+2
	\Delta _{\phi }+2 J-3\right) \left(2 \Delta _{\psi }+2 \Delta _{\phi }+4 J-3\right)
	\left(2 \Delta _{\psi }+2 \Delta _{\phi }+4 J-1\right)}
\mathfrak{dp}_1^{J-\half,0}(\psi^{\half},\f,\cO_6)
\bigg).
\end{multline}
\normalsize
We would like to point out the appearance of  $\mathfrak{dp}_1^{J-\half,1}(\psi^{-\half},\f,\cO_6)$ which indicates that we may need to extract the data of \emph{non-leading twist} scalar towers from the scalar $6j$ symbol in order to obtain the \emph{leading twist} spinning towers. This happens for the cases where the scaling dimensions of $\cO_{1,2}$ and $l_5$ are shifted downwards while the scaling dimension of $\cO_5$ is shifted upwards. We also note that the absence of $\lambda_{\psi\psi\cO_6}^{3,4}$ follows from our parity-definite choice of three-point structures.\footnote{The OPE coefficients $\lambda_{\psi\psi\cO_6}^{3,4}$ can only appear through the product of $3-$point structures $\<\f\f\cO_6\>\<\psi\psi\cO_6\>^{3,4}$, which are parity-odd under $X\rightarrow -X$ in embedding space (see footnote \ref{footnote: parity of three-point structures} as well). In the s-channel this can only match to the non-diagonal pieces $\<\psi\f\cO_5\>^{1,2}\<\psi\f\cO_5\>^{2,1}$ for which no double pole appears: \mbox{$(\delta h P)_{12}\left([\f\psi]_{0,J}\right)=(\delta h P)_{21}\left([\f\psi]_{0,J}\right)=0$}.
}

Using the $\cK$ coefficients given explicitly in the attached \texttt{Mathematica} notebook, one can similarly obtain all cases for four fermions as well. For brevity, we reproduce here a few cases for $l_6=0$:
\begin{multline}
\label{eq: parity even tower of four fermions}
(\delta h P)_{22}\left([\psi\psi]^+_{0,J}\right)\evaluated_{G^{(t)}_{\f_6}}= \lambda_{\psi\psi\cO_6}^1\lambda_{\psi\psi\cO_6}^1\mathfrak{dp}_1^{J-1,0}(\psi^{\half},\psi^{\half},\f_6)
\\+\lambda_{\psi\psi\cO_6}^3\lambda_{\psi\psi\f_6}^3
\bigg(
\frac{2 J}{\left(\Delta _6-1\right)^2}\mathfrak{dp}_2^{J,0}(\psi^{-\half},\psi^{\half},\f_6)
\\+\frac{\left(\Delta _{\psi }+\frac{J-2}{2}\right) \left(\Delta _{\psi
	}+J-\frac{3}{2}\right)}{\left(\Delta _6-1\right){}^2 \left(\Delta _{\psi }+J-1\right)}\mathfrak{dp}_2^{J-1,0}(\psi^{-\half},\psi^{\half},\f_6)
\bigg),
\end{multline}
and 
\be 
(\delta h P)_{11}\left([\psi\psi]^+_{0,J}\right)=(\delta h P)_{12}\left([\psi\psi]^+_{0,J}\right)=(\delta h P)_{21}\left([\psi\psi]^+_{0,J}\right)=0 .
\ee 
Unlike the parity-even case, parity-odd families have non-zero off-diagonal components:
\footnotesize
\be 
{}&(\delta h P)_{34}\left([\psi\psi]^+_{0,J}\right)\evaluated_{G^{(t)}_{\f_6}}=\frac{\lambda_{\psi\psi\f_6}^3\lambda_{\psi\psi\f_6}^3}{(\De_6-1)^2}\bigg(
-\mathfrak{dp}_2^{J-1,1}(\psi^{-\half},\psi^{\half},\f_6) 
+\frac{J\left(J+1\right) \left(2 J+3\right)}{2 J+1}
\mathfrak{dp}_2^{J+1,0}(\psi^{-\half},\psi^{\half},\f_6) 
\\
{}&\qquad
+\frac{\left(2 \Delta _{\psi }+J-2\right) \left(2 \Delta _{\psi }+J-1\right) \left(2 \Delta
	_{\psi }+2 J-3\right) \left(2 \Delta _{\psi }+2 J-1\right) \left(4 \Delta _{\psi }+2
	J-5\right)}{64 \left(\Delta _{\psi }+J-1\right){}^2 \left(4 \Delta _{\psi }+2
	J-3\right)}\mathfrak{dp}_2^{J-1,0}(\psi^{-\half},\psi^{\half},\f_6) 
\\
{}&\qquad
+\frac{J \left(2 \Delta _{\psi }+J-1\right) \left(8 \Delta _{\psi }^3+2 \left(4 J^2-8
	J+3\right) \Delta _{\psi }-4 J^2+4 (4 J-3) \Delta _{\psi }^2+4 J-3\right)}{8 \left(2
	\Delta _{\psi }-1\right) \left(\Delta _{\psi }+J-1\right) \left(\Delta _{\psi
	}+J\right)}\mathfrak{dp}_2^{J,0}(\psi^{-\half},\psi^{\half},\f_6)
\bigg),
\\
{}&(\delta h P)_{43}\left([\psi\psi]^+_{0,J}\right)\evaluated_{G^{(t)}_{\f_6}}=(\delta h P)_{34}\left([\psi\psi]^+_{0,J}\right)\evaluated_{G^{(t)}_{\f_6}}\,,
\ee  
\normalsize
where one can write down $(\delta h P)_{33}\left([\psi\psi]^+_{0,J}\right)\ne 0$ and $(\delta h P)_{44}\left([\psi\psi]^+_{0,J}\right)\ne 0$ as well. However, all non-block-diagonal terms such as $(\delta h P)_{13}\left([\psi\psi]^+_{0,J}\right)$ are zero as the corresponding structures do not develop double poles. This is why we do not have terms with the mixed coefficient $\lambda_{\psi\psi\cO_6}^1\lambda_{\psi\psi\cO_6}^3$.

To calculate the scalar coefficients $\mathfrak{dp}_{i}^{J,0}$ we will use the Lorentzian inversion formula~\cite{Caron-Huot:2017vep,Liu:2018jhs,Albayrak:2019gnz}, combined with either dimensional reduction of the 3d block~\cite{Hogervorst:2016hal,Albayrak:2019gnz} or resummations of the lightcone expansion~\cite{Li:2019dix}. Using dimensional reduction we find for the $n=0$ double-twist operators:
\begin{align}
\label{eq: doublePoles1}
 \mathfrak{dp}_1^{J,0}(\phi_1,\phi_2,\mathcal{O}_{6})&=-\sum\limits_{p=0}^{\infty}\sum\limits_{q=\text{max}(-p,p-2J_6)}^{p}2\kappa^{0,0}_{2\bar{h}}\sin(\pi(h_6-2h_1))\sin(\pi(h_6-2h_2))
\nonumber \\ &\hspace{4cm}  \times \mathcal{A}^{0,0}_{p,q}\Omega^{h_1h_2h_2h_1}_{\bar{h},h_{6}+p,2h_2}\frac{\Gamma(2(\bar{h}+q))}{\Gamma^{2}(\bar{h}+q)}\bigg|_{\bar{h}=h_1+h_2+J},
\\
\label{eq: doublePoles2}
 \mathfrak{dp}_2^{J,0}(\phi_1,\phi_2,\mathcal{O}_{6})&=-\sum\limits_{p=0}^{\infty}\sum\limits_{q=\text{max}(-p,p-2J_6)}^{p}2\kappa^{h_{21},h_{12}}_{2\bar{h}}\sin^{2}(\pi(h_6-h_1-h_2))
\nonumber \\ & \hspace{4cm} \times \mathcal{A}^{h_{21},h_{12}}_{p,q}\Omega^{h_1h_2h_1h_2}_{\bar{h},h_{6}+p,h_1+h_2}\frac{\Gamma(2(\bar{h}+q))}{\Gamma(h_{12}+\bar{h}+q)\Gamma(h_{21}+\bar{h}+q)}\bigg|_{\bar{h}=h_1+h_2+J}.
\end{align} 
Here we have defined $h=\frac{1}{2}(\Delta-J)$, $\bar{h}=\frac{1}{2}(\Delta+J)$, $h_{ij}=h_i-h_j$, and 
\begin{align}
\kappa^{a,b}_{2\bar{h}}=\frac{\Gamma(\bar{h}+a)\Gamma(\bar{h}-a)\Gamma(\bar{h}+b)\Gamma(\bar{h}-b)}{2\pi^{2}\Gamma(2\bar{h})\Gamma(2\bar{h}-1)}\,.
\end{align}

 The coefficients $\mathcal{A}^{a,b}_{p,q}$ come from performing dimensional reduction for 3d blocks in terms of the chiral, 2d blocks and were found for $a=b=0$ in  \cite{Hogervorst:2016hal}.\footnote{In comparison to \cite{Hogervorst:2016hal} we use $\mathcal{A}^{\text{here}}_{p,q}=\mathcal{A}^{\text{there}}_{\frac{p_q}{2},\bar{h}-h+q-p}$, and for Eq. (2.35) there we use $c^{(d)}_{\ell}=\frac{(d-2)_{\ell}}{\left(\frac{d-2}{2}\right)_{\ell}}$.} For general $a$ and $b$ we can compute $\mathcal{A}^{a,b}_{h,\bar{h}}$ recursively using the Casimir equation. For explicit results, we will mainly be interested in the large spin asymptotics, in which case we can restrict to $p=q=0$ and use $\mathcal{A}^{a,b}_{0,0}=1$. Finally, the function $\Omega$ is given by \cite{Liu:2018jhs}:
 \begin{align}
 \label{eq: definition of omega}
 \Omega^{h_1,h_2,h_3,h_4}_{h_5,h_6,p} &= \frac{\Gamma (2 h_5) \Gamma (h_6-p+1) \Gamma (h_5+h_{12}-h_6+p-1) \Gamma (-h_{12}+h_{34}+h_6-p+1)}{\Gamma (h_5+h_{12}) \Gamma (h_5+h_{34}) \Gamma (h_5-h_{12}+h_6-p+1)}
 \nonumber \\ &
\hspace{1cm}   \pFq{4}{3}{h_{23}+h_6,h_6-h_{14},-h_{12}+h_{34}+h_6-p+1,h_6-p+1}{2 h_6,h_5-h_{12}+h_6-p+1,-h_5-h_{12}+h_6-p+2}{1}
\nonumber\\ \nonumber &
+\frac{\Gamma (2 h_6) \Gamma (h_5+h_{13}+p-1) \Gamma (h_5+h_{42}+p-1) \Gamma (-h_5-h_{12}+h_6-p+1)}{\Gamma (h_6-h_{14}) \Gamma (h_{23}+h_6) \Gamma (h_5+h_{12}+h_6+p-1)} 
\nonumber \\ &
\hspace{1cm} \pFq{4}{3}{h_5+h_{13}+p-1,h_5+h_{42}+p-1,h_5+h_{34},h_5+h_{12}}{h_5+h_{12}+h_6+p-1,2 h_5,h_5+h_{12}-h_6+p}{1}\,.
 \end{align}
When we study double-twist operators with large spin, or equivalently large $\bar{h}$, the first $_4F_3$ hypergeometric yields the asymptotic, large-spin prediction while the second $_4F_3$ gives effects which are exponentially suppressed.  

By inserting \equref{eq: doublePoles1} and \equref{eq: doublePoles2} into  \equref{eq: general form for deltahP}, we can obtain $\delta h P$ for various double-twist operators of fermions. Below we will consider some examples.

\subsubsection*{$\delta h P$ of double-twist towers $[\phi\psi_\a]^\pm_{0,l_5}$ due to scalar exchange:}
From \equref{eq: parity even tower of two fermions and two scalars} and \equref{eq: parity odd tower of two fermions and two scalars}, we find that
\begin{subequations}
\begin{multline}
(\delta h P)_{11}^{(p)}=
-\lambda_{\f\f\f_6}\lambda_{\psi\psi\f_6}^1
\frac{ (-1)^{l_5+1} \Gamma \left(\Delta _6\right) \left(\sin \left(\pi  \left(-\Delta _{\psi }-\Delta _{\phi }+\Delta
	_6\right)\right)+\sin \left(\pi  \left(\Delta _{\psi }-\Delta _{\phi }\right)\right)\right)}{\Gamma \left(\frac{\Delta _6}{2}\right)}
\\ \x \Omega^{\frac{2 \Delta _{\psi}+1}{4} ,\frac{\Delta _{\phi }}{2},\frac{\Delta _{\phi }}{2},\frac{2 \Delta _{\psi }+1}{4}} _{\frac{2\Delta _{\psi }+2 \Delta _{\phi }+4l_5-1}{4},\frac{\Delta _6}{2},\Delta _{\phi }}
\end{multline}
and 
\begin{multline}
(\delta h P)_{22}^{(p)}=
-\lambda_{\f\f\f_6}\lambda_{\psi\psi\f_6}^1
\frac{ (-1)^{l_5} \Gamma \left(\Delta _6\right) \cos \left(\pi  \left(\Delta _{\psi }-\frac{\Delta _6}{2}\right)\right) \sin
	\left(\pi  \left(\Delta _{\phi }-\frac{\Delta _6}{2}\right)\right)}{\Gamma \left(\frac{\Delta _6}{2}\right)^2}
\\
\x\Bigg(
\frac{\left(2 l_5+1\right)}{l_5+1}
\Omega^{\frac{2 \Delta_{\psi }+1}{4},\frac{\Delta _{\phi }}{2},\frac{\Delta _{\phi }}{2},\frac{2 \Delta _{\psi }+1}{4}}
_{\frac{2\Delta _{\psi }+2\Delta _{\phi }+4l_5+3}{4},\frac{\Delta _6}{2},\Delta _{\phi }}
\\	
-\frac{4 \left(2 \Delta _{\phi
		}+2 l_5-1\right){}^2 \left(\Delta _{\psi }+\Delta _{\phi }+l_5-1\right)}{\left(2 \Delta _{\psi }+2 \Delta
		_{\phi }+2 l_5-3\right) \left(2 \Delta _{\psi }+2 \Delta _{\phi }+4 l_5-3\right) \left(2 \Delta _{\psi }+2 \Delta _{\phi }+4
		l_5-1\right)}
\Omega^{\frac{2 \Delta _{\psi}+1}{4},\frac{\Delta _{\phi }}{2},\frac{\Delta _{\phi }}{2},\frac{2 \Delta _{\psi }+1}{4}}_{\frac{2\Delta _{\psi }+2\Delta _{\phi }+4l_5-1}{4},\frac{\Delta _6}{2},\Delta _{\phi }} 
\Bigg)\,.
\end{multline}
\end{subequations}

By expanding the perturbative terms at large $l$, we can obtain the leading order behavior
\bea
(\delta h P)_{11}([\f\psi]^+_{0,l_5})\evaluated_{G^{(t)}_{\f_6}}\sim{}&{} \lambda_{\f\f\f_6}\lambda_{\psi\psi\f_6}^1 \frac{\sqrt{\pi } (-1)^{l_5+1} \Gamma \left(\Delta _6\right) 2^{-\Delta _{\psi }-\Delta _{\phi
		}-2 l_5+\frac{5}{2}} l_5^{\Delta _{\psi }+\Delta _{\phi }-\Delta _6-1}}{\Gamma \left(\frac{\Delta
		_6}{2}\right){}^2 \Gamma \left(-\frac{\Delta _6}{2}+\Delta _{\psi }+\frac{1}{2}\right) \Gamma \left(\Delta
	_{\phi }-\frac{\Delta _6}{2}\right)},
\\
(\delta h P)_{22}([\f\psi]^-_{0,l_5})\evaluated_{G^{(t)}_{\f_6}}\sim{}&{}
\frac{\lambda_{\f\f\f_6}\lambda_{\psi\psi\f_6}^1(-1)^{l_5} \left(-2 \Delta _{\psi }+\Delta _6+2\right) \Gamma \left(\frac{1}{2}
	\left(\Delta _6+1\right)\right) l_5^{\Delta
		_{\psi }+\Delta _{\phi }-\Delta _6-2}}{2^{\Delta _{\psi }+\Delta _{\phi }-\Delta _6+2 l_5+\frac{1}{2}}\Gamma \left(\frac{\Delta _6}{2}\right) \Gamma \left(-\frac{\Delta
		_6}{2}+\Delta _{\psi }+\frac{1}{2}\right) \Gamma \left(\Delta _{\phi }-\frac{\Delta _6}{2}\right)}.
\eea
By dividing these by the MFT coefficients at large $l$ give in \equref{eq: MFT coefficients at large spin for two fermion and two scalars}, we obtain the anomalous dimensions at leading order:
\bea 
\gamma_{[\f\psi]^+_{0,l_5}}\evaluated_{G^{(t)}_{\f_6}}={}&{}\frac{2}{P_{11}^{(s)}} (\delta h P)_{11}\evaluated_{G^{(t)}_{\f_6}}=
\frac{i\lambda_{\f\f\f_6}\lambda_{\psi\psi\f_6}^1}{l_5^{\Delta _6}}\frac{2^{\Delta _6} \Gamma \left(\frac{1}{2} \left(\Delta _6+1\right)\right) \Gamma
	\left(\Delta _{\psi }+\frac{1}{2}\right) \Gamma \left(\Delta _{\phi }\right)}{\sqrt{\pi } \Gamma
	\left(\frac{\Delta _6}{2}\right) \Gamma \left(\frac{1}{2} \left(-\Delta _6+2 \Delta _{\psi }+1\right)\right)
	\Gamma \left(\frac{1}{2} \left(2 \Delta _{\phi }-\Delta _6\right)\right)},
\\
\gamma_{[\f\psi]^-_{0,l_5}}\evaluated_{G^{(t)}_{\f_6}}={}&{}\frac{2}{P_{22}^{(s)}}(\delta h P)_{22}\evaluated_{G^{(t)}_{\f_6}} = -\frac{i\lambda_{\f\f\f_6}\lambda_{\psi\psi\f_6}^1}{l_5^{\Delta _6}}\frac{2^{\Delta _6-1} \left(\frac{\Delta _6}{2}-\Delta _{\psi }+1\right) \Gamma \left(\frac{\Delta
		_6+1}{2}\right)  \Gamma \left(\Delta _{\psi }+\frac{1}{2}\right) \Gamma \left(\Delta _{\phi
	}\right)}{\sqrt{\pi } \left(\Delta _{\psi }-1\right) \Gamma \left(\frac{\Delta _6}{2}\right) \Gamma
	\left(\frac{-\Delta _6+2 \Delta _{\psi }+1}{2}\right) \Gamma \left(\frac{2 \Delta
		_{\phi }-\Delta _6}{2}\right)}.
\eea 
\subsubsection*{$\delta h P$ of parity-even double-twist tower $[\phi\psi_\a]^+_{0,l_5}$ due to stress tensor exchange:}
By inserting \equref{eq: doublePoles1} and \equref{eq: doublePoles2} into \equref{eq: parity even tower of two fermions and two scalars}, we obtain
\begin{multline}
(\delta h P)_{11}([\f\psi]^+_{0,\ell_5})\evaluated_{G^{(t)}_{T}}=
\frac{3 \Delta _{\phi } \sin \left(\pi  l_5\right) \Gamma \left(l_5+\Delta _{\psi }\right) \Gamma
	\left(l_5+\Delta _{\phi }-\frac{1}{2}\right) \Gamma \left(l_5+\Delta _{\phi }+\Delta _{\psi
	}-\frac{3}{2}\right)}{2 \sqrt{2} \pi ^2 c_{T} \Gamma \left(l_5+1\right) \Gamma \left(\Delta _{\psi
	}-1\right) \Gamma \left(\Delta _{\phi }-\frac{1}{2}\right) \Gamma \left(2 l_5+\Delta _{\phi }+\Delta _{\psi
	}-\frac{1}{2}\right)}
\\\x\Bigg(
\left(2 \Delta _{\psi }+1\right) \left(2 \Delta _{\psi }+2 \Delta _{\phi }+4 l_5-3\right)\pFq{4}{3}{\frac{1}{2},\frac{1}{2},\frac{3}{2}-\Delta _{\phi },1-\Delta _{\psi }}{1,l_5+1,-l_5-\Delta _{\phi
	}-\Delta _{\psi }+\frac{5}{2}}{1}
\\
-4 \left(\Delta _{\psi }+l_5-1\right) \left(\Delta _{\psi }+\Delta _{\phi }+l_5-2\right)\pFq{4}{3}{\frac{1}{2},\frac{1}{2},\frac{3}{2}-\Delta _{\phi },2-\Delta _{\psi }}{1,l_5+1,-l_5-\Delta _{\phi
	}-\Delta _{\psi }+\frac{7}{2}}{1}
\\
+4 \left(l_5+\frac{1}{2}\right) \left(\Delta _{\phi }+l_5-\frac{1}{2}\right)
\pFq{4}{3}{\frac{1}{2},\frac{1}{2},\frac{3}{2}-\Delta _{\phi },2-\Delta _{\psi }}{1,l_5+2,-l_5-\Delta _{\phi
	}-\Delta _{\psi }+\frac{5}{2}}{1}
\Bigg)
\\+\text{ (non-perturbative terms)},
\end{multline}
where we set
\be 
\label{eq: conditions for stress tensor}
l_6=2\;,\quad \De_6=3\;,\quad \lambda_{\f\f T}=-\frac{3 \Delta_\phi  \Gamma \left(\frac{3}{2}\right)}{2 (2 \pi )^{3/2} \sqrt{c_T}}\;,\quad \lambda_{\psi\psi T}^1=\frac{3 i (\Delta_\psi -1)}{4 \sqrt{c_T}}\;,\quad  \lambda_{\psi\psi T}^2=-\frac{3 i}{2 \sqrt{c_T}},
\ee 
where $c_{T}$ is the central charge, which here is defined as the normalization of the stress tensor two-point function. At large spin, the leading order term is then
\be
(\delta h P)_{11}([\f\psi]^+_{0,l_5})\evaluated_{G^{(t)}_{T}}\sim \frac{3 i (-1)^{l_5} \Delta _{\phi } 2^{-\Delta _{\psi }-\Delta _{\phi }-2 l_5+4} l_5^{\Delta _{\psi }+\Delta
		_{\phi }-2}}{\pi ^{3/2} c_T\Gamma \left(\Delta _{\psi }-1\right) \Gamma \left(\Delta _{\phi
	}-\frac{1}{2}\right)},
\ee
and we obtain the anomalous dimension via \equref{eq: MFT coefficients at large spin for two fermion and two scalars}:
\be 
\gamma_{[\f\psi]^+_{0,l_5}}\evaluated_{G^{(t)}_{T}}=\frac{2}{P_{11}^{(s)}} (\delta h P)_{11}\evaluated_{G^{(t)}_{T}}= \frac{1}{ l_5 }\frac{3 \sqrt{2} \Gamma \left(\Delta _{\psi }+\frac{1}{2}\right) \Gamma \left(\Delta _{\phi }+1\right)}{\pi ^2
	c_T\Gamma \left(\Delta _{\psi }-1\right) \Gamma \left(\Delta _{\phi }-\frac{1}{2}\right)}.
\ee 

\subsubsection*{$\delta h P$ of parity-even double-twist tower $[\psi\psi]^+_{0,l_5}$ due to the exchange of a generic parity-even operator or parity-odd scalar:}
Similar to the previous examples, we obtain the relevant $\delta h P$ by inserting \equref{eq: doublePoles1} and \equref{eq: doublePoles2} into \equref{eq: parity even tower of four fermions}. As the expressions are quite lengthy, we will not reproduce the full results but instead present their asymptotic forms at large spin. We find that 
\be
\label{eq: delta hP for four fermion}
(\delta h P)_{11}([\psi\psi]^+_{0,l_5})\evaluated_{G^{(t)}_{\cO_{6}}}={}&{}(\delta h P)_{12}([\psi\psi]^+_{0,l_5})\evaluated_{G^{(t)}_{\cO_6}}=(\delta h P)_{21}([\psi\psi]^+_{0,l_5})\evaluated_{G^{(t)}_{\cO_6}}=0,
\\
(\delta h P)_{22}([\psi\psi]^+_{0,l_5})\evaluated_{G^{(t)}_{\cO_6}}\sim{}&{}\left(\lambda_{\psi\psi\cO_6}^1\right)^2\frac{(-1)^{l_5+1} 2^{-2 \Delta _{\psi }+\Delta _6-2 l_5+l_6+2} l_5^{2 \Delta _{\psi }-\Delta _6+l_6-\frac{1}{2}}
	\Gamma \left(\frac{1}{2} \left(l_6+\Delta _6+1\right)\right)}{\Gamma \left(\frac{1}{2} \left(l_6+\Delta
	_6\right)\right) \Gamma \left(\frac{1}{2} \left(l_6-\Delta _6+1\right)+\Delta _{\psi }\right)^2},
\ee
where contributions due to $\lambda_{\psi\psi\cO_6}^{2,3,4}$ come at subleading order.\footnote{We see that this result matches the one calculated using lightcone bootstap methods in \cite{Albayrak:2019gnz}, once the change of basis and the difference in conformal block normalization is taken into account; compare \equref{eq: delta hP for four fermion} here with (3.30c) there.} By dividing by the MFT coefficients given in \equref{eq: MFT coefficients for four fermions}, we obtain the anomalous dimension at large spin:
\be 
\gamma_{[\psi\psi]^+_{0,l_5}}\evaluated_{G^{(t)}_{\cO_6}}=\frac{2}{P_{22}^{(s)}}(\delta h P)_{22} \evaluated_{G^{(t)}_{\cO_6}}=
\frac{\left(\lambda_{\psi\psi\cO_6}^1\right)^2}{l_5^{\Delta _6-l_6}}\frac{ 2^{\Delta _6+l_6} \Gamma \left(\Delta _{\psi
	}+\frac{1}{2}\right){}^2 \Gamma \left(\frac{1}{2} \left(l_6+\Delta _6+1\right)\right)}{\sqrt{\pi } \Gamma
	\left(\frac{1}{2} \left(l_6+\Delta _6\right)\right) \Gamma \left(\frac{1}{2} \left(l_6-\Delta
	_6+1\right)+\Delta _{\psi }\right)^2}.
\ee 
For example, for stress tensor exchange we can impose \equref{eq: conditions for stress tensor} which yields
\be 
\label{eq: anomalous dimension of parity even fermion fermion double-twist operator due to stress tensor}
\gamma_{[\psi\psi]^+_{0,l_5}}\evaluated_{G^{(t)}_{T}}=
-\frac{1}{l_5}\frac{48 \Gamma \left(\Delta _{\psi }+\frac{1}{2}\right)^2}{\pi c_T \Gamma \left(\Delta _{\psi
	}-1\right)^2},
\ee 
whereas for parity-even scalar exchange it becomes
\be 
\label{eq: anomalous dimension of parity even fermion fermion double-twist operator due to parity even scalar}
\gamma_{[\psi\psi]^+_{0,l_5}}\evaluated_{G^{(t)}_{\f_6}}=
\frac{\left(\lambda_{\psi\psi\cO_6}^1\right)^2}{l_5^{\De_6}}\frac{2^{\Delta _6} \Gamma \left(\frac{\Delta _6+1}{2}\right) \Gamma \left(\Delta _{\psi
	}+\frac{1}{2}\right)^2}{\sqrt{\pi } \Gamma \left(\frac{\Delta _6}{2}\right) \Gamma \left(-\frac{\Delta
		_6}{2}+\Delta _{\psi }+\frac{1}{2}\right)^2}.
\ee 

For the exchange of a parity-odd scalar in the crossed channel, we still have $(\delta h P)_{ij}=0$ unless $i=j=2$, which now becomes
\be
\label{eq: delta hP for four fermion 2}
(\delta h P)_{22}([\psi\psi]^+_{0,l_5})\evaluated_{G^{(t)}_{\f_6}}\sim{}&{}\left(\lambda_{\psi\psi\f_6}^3\right)^2\frac{(-1)^{l_5} \Gamma \left(\frac{\Delta _6}{2}\right) 2^{-2 \Delta _{\psi }+\Delta _6-2 l_5+1} l_5^{2 \Delta
		_{\psi }-\Delta _6-\frac{3}{2}}}{\Gamma \left(\frac{1}{2} \left(\Delta _6+1\right)\right) \Gamma \left(\Delta
	_{\psi }-\frac{\Delta _6}{2}\right){}^2},
\ee
from which we can extract the anomalous dimension as
\be 
\label{eq: anomalous dimension of parity even fermion fermion double-twist operator due to parity odd scalar}
\gamma_{[\psi\psi]^+_{0,l_5}}\evaluated_{G^{(t)}_{\f_6}}=\frac{2}{P_{22}^{(s)}}(\delta h P)_{22} \evaluated_{G^{(t)}_{\f_6}}= \frac{\left(\lambda_{\psi\psi\f_6}^3\right)^2}{l_5^{\De_6+1}} \frac{2^{\Delta _6-1} \Gamma \left(\frac{\Delta _6}{2}\right) \Gamma \left(\Delta _{\psi
	}+\frac{1}{2}\right){}^2}{\sqrt{\pi } \Gamma \left(\frac{\Delta _6+1}{2} \right) \Gamma
	\left(\Delta _{\psi }-\frac{\Delta _6}{2}\right)^2}.
\ee 
We see that the anomalous dimensions in \equref{eq: anomalous dimension of parity even fermion fermion double-twist operator due to stress tensor}, \equref{eq: anomalous dimension of parity even fermion fermion double-twist operator due to parity even scalar}, and \equref{eq: anomalous dimension of parity even fermion fermion double-twist operator due to parity odd scalar} match precisely to the results computed using large-spin expansions in \cite{Albayrak:2019gnz}.
\section{Conclusion}
\label{sec:conclusion}

In this paper we combined ideas from harmonic analysis for the Euclidean conformal group \cite{Dobrev:1977qv,Gadde:2017sjg,Karateev:2017jgd,Kravchuk:2018htv,Karateev:2018oml} and the Lorentzian inversion formula \cite{Caron-Huot:2017vep,Simmons-Duffin:2017nub,Liu:2018jhs} to derive the $6j$ symbol for fermionic operators in $3d$ CFTs. That is, by using the Euclidean representation of the $6j$ symbol we were able to spin-down the fermionic $6j$ symbols to the scalar $6j$ symbols, which in practice are computed with the Lorentzian formula. As an application we computed the Mean Field Theory OPE coefficients using the Euclidean inversion formula. We also used the relation between conformal partial waves and conformal blocks to study the inversion of a single block in a fermionic correlator. This determines corrections to the anomalous dimensions and OPE coefficients of the double-twist operators, including both perturbative and non-perturbative terms in the large spin expansion.

We believe there are many interesting open questions still to consider in this program. Since the inclusion of non-perturbative effects in the large spin expansion improves analytic predictions for the Ising and $O(2)$ model it is natural to ask if the same improvement can be seen in fermionic CFTs. For example, can one make new precise analytic predictions for the Gross-Neveu-Yukawa models which can be compared with the numerical bootstrap~\cite{Iliesiu:2015qra,Iliesiu:2017nrv}? In 3d $\mathcal{N}=1$ SCFTs such as the supersymmetric Ising~\cite{Rong:2018okz,Atanasov:2018kqw} or Wess-Zumino~\cite{Rong:2019qer} models, can one predict analytical trajectories that cannot be accessed using scalar correlators? More generally in SCFTs it will be interesting to understand the implications of imposing analytical bootstrap constraints for all external operators in the same supermultiplet.

Furthermore, by understanding how to spin down a fermionic $6j$ symbol in $3d$ it is also now straightforward to go to higher spin. As a simple example, our results could then also be used to study correlation functions of conserved currents $J_{\mu}$ in the $O(N)$ vector model. There are many physically relevant observables, such $\<JJT\>$, which are only accessible with spinning correlators. Based on \cite{Simmons-Duffin:2016wlq,Caron-Huot:2017vep,Albayrak:2019gnz} we now know that the current and stress tensor lie on the double-twist trajectories composed of the fundamental scalars, $\phi$, so these correlators are now within reach of analytic methods.

Finally, we note that our results are directly applicable to the study of Witten diagrams with external fermionic operators. For example, by studying the contribution of the stress tensor $T^{\mu\nu}$ to a fermionic correlator, e.g. $\<\phi \psi \psi \phi\>$, we can derive the binding energy for a two-particle state dual to $[\phi \psi]_{n,\ell}$ due to tree-level graviton exchange. The anomalous dimension, or corresponding $6j$ symbol, can then be used to bootstrap a graviton loop in $AdS_{4}$ \cite{Aharony:2016dwx,Liu:2018jhs}. In general, if one wants to study an AdS theory with fermions, we need to understand the tree-level fermionic correlators to fully determine a one-loop scalar four-point function. We therefore hope the results presented here are useful in the wider study of AdS$_{4}$ correlators.

\section*{Acknowledgments}

We thank Petr Kravchuk, Eric Perlmutter, Valentina Prilepina, and David Simmons-Duffin for relevant discussions. SA also thank Petr Kravchuk and David Simmons-Duffin for sharing their previous work on  computations with weight shifting operators.  The research of DP and SA is supported  by DOE grant no.\ DE-SC0020318 and Simons  Foundation grant 488651 (Simons Collaboration on the Nonperturbative Bootstrap). DP and SA also thank the Walter Burke Institute for Theoretical Physics at Caltech for its support and hospitality during the completion of this work. The research of DM is supported by the Walter Burke Institute for Theoretical Physics at Caltech and the Sherman Fairchild Foundation. This material is based upon work supported by the U.S. Department of Energy, Office of Science, Office of High Energy Physics, under Award Number DE-SC0011632.

\appendix

\section{Conventions}
\label{sec:conventions_fermions}
\subsection{Review of Embedding Formalism}
\label{sec:EmbeddingReview}

Our conventions are identical to those of our previous paper \cite{Albayrak:2019gnz} so we will be brief here. First, we will use the mostly plus metric. Since we are studying fermions, we will need the double cover of the $3d$ conformal group $\SO(3,2)$, which is $\Sp(2,\mathbb{R})$. As is well-known, the conformal group can be realized linearly in 2 higher dimensions \cite{Dirac:1936fq,Weinberg:2010fx,Costa:2011mg}. Here this means we embed representations of $\Sp(2,\mathbb{R})$  into projective, null representations of $\Sp(4,\R)$ in embedding space. By imposing the appropriate constraints on the higher-dimensional fields and structures we recover the action of the conformal group on the Poincar\'e section,
\be 
X^A \rightarrow (x^\mu,1,x^2)\;,
\ee 
where we are working in the lightcone coordinates $X^A=(X^\mu,X^+,X^-)$, and $X^\pm$ are related to the Cartesian coordinates as $X^\pm=X^4\pm X^3$. 

Here we introduce auxiliary spinors \cite{Iliesiu:2015qra},
\be \cO(x,s)={}&s_{\a_1}s_{\a_2}\dots s_{\a_{2l}} \cO^{\a_1\a_2\dots \a _{2l}}(x),\\
\cO^{\a_1\a_2\dots \a _{2l}}(x)={}&\frac{1}{(2l)!}\frac{\partial^{2l}}{\partial_{s_{a_1}}\partial_{s_{a_2}}\dots \partial_{s_{a_{2l}}}}\cO(x,s),
\label{eq: spinor convention}
\ee 
and we recover the original field
\begin{equation}
\Psi(X,S)=\frac{1}{(X^+)^{\Delta_\psi}}\psi(x,s)
\end{equation}
by going to the Poincar\'e section and setting
\begin{equation}
S_I=\sqrt{X^+}\begin{pmatrix}
s_\alpha\\ x^\alpha_{\;\;\beta}s^\beta
\end{pmatrix}\;.
\end{equation}
Here we use the matrices $\gamma$ and $\Gamma$ to convert the indices,
\begin{equation}
X^I_{\;\;J}\equiv X^A(\Gamma_A)^I_{\;\;J}\;,\qquad x^\alpha_{\;\;\beta}\equiv x^\mu(\gamma_\mu)^\alpha_{\;\;\beta}\;.
\end{equation}
The $3d$ gamma matrices are defined as
\be 
(\gamma_0)^\a _{\;\;\b}=i(\sigma_2)_{\a \b}\;,\quad
(\gamma_1)^\a _{\;\;\b}=(\sigma_1)_{\a \b}\;,\quad
(\gamma_2)^\a _{\;\;\b}=(\sigma_3)_{\a \b}\;,
\ee 
where $\sigma_i$ are the standard Pauli matrices:
\bea 
\sigma_1=\left(
\begin{array}{cc}
	0 & 1 \\
	1 & 0 \\
\end{array}
\right)
\quad,\quad\sigma_2=\left(
\begin{array}{cc}
	0 & -i \\
	i & 0 \\
\end{array}
\right)
\quad,\quad\sigma_3=\left(
\begin{array}{cc}
	1 & 0 \\
	0 & -1 \\
\end{array}
\right).
\eea 
We raise and lower the spinor indices with $\e_{\a\b}=\e ^{\a \b}=i(\sigma_2)_{\a \b}$, e.g.~$x_\a=\epsilon_{\a\b}x^\b$ and $x^\a=x_\b\epsilon^{\b\a}$. The embedding space gamma matrices and $\Sp(4,\mathbb{R})$ invariant is then defined as:
\be 
\Omega=\epsilon\otimes\mathbbm{I}
\;,\quad
\Gamma_0=\gamma_2\otimes\gamma_0
\;,\quad
\Gamma_1=\mathbbm{I}\otimes\gamma_1
\;,\quad
\Gamma_2=\mathbbm{I}\otimes\gamma_2
\;,\quad
\Gamma_3=\gamma_0\otimes\gamma_0
\;,\quad
\Gamma_4=\gamma_1\otimes\gamma_0
\ee 
with the embedding space metric $g_{IJ}=\text{diag}\left(-,+,+,+,-\right)$. 

In lightcone coordinates, the gamma matrices take the form
\be 
(\G^\mu)^I_{\;\;J}=\begin{pmatrix}
	\left(\gamma^\mu\right)^{\a}_{\;\;\b} & 0 \\
	0 & \left(\gamma^\mu\right)_{\a}^{\;\;\b}
\end{pmatrix}\;,\quad 
(\G^+)^I_{\;\;J}=\begin{pmatrix}
	0 & 2\e^{\a\b}\\ 0 & 0
\end{pmatrix}\;,\quad 
(\G^-)^I_{\;\;J}=\begin{pmatrix}
	0 & 0\\
	2\e_{\a\b} & 0
\end{pmatrix}.
\ee 

\subsection{Two and Three Point Functions}
To describe the embedding space spinor structures, we define
\be
\<S_1X_2X_3\dots X_{n-1}S_n\>\equiv&-S_1\cdot X_2\cdot X_3\cdots X_{n-1}\cdot S_n\\=& -(S_{1})_I(X_2)^I_{\;\;J}(X_3)^J_{\;\;K}\cdots(X_{n-1})^L_{\;\;M}(S_n)^M.
\ee
We normalize the operators such that the two-point function takes the form
\be
\<\cO^{\De,l}(X_1,S_1)\cO^{\De,l}(X_2,S_2)\>= i^{2l}\frac{\<S_1S_2\>^{2l}}{X_{12}^{2\De+2l}}. \label{eq: 2pt convention}
\ee 
This is the unique result, but once we go to three points there are multiple structures to consider. We follow the conventions of \cite{Iliesiu:2015qra,Iliesiu:2015akf} and write them as:
\bea[eq: 3pt structures]
\<\phi_1\phi_2\cO_3\>&=\frac{\<S_3X_1X_2S_3\>^l}{X_{12}^{(\De_{123}+l)/2}X_{23}^{(\De_{231}+l)/2}X_{31}^{(\De_{312}+l)/2}},\label{eq: 3pt structure scalar scalar spin l}
\\
\<\psi_1\phi_2\cO_3\>^1&=\frac{\<S_1S_3\>\<	S_3X_1X_2S_3\>^{l-\frac{1}{2}}}{X_{12}^{(\De_{123}+l-\half)/2}X_{23}^{(\De_{231}+l-\half)/2}X_{31}^{(\De_{312}+l+\half)/2}},
\\
\<\psi_1\phi_2\cO_3\>^2&=\frac{\<S_1X_2S_3\>\<	S_3X_1X_2S_3\>^{l-\frac{1}{2}}}{X_{12}^{(\De_{123}+l+\half)/2}X_{23}^{(\De_{231}+l+\half)/2}X_{31}^{(\De_{312}+l-\half)/2}},
\\
\<\psi_1\psi_2\cO_3\>^1&=
\frac{\<S_1S_2\>\<	S_3X_1X_2S_3\>^{l}}{X_{12}^{(\De_{123}+l+1)/2}X_{23}^{(\De_{231}+l)/2}X_{31}^{(\De_{312}+l)/2}},
\\
\<\psi_1\psi_2\cO_3\>^2&=
\frac{\<S_1S_3\>\<S_2S_3\>\<	S_3X_1X_2S_3\>^{l-1}}{X_{12}^{(\De_{123}+l-1)/2}X_{23}^{(\De_{231}+l)/2}X_{31}^{(\De_{312}+l)/2}},
\\
\<\psi_1\psi_2\cO_3\>^3&=
\frac{\left(X_{23}\<S_1S_3\>\<S_2X_1S_3\>+X_{13}\<S_2S_3\>\<S_1X_2S_3\>\right)\<	S_3X_1X_2S_3\>^{l-1}}{X_{12}^{(\De_{123}+l-1)/2}X_{23}^{(\De_{231}+l)/2}X_{31}^{(\De_{312}+l)/2}},
\\
\<\psi_1\psi_2\cO_3\>^4&=
\frac{\left(X_{23}\<S_1S_3\>\<S_2X_1S_3\>-X_{13}\<S_2S_3\>\<S_1X_2S_3\>\right)\<	S_3X_1X_2S_3\>^{l-1}}{X_{12}^{(\De_{123}+l-1)/2}X_{23}^{(\De_{231}+l)/2}X_{31}^{(\De_{312}+l)/2}},
\eea 
where $X_{ab}\equiv-2X_a\. X_b$ and $\De_{abc}\equiv\De_a+\De_b-\De_c$.

For integer spin we can also convert to vector notation,
\be 
\label{eq: spinor-vector transition}
\cO^{\a_1\dots \a _{2J}}=&\cO^{\mu_1\dots\mu_J}\gamma_{\mu_1}^{\a _1 \a _2}\cdots \gamma_{\mu_J}^{\a _{2J-1} \a _{2J}},
\\
\cO^{\mu_1\dots\mu_J}=&\left(-\half\right)^J\gamma^{\mu_1}_{\a _1 \a _2}\cdots \gamma^{\mu_J}_{\a _{2J-1} \a _{2J}}\cO^{\a_1\dots \a _{2J}},
\ee 
where $\cO^{\mu_1\dots\mu_J}$ is a symmetric traceless tensor. If we introduce the auxiliary polarization vectors,
\be 
z_{\mu_1}\dots z_{\mu_l}\cO^{\mu_1\dots\mu_l}=s_{\a_1}\dots s_{\a_{2l}}\cO^{\a_1\dots\a_{2l}},
\ee 
we find the following relation:
\be 
z_\mu=s_\a s_\b \gamma^{\a\b}.
\ee 
We can now use these relations to convert two and three-point functions to vector notation for integer spin.  In particular, \equref{eq: 2pt convention} and \equref{eq: 3pt structure scalar scalar spin l} become
\bea
\<\cO^{\De,l}(X_1,Z_1)\cO^{\De,l}(X_2,Z_2)\>=&\frac{1}{2^l}\frac{H_{12}^l}{X_{12}^{\De_l}},\\
\<\f(X_1,Z_1)\f(X_2,Z_2)\cO(X_3,Z_3)\>=&\frac{V_3^l}{X_{12}^{(\De_{123}-l)/2}X_{23}^{(\De_{231}+l)/2}X_{31}^{(\De_{312}+l)/2}},
\eea 
where we define
\be 
H_{12}\equiv{} &{}-2\left[(Z_1\.Z_2)(X_1\.X_2)-(X_1\.Z_2)(Z_1\.X_2)\right],\\
V_{3}\equiv {}& {}\frac{(Z_3\.X_1)(X_2\.X_3)-(Z_3\.X_2)(X_1\.X_3)}{X_1\.X_2},
\ee 
in the conventions of \cite{Costa:2011mg}.

\section{Partial Waves and Conformal Blocks}
\label{app:Scalar_Conventions}
In this appendix we will briefly review the relation between the conformal partial wave expansion and the conformal block expansion. The goal is to establish the general dictionary between the two for general four-point functions. The method we use is not new, but it will be useful to present the results in our conventions, taking care of signs with fermionic operators. As in the rest of the paper, we work in $d=3$ and suppressed indices will always go from southwest to northeast. 

First recall we define our definition for the partial wave and shadow transform is:
\begin{align}
\Psi^{(s);ab}_{\cO_5}&=\int d^{d}x_{5}\<\cO_1\cO_2\cO_5\>^a\<\cO_3\cO_4\widetilde{\cO}_{5}\>^b, \label{eqn:DefCPWApp}
\\
\mathbf{S}[\cO](x)&=\int d^dy \cO(y)\<\cO(y)\cO(x)\>.
\end{align}

We will also find it useful to define the kinematic functions $C$:
\begin{align}
\lim\limits_{x_1\rightarrow x_2}\<\cO_1(x_1)\cO_2(x_2)\cO_5(x_5)\>^a\sim C^{a}_{\cO_1\cO_2\cO_5}(x_{12})\<\cO_2(x_2)\cO_2(x_5)\>.
\end{align}

We can then define s-channel conformal blocks for $\<\cO_1\cO_2\cO_3\cO_4\>$ as solutions to the conformal Casimir equation with the following behavior in the limit $x_3\rightarrow x_4$ and $x_1\rightarrow x_2$:
\begin{align}
G^{(s);ab}_{\cO_5}(x_i)\approx C^{p}_{\cO_1\cO_2\cO_5}(x_{12})C^{q}_{\cO_3\cO_4\cO_5}(x_{34})\<\cO_{5}(x_2)\cO_{5}(x_4)\>.
\end{align}
Here we work in Euclidean space and the order of limits does not matter.

With this definition the four-point function has the following conformal block expansion:
\begin{align}
\<\cO_1\cO_2\cO_3\cO_4\>=\sum\limits_{\cO}\lambda_{\cO_1\cO_2\cO}^a\lambda_{\cO_3\cO_4\cO}^bG^{(s);ab}_{\cO}(x_i),
\end{align}
where we define the OPE coefficients by:
\begin{align}
\<\cO_1\cO_2\cO_3\>_{\Omega}=\lambda_{\cO_1\cO_2\cO_3}^a\<\cO_1\cO_2\cO_3\>^{a}.
\end{align}

Now we have to expand the conformal partial wave as a sum of two conformal blocks. To extract their coefficients, we just need to study the integrand in certain limits. We start by taking the limit $x_1\rightarrow x_2$ under the integrand in (\ref{eqn:DefCPWApp}) and then performing the $x_5$ integral. In this limit we have:
\begin{align}
\Psi^{(s);ab}_{\cO_5}(x_i)&\supset \int d^{d}x_{5}C^a_{\cO_1\cO_2\cO_5}(x_{12})\<\cO_5(x_2)\cO_5(x_5)\>\<\cO_3\cO_5\widetilde{\cO}_{5}(x_5)\>^b \nonumber
\\ &\supset C^a_{\cO_1\cO_2\cO_5}(x_{12})S^{b}_{c}(\cO_3\cO_4[\widetilde{\cO}_{5}])\<\cO_3\cO_4\cO_5(x_2)\>^c.
\end{align}
To get the second line we have to reorder the operators in the two-point function and implicitly raise and lower the spinor indices, so the two possible signs cancel. Taking the $x_3\rightarrow x_4$ limit we find:
\begin{align}
\Psi^{(s);ab}_{\cO_5}(x_i)&\supset C^a_{\cO_1\cO_2\cO_5}(x_{12})S^{b}_{c}(\cO_3\cO_4[\widetilde{\cO}_{5}])C^{c}_{\cO_3\cO_4\cO_5}(x_{34})(-1)^{\Sigma_{55}}\<\cO_{5}(x_2)\cO_5(x_4)\>.
\end{align}

To get the coefficient for the other block we take the limit $x_3\rightarrow x_4$ under the integrand, perform the $x_5$ integral and then take the limit $x_1\rightarrow x_2$:
\begin{align}
\Psi^{(s);ab}_{\cO_5}(x_I)&\supset S^{a}_{c}(\cO_3\cO_4[\cO_{5}])C^{c}_{\cO_1\cO_2\cO_5}(x_{12})C^b_{\cO_3\cO_4\cO_5}(x_{34})(-1)^{\Sigma_{55}}\<\cO_{5}(x_2)\cO_5(x_4)\>.
\end{align}

We therefore find the full partial wave is:
\begin{align}
\Psi^{(s);ab}_{\cO_5}(x_i)=(-1)^{\Sigma_{55}}\left[S^{b}_{c}(\cO_3\cO_4[\widetilde{\cO}_{5}])G^{ac}_{\cO_5}(x_i)+S^{a}_{c}(\cO_1\cO_2[\cO_5])G^{cb}_{\widetilde{\cO}_{5}}(x_i)\right].
\end{align}

Using the form of the partial wave expansion:
\begin{align}
\<\cO_1\cO_2\cO_3\cO_4\>=\<\cO_1\cO_2\>\<\cO_3\cO_4\>+\int\limits_{\mathcal{C}}d\cO \rho^{(s)}_{ab}(\Delta,J)\Psi^{ab}_{\cO},
\end{align}
we find the following relation between the OPE function and the OPE coefficients:
\begin{align}
\lambda_{\cO_1\cO_2\cO_5}^a\lambda_{\cO_3\cO_4\cO_5}^b=-\text{Res}_{\Delta=\Delta_{5}}\rho^{(s)}_{ac}(-1)^{\Sigma_{JJ}}S^{c}_{b}(\cO_3\cO_4[\widetilde{\cO}_{5}])\bigg|_{J=J_5},
\end{align}
where we first set $J=J_5$ and then evaluate the residue.

\section{Two and Three Point Pairings}

\subsection{Two Point Pairings and Plancherel Measure}
\label{sec: two point pairing}

Let us first consider the pairing of scalar two-point functions. It reads as 
\be 
{}\left(\<\phi(x_1)\phi(x_2)\>,\<\tl\phi(x_1)\tl\phi(x_2)\>\right)
=\,&\int \frac{d^dx_1d^dx_2}{\vol(\SO(d+1,1))}\<\phi(x_1)\phi(x_2)\>\<\tl\phi(x_1)\tl\phi(x_2)\>
\\
=\,& \frac{1}{2^d\vol(\SO(1,1))\x\vol(\SO(d))}\<\phi(0)\phi(\infty)\>\<\tl\phi(0)\tl\phi(\infty)\>,
\ee 
where $\vol(\SO(1,1))\x\vol(\SO(d))$ is the stabilizer group for two points and the factor $2^d$ is the Fadeev-Popov determinant.

As we define an operator at infinity as 
\be 
\cO(\infty)\equiv \lim\limits_{L\rightarrow\infty} L^{2\De}\cO(\hat{e}L)
\ee 
for a unit vector $\hat e$. We have $\<\tl\phi(0)\tl\phi(\infty)\>=1$ in our conventions, meaning
\be 
\left(\<\phi(x_1)\phi(x_2)\>,\<\tl\phi(x_1)\tl\phi(x_2)\>\right)
=\frac{1}{64\pi^2\vol(\SO(1,1))}.
\label{eq: scalar two point pairing}
\ee 

Like we have done for the three-point structures, we can use weight-shifting operators to relate two-point functions as well. We can rewrite the two-point function of $\<\cO\cO\>^{\Delta,J}$ in terms of weight-shifting operators $\cD^{a,b}$ acting on $\<\cO\cO\>^{\Delta-a,J-b}$, integrate by parts, and act with the adjoint weight-shifting operators $\left(\cD^{a,b}\right)^*\propto \cD^{a,-b}$ on the other two-point function. Schematically,
\begin{center}
	\includegraphics[scale=1]{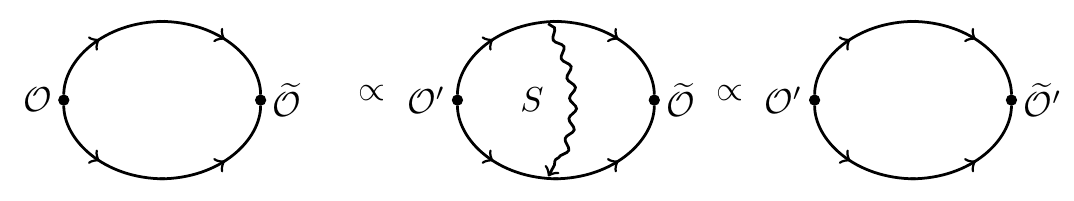}.
\end{center}
We can find the coefficient between the first two diagrams above by direct calculation. For example, if we choose $a=b=\half$, we have
\be 
\<\cO_1\cO_2\>^{\Delta,J}=
-\frac{i \cD^{(+,+)}_{1A}\cD^{(+,+)A}_{2}}{16 \left(\Delta -2\right) \left(\Delta -\frac{3}{2}\right) \left(\Delta
	+J-2\right) \left(\Delta +J-1\right)}
\<\cO_1\cO_2\>^{\De-\half,J-\half},
\ee 
where we can integrate by parts and carry these differential operators to the other two-point function using
\be
\left(\cD^{+,+}_\a \cO^{\De-\half,J-\half}\right)\. \cO^{3-\De,J}= \cO^{\De-\half,J-\half} \. \left(\left(\cD^{+,+}_\a\right)^* \cO^{3-\De,J}\right)
=-\frac{1}{2J} \cO^{\De-\half,J-\half} \. \left(\cD^{+,-}_\a \cO^{3-\De,J}\right).
\ee

Carrying out the calculation, we find that
\be 
\left(\<\cO_1\cO_2\>^{\Delta,J}, \<\cO_1\cO_2\>^{3-\Delta,J}\right)
= \frac{2J+1}{2J}
\left(\<\cO_1\cO_2\>^{\Delta,J-\half},\<\cO_1\cO_2\>^{3-\Delta,J-\half}\right).
\ee 
Note that this recursion relation is independent of which weight-shifting operator we choose: we get exactly the same relation for all $a,b=\pm\half$ choices.

Using \equref{eq: scalar two point pairing}, we get
\be 
\left(\<\cO_1\cO_2\>^{\Delta,J},\<\cO_1\cO_2\>^{3-\Delta,J}\right)
=\frac{2J+1}{64\pi^2\vol(\SO(1,1))}.
\label{eq: spinning two point pairing}
\ee 

We can use two explicit expressions for the two-point function to compute the Plancherel measure. It is easy to see this diagrammatically:
\begin{center}
	\includegraphics[scale=1.8]{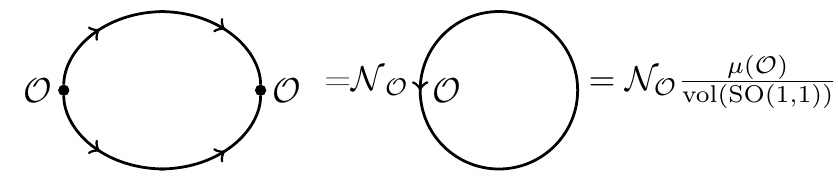},
\end{center}
where we first make use of $\mathbf{S}^2=\cN$ to convert the pairing into a circle and then identify the circle as the Plancherel measure up to the volume factor; see \cite{Karateev:2018oml} for details. Therefore, we conclude that
\be 
\mu(\Delta,J)=\frac{\vol(\SO(1,1))}{\cN_{\Delta,J}}
\<\cO_1\cO_2\>^{\Delta,l}\.\<\cO_1\cO_2\>^{3-\Delta,l},
\ee 
and we compute it as
\be 
\mu(\Delta,J)=\frac{(2 \Delta -3) (-\Delta +l+2) (\Delta +l-1) \Gamma (2 l+2) \cot (\pi  (\Delta +l))}{128
	\pi ^5}.
\ee 

\subsection{Three Point Pairings}
In our conventions we have
\begin{multline}
\left(\<\phi_1\phi_2\cO\>,\<\tl\phi_1\tl\phi_2\tl\cO\>\right)={}(-2)^J\int\frac{d^dx_1d^dx_2d^dx_3}{\vol(\SO(d+1,1))}\<\phi_1(x_1)\phi_2(x_2)\cO_{\mu_1\dots\mu_J}(x_3)\>\\\x\<\tl\phi_1(x_1)\tl\phi_2(x_2)\tl\cO^{\mu_1\dots\mu_J}(x_3)\>,
\end{multline}
which can be calculated by gauge fixing
\be 
\left(\<\phi_1\phi_2\cO_J\>,\<\tl\phi_1\tl\phi_2\tl\cO_J\>\right)=\frac{(-1)^J\widehat C_J(1)}{2^{d-J}\;\vol(\SO(d-1))},
\ee
where $2^d$ is the appropriate Fadeev-Popov determinant. In $3d$ this reads as
\be 
\left(\<\phi_1\phi_2\cO_J\>,\<\tl\phi_1\tl\phi_2\tl\cO_J\>\right)=\frac{(-1)^J\G\left(J+1\right)}{16\sqrt{\pi}\G\left(J+\frac{1}{2}\right)},
\label{eq: scalar scalar three point pairing}
\ee 
where we have the convention $\vol(\SO(n))=\vol(\SO(n-1))\vol(S^{n-1})$ and we used $\vol(\SO(2))=2\pi$. As what really matters is only the ratios of group volumes, this choice does not affect any physical result.

The pairing of spinning three-point functions can be calculated by reducing them via the weight-shifting operators and using the scalar pairing above. Schematically,
\begin{center}
	\includegraphics[scale=1.8]{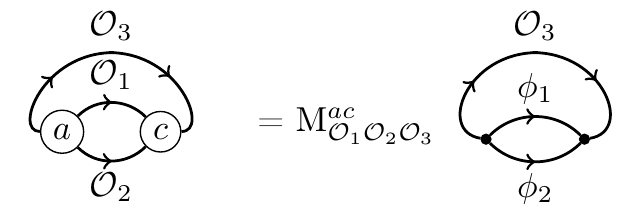}.
\end{center}

The procedure to calculate the matrix $M^{ac}$ is as follows. We first expand the $\<\psi\psi\cO\>^a$ and $\<\psi\phi\cO\>^a$ three-point functions in terms of $\<\phi\phi\cO\>$:
\begin{center}
	\includegraphics[scale=.8]{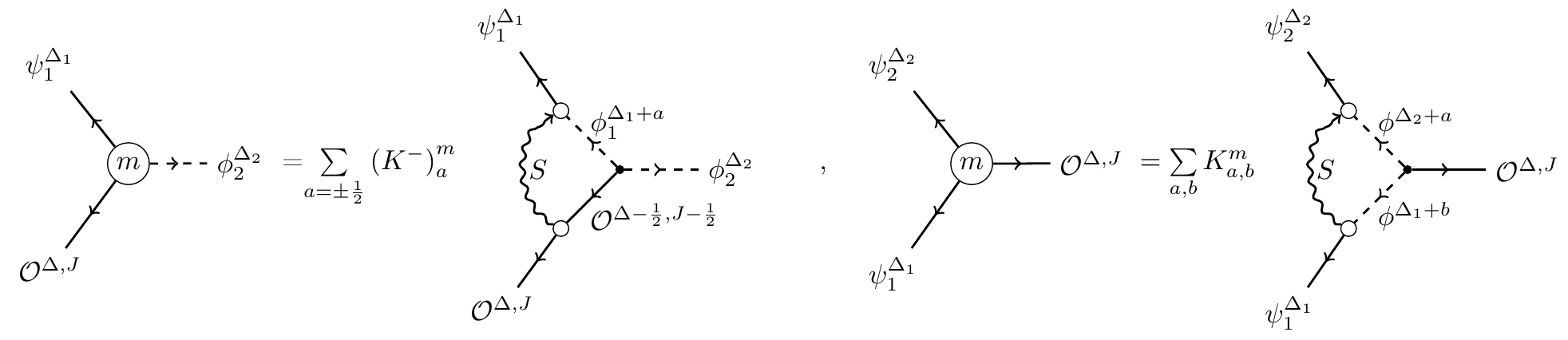}.
\end{center}
We then integrate by parts and act with the adjoint of these weight-shifting operators on the other spinning three-point function, which produces $\<\phi\phi\cO\>$ up to overall coefficients.

By this procedure, we find that
\begin{footnotesize}
\bea 
\left(\<\psi_1^{\Delta_1}\phi_2^{\Delta_2}\cO^{\Delta,J}\>^m,\<\psi_1^{3-\Delta_1}\phi_2^{3-\Delta_2}\cO^{3-\Delta,J}\>^n\right)={}&{}
\frac{(-1)^{J-\half} \Gamma \left(J+\frac{3}{2}\right)}{16 \sqrt{\pi } \Gamma (J+1)}
\begin{pmatrix}
	-1&0\\0&1
\end{pmatrix},
\\
\left(\<\psi_1^{\Delta_1}\psi_2^{\Delta_2}\cO^{\Delta,J}\>^m,\<\psi_1^{3-\Delta_1}\psi_2^{3-\Delta_2}\cO^{3-\Delta,J}\>^n\right)={}&{}
\left\{
\begin{aligned}
&\frac{(-1)^J \Gamma (J+1)}{8 \sqrt{\pi } \Gamma \left(J+\frac{1}{2}\right)}
\left(
\begin{array}{cccc}
	-1 & \frac{1}{2} & 0 & 0 \\
	\frac{1}{2} & -\frac{2 J+1}{4 J} & 0 & 0 \\
	0 & 0 & 1 & 0 \\
	0 & 0 & 0 & -\frac{J+1}{J} \\
\end{array}
\right)
&& J>0
\\
	&\frac{1}{8\pi}\begin{pmatrix}
	-1&0\\0&1
\end{pmatrix}
&& J=0
\end{aligned}.
\right.
\label{eq: fermion fermion three point pairing}
\eea
\end{footnotesize}

\section{Shadow Coefficients, Partial Waves, and Euclidean Inversion}
\subsection{Shadow Coefficients}
We define the shadow coefficients as
\begin{align}
\<\cO_1\cO_2\mathbf{S}[\cO_3]\>^a=S^{a}_{b}(\cO_1\cO_2[\cO_3])\<\cO_1\cO_2\widetilde{\cO}_{3}\>^b,
\end{align}
which can be seen in Figure~\ref{fig:shadow} in diagrammatic language as well. We reviewed how these matrices can be computed via weight shifting operators in  \secref{\ref{sec: shadow matrices}}; here we will simply present the explicit results.

In our conventions, we have
\footnotesize
\be 
S^1_2([\psi_{\De_\psi}]\phi_{\De_\phi}\cO_{\De,l})=\,& -\frac{i \pi ^{3/2} \Gamma \left(\Delta _{\psi }-1\right) \Gamma \left(\frac{1}{2} \left(l+\Delta -\Delta _{\phi }-\Delta _{\psi
	}+\frac{5}{2}\right)\right) \Gamma \left(\frac{1}{2} \left(l-\Delta +\Delta _{\phi }-\Delta _{\psi }+\frac{7}{2}\right)\right)}{\Gamma
	\left(\frac{7}{2}-\Delta _{\psi }\right) \Gamma \left(\frac{1}{2} \left(l+\Delta -\Delta _{\phi }+\Delta _{\psi }+\frac{1}{2}\right)\right)
	\Gamma \left(\frac{1}{2} \left(l-\Delta +\Delta _{\phi }+\Delta _{\psi }-\frac{1}{2}\right)\right)},
\\
S^1_1(\psi_{\De_\psi}[\phi_{\De_\phi}]\cO_{\De,l})=\,& \frac{\pi  \sin \left(\pi  \Delta _{\phi }\right) \Gamma \left(2 \left(\Delta _{\phi }-2\right)\right) \Gamma
	\left(\frac{1}{2} \left(l+\Delta -\Delta _{\phi }-\Delta _{\psi }+\frac{5}{2}\right)\right) \Gamma \left(\frac{1}{2} \left(l-\Delta -\Delta
	_{\phi }+\Delta _{\psi }+\frac{5}{2}\right)\right)}{2^{2 \Delta _{\phi }-5}\Gamma \left(\frac{1}{2} \left(l+\Delta +\Delta _{\phi }-\Delta _{\psi
	}-\frac{1}{2}\right)\right) \Gamma \left(\frac{1}{2} \left(l-\Delta +\Delta _{\phi }+\Delta _{\psi }-\frac{1}{2}\right)\right)},
\\
S^1_2(\psi_{\De_\psi}\phi_{\De_\phi}[\cO_{\De,l}])=\,&
\frac{\pi ^{3/2} (-1)^{l+1} \Gamma (\Delta -1) \Gamma (l+\Delta -1) \Gamma \left(\frac{1}{2} \left(l-\Delta +\Delta _{\phi }-\Delta _{\psi
	}+\frac{7}{2}\right)\right) \Gamma \left(\frac{1}{2} \left(l-\Delta -\Delta _{\phi }+\Delta _{\psi }+\frac{5}{2}\right)\right)}{\Gamma
	\left(\Delta -\frac{1}{2}\right) \Gamma (l-\Delta +3) \Gamma \left(\frac{1}{2} \left(l+\Delta +\Delta _{\phi }-\Delta _{\psi
	}-\frac{1}{2}\right)\right) \Gamma \left(\frac{1}{2} \left(l+\Delta -\Delta _{\phi }+\Delta _{\psi }+\frac{1}{2}\right)\right)},
\\
S^1_3([\psi_{\De_1}]\psi_{\De_2}\cO_{\De,l})=\,&
\frac{i \pi ^{3/2} \left(-\Delta +\Delta _1+\Delta _2-2\right) \Gamma \left(\Delta _1-1\right) \Gamma \left(\frac{1}{2} \left(l+\Delta -\Delta
	_1-\Delta _2+2\right)\right) \Gamma \left(\frac{1}{2} \left(l-\Delta -\Delta _1+\Delta _2+3\right)\right)}{2 \Gamma \left(\frac{7}{2}-\Delta
	_1\right) \Gamma \left(\frac{1}{2} \left(l+\Delta +\Delta _1-\Delta _2\right)\right) \Gamma \left(\frac{1}{2} \left(l-\Delta +\Delta _1+\Delta
	_2+1\right)\right)},
\\
S^1_3(\psi_{\De_1}[\psi_{\De_2}]\cO_{\De,l})=\,&
S^1_3([\psi_{\De_2}]\psi_{\De_1}\cO_{\De,l}),
\\
S^1_1(\psi_{\De_1}\psi_{\De_2}[\cO_{\De,l}])=\,&
\frac{\pi ^{3/2} (-1)^l \Gamma \left(\Delta -\frac{3}{2}\right) \Gamma (l+\Delta -1) \Gamma \left(\frac{1}{2} \left(l-\Delta +\Delta _1-\Delta
	_2+3\right)\right) \Gamma \left(\frac{1}{2} \left(l-\Delta -\Delta _1+\Delta _2+3\right)\right)}{\Gamma (\Delta -1) \Gamma (l-\Delta +3) \Gamma
	\left(\frac{1}{2} \left(l+\Delta +\Delta _1-\Delta _2\right)\right) \Gamma \left(\frac{1}{2} \left(l+\Delta -\Delta _1+\Delta _2\right)\right)},
\ee 
\normalsize
where we can get all other nonzero components from the relations
\footnotesize
\be 
S^1_2([\psi_{\De_\psi}]\phi_{\De_\phi}\cO_{\De,l})=-S^2_1([\psi_{\De_\psi}]\phi_{-\De_\phi}\cO_{-\De,l}),
\\
S^1_1(\psi_{\De_\psi}[\phi_{\De_\phi}]\cO_{\De,l})=S^2_2(\psi_{\De_\psi}[\phi_{\De_\phi}]\cO_{\De,l-1}),
\\
S^1_2(\psi_{\De_\psi}\phi_{\De_\phi}[\cO_{\De,l}])=-S^2_1(\psi_{-\De_\psi}\phi_{-\De_\phi}[\cO_{\De,l}]),
\\
\frac{S^1_4([\psi_{\De_1}]\psi_{\De_2}\cO_{\De,l})}{\frac{l}{\Delta -\Delta _1-\Delta _2+2}}=\frac{S^2_3([\psi_{\De_1}]\psi_{\De_2}\cO_{\De,l})}{\frac{-\Delta +\Delta _1+\Delta _2+l-1}{2 \left(\Delta -\Delta _1-\Delta _2+2\right)}}=\frac{S^2_4([\psi_{\De_1}]\psi_{\De_2}\cO_{\De,l})}{-\frac{-\Delta +\Delta _1+\Delta _2+l-1}{2 \left(\Delta -\Delta _1-\Delta _2+2\right)}}=S^1_3([\psi_{\De_1}]\psi_{\De_2}\cO_{\De,l}),
\\
\frac{S^3_1([\psi_{\De_1}]\psi_{\De_2}\cO_{\De,l})}{\frac{-\Delta -\Delta _1+\Delta _2+l+2}{\Delta +\Delta _1-\Delta _2-2}}=\frac{S^3_2([\psi_{\De_1}]\psi_{\De_2}\cO_{\De,l})}{\frac{2 l}{\Delta +\Delta _1-\Delta _2-2}}=\frac{S^4_1([\psi_{\De_1}]\psi_{\De_2}\cO_{\De,l})}{\frac{-\Delta -\Delta _1+\Delta _2+l+2}{\Delta +\Delta _1-\Delta _2-2}}=\frac{S^4_2([\psi_{\De_1}]\psi_{\De_2}\cO_{\De,l})}{-\frac{2 \left(\Delta +\Delta _1-\Delta _2-1\right)}{\Delta +\Delta _1-\Delta _2-2}}=S^1_3([\psi_{\De_1}]\psi_{-\De_2}\cO_{-\De,l}),
\\
\frac{S^1_4(\psi_{\De_1}[\psi_{\De_2}]\cO_{\De,l})}{-\frac{l}{\Delta -\Delta _1-\Delta _2+2}}=\frac{S^2_3(\psi_{\De_1}[\psi_{\De_2}]\cO_{\De,l})}{\frac{-\Delta +\Delta _1+\Delta _2+l-1}{2 \left(\Delta -\Delta _1-\Delta _2+2\right)}}=\frac{S^2_4(\psi_{\De_1}[\psi_{\De_2}]\cO_{\De,l})}{\frac{-\Delta +\Delta _1+\Delta _2+l-1}{2 \left(\Delta -\Delta _1-\Delta _2+2\right)}}=S^1_3(\psi_{\De_1}[\psi_{\De_2}]\cO_{\De,l}),
\\
\frac{S^3_1(\psi_{\De_1}[\psi_{\De_2}]\cO_{\De,l})}{\frac{-\Delta +\Delta _1-\Delta _2+l+2}{\Delta -\Delta _1+\Delta _2-2}}=\frac{S^3_2(\psi_{\De_1}[\psi_{\De_2}]\cO_{\De,l})}{\frac{2 l}{\Delta -\Delta _1+\Delta _2-2}}=\frac{S^4_1(\psi_{\De_1}[\psi_{\De_2}]\cO_{\De,l})}{\frac{\Delta -\Delta _1+\Delta _2-l-2}{\Delta -\Delta _1+\Delta _2-2}}=\frac{S^4_2(\psi_{\De_1}[\psi_{\De_2}]\cO_{\De,l})}{\frac{2 \left(\Delta -\Delta _1+\Delta _2-1\right)}{\Delta -\Delta _1+\Delta _2-2}}=S^1_3(\psi_{-\De_1}[\psi_{\De_2}]\cO_{-\De,l}),
\\
\frac{S^2_1(\psi_{\De_1}\psi_{\De_2}[\cO_{\De,l}])}{-\frac{2 \Delta -3}{2 (\Delta -1)}}=\frac{S^2_2(\psi_{\De_1}\psi_{\De_2}[\cO_{\De,l}])}{-\frac{\Delta -2}{\Delta -1}}=S^1_1(\psi_{\De_1}\psi_{\De_2}[\cO_{\De,l}]),
\\
\frac{S^3_3(\psi_{\De_1}\psi_{\De_2}[\cO_{\De,l}])}{(\Delta -1) \left(\Delta +\Delta _1-\Delta _2-2\right) \left(\Delta -\Delta _1+\Delta _2-2\right)-(\Delta -2) l^2-(\Delta -2) l}=\frac{S^1_1(\psi_{-\De_1}\psi_{-\De_2}[\cO_{\De+1,l}])}{2 (2 \Delta -3) (-\Delta +l+2) (\Delta +l-1)},
\\
\frac{S^4_4(\psi_{\De_1}\psi_{\De_2}[\cO_{\De,l}])}{-(\Delta -2) \left((\Delta -1)^2-\Delta _1^2-\Delta _2^2+2 \Delta _1 \Delta _2\right)+(\Delta -1) l^2+\Delta  l-l}=\frac{S^1_1(\psi_{-\De_1}\psi_{-\De_2}[\cO_{\De+1,l}])}{2 (2 \Delta -3) (-\Delta +l+2) (\Delta +l-1)},
\\
\frac{S^3_4(\psi_{\De_1}\psi_{\De_2}[\cO_{\De,l}])}{\left(2\Delta -3\right) \left(\Delta _2-\Delta _1\right) l}=\frac{S^4_3(\psi_{\De_1}\psi_{\De_2}[\cO_{\De,l}])}{(2 \Delta -3) \left(\Delta _1-\Delta _2\right) (l+1)}=\frac{S^1_1(\psi_{-\De_1}\psi_{-\De_2}[\cO_{\De+1,l}])}{2 (2 \Delta -3) (-\Delta +l+2) (\Delta +l-1)}.
\ee 
\normalsize

The block (anti-)diagonal form of shadow matrices reflects the property that shadow transformation is parity-definite and that we have chosen our three-point structures with definite parity. As the two point function in \equref{eq: shadow definition} carries a definite parity, the shadow matrix relates the same (opposite) parity structures if the shadowed operator is of integer (half-integer) spin; this is why, say, $S^a_b([\psi]\psi\cO)$ is block anti-diagonal whereas $S^a_b(\psi\psi[\cO])$ is block diagonal.\footnote{We remind the reader that what we refer to here as parity is simply the inversion $X_i\rightarrow -X_i$ in embedding space.}

\subsection{Bubble Coefficients and Partial Wave Normalization}

\begin{figure}
	\centering
	\includegraphics[scale=1.6]{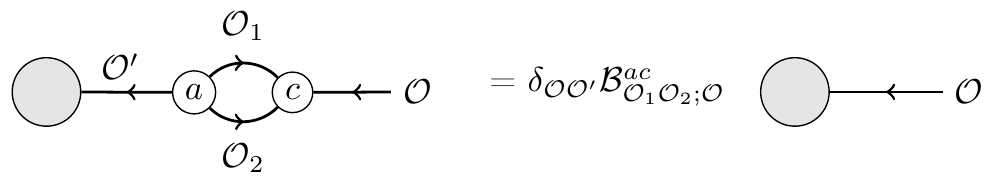}
	\caption{\label{fig:bubble} We define the bubble annihilation matrix $\cB$ as shown in the figure, where we suppress its possible dependence on $\cO_1$ and $\cO_2$. One can explicitly calculate $\cB$ by removing the gray blob above and connecting both ends: This relates $\cB$ times Plancherel measure to pairing of two three-point functions, which can then be calculated by
		going to a fixed conformal frame and carrying out the explicit calculations. A similar calculation is carried out in $4d$ in \cite{Karateev:2018oml}, see appendix C there. Here we repeated it for $3d$.}
\end{figure}

One of the interesting pairings that we can consider is
the so-called bubble integral\footnote{This follows from the irreducibility of representations. See (2.32) of \cite{Karateev:2018oml}.} 
\begin{multline}
\label{eq: bubble coefficient}
\<\cdots\tl\cO'(x)\cdots\>\.
\int d^{d}x_1d^{d}x_2
\<\cO_{1}(x_1)\cO_{2}(x_2)\cO'(x)\>^a\.
\<\tl\cO_{1}(x_1)\tl\cO_{2}(x_2)\tl\cO(y)\>^b
\\= \delta_{\cO\cO'}
\delta(x-y)\cB_{\cO_1\cO_2;\cO}^{ab}\<\cdots\tl\cO(y)\cdots\>,
\end{multline}
which we can see in figure \ref{fig:bubble} in diagrammatic language. By imposing $\cO'=\cO$ and taking the trace of both sides without acting on $\<\cdots\tl\cO'(x)\cdots\>$, we can relate the bubble coefficient $\cB_{\cO_1\cO_2;\cO}^{ab}$ to the three point pairing and the Plancherel measure:
\be
\label{eq: bubble coefficient in explicit form}
\cB_{\cO_1\cO_2;\cO}^{ab}=&
\frac{\left(\<\cO_{1}\cO_{2}\cO\>^a,\<\tl\cO_{1}\tl\cO_{2}\tl\cO\>^b\right)}{\mu\left(\cO\right)}.
\ee

One straightforward usage of these bubble matrices is the calculation of the partial wave normalization. We define the s-channel partial wave as the gluing of two three-point functions
\be 
\label{eq: partial wave convention}
\Psi^{(s)ac}_{\cO}(x_i)=\int d^dx\<\cO_1\cO_2\cO(x)\>^a\.\<\cO_3\cO_4\tl\cO(x)\>^c=
\begin{aligned}
	\includegraphics[scale=1.2]{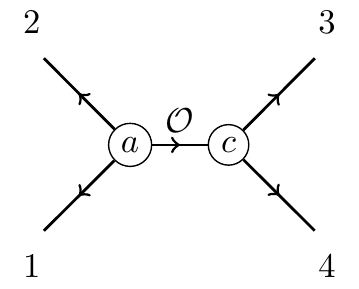}.
\end{aligned}
\ee
The normalization of the partial wave is then given by its pairing with itself, that is 
\be 
\left(\Psi^{\tl{(s)}ab}_{\cO_5},\Psi^{(s)cd}_{\cO_6}\right)=
\int \frac{d^{d}x_1...d^dx_6}{\SO(d+1,1)}\<\tl\cO_1\tl\cO_2\cO_5\>^a\.\<\tl\cO_3\tl\cO_4\widetilde{\cO}_5\>^b\.\<\cO_1\cO_2\cO_6\>^c\.\<\cO_3\cO_4\widetilde{\cO}_6\>^d,
\ee 
where $\sim$ in the first argument of the pairing indicates that external operators $\cO_i$ are replaced by $\tl\cO_i$. The integral is invariant under conformal transformations, so we also divide by its volume to obtain a finite result. We will follow the presentation of \cite{Karateev:2018oml}, specifically section 2.7, although there are additional subtleties because we are working with fermions.

First we perform the $x_{3,4}$ integrals to obtain
\be 
\left(\Psi^{\tl{(s)}ab}_{\cO_5},\Psi^{(s)cd}_{\cO_6}\right)=\delta_{\cO_5\tl\cO_6}
(-1)^{2l_{\cO_6}}
\cB_{\tl\cO_3\tl\cO_4;\cO_6}^{bd}
\left(\<\tl\cO_1\tl\cO_2\tl\cO_6\>^a, \<\cO_1\cO_2\cO_6\>^c\right),
\ee 
where $(-1)^{2l_{\cO_6}}$ follows from the change of the order of the three-point functions. Next we use \equref{eq: bubble coefficient in explicit form} to find the more symmetric form:
\begin{multline}
\left(\Psi^{\tl{(s)}ab}_{\cO_5},\Psi^{(s)cd}_{\cO_6}\right)=\frac{\delta_{\cO_5\tl\cO_6}(-1)^{2l_6}}{\mu\left(\cO_6\right)}\left(\<\tl\cO_1\tl\cO_2\tl\cO_{6}\>^a\.\<\cO_1\cO_2\cO_{_6}\>^c\right)\left(\<\tl\cO_3\tl\cO_4\cO_{6}\>^b\.\<\cO_3\cO_4\tl\cO_{6}\>^d\right).
\label{eq: partial wave normalization}
\end{multline}

Note that changing the order of the three-point functions brings an overall sign\linebreak \mbox{$(-1)^{2(l_1+l_2+l_3+l_4)}=1$}.\footnote{That this term is 1 follows from Lorentz invariance as we need an even number of fermions in a non-zero vacuum expectation value.} Additionally, for our relevant cases, we will only deal with pairings of $\<\psi_1\psi_2\cO\>$ and $\<\psi\phi\cO\>$ which are all independent of scaling dimensions. Since we also have $\mu\left(\cO\right)=\mu(\tl\cO)$, the partial wave normalization satisfies the following symmetries:
\bea[eq: symmetry of partial wave normalization]
\left(\Psi^{\tl{(s)}ab}_{\tl\cO},\Psi^{(s)cd}_{\cO}\right)=
\left(\Psi^{(s)cd}_{\cO},\Psi^{\tl{(s)}ab}_{\tl\cO}\right),
\\
\left(\Psi^{\tl{(s)}ab}_{\tl\cO},\Psi^{(s)cd}_{\cO}\right)=
\left(\Psi^{\tl{(s)}cd}_{\tl\cO},\Psi^{(s)ab}_{\cO}\right).
\eea

\subsection{Partial Wave Expansion and Euclidean Inversion Formula}
An $n-$point correlator can be expanded as a tensor product of two irreducible representations of the Euclidean conformal group, which basically provides us with an integral representation of a higher-point correlator in terms of lower point ones. This has been known for almost half a century since the early work of Dobrev et. al. \cite{Dobrev:1977qv} and was revived in recent years~\cite{Gadde:2017sjg,Caron-Huot:2017vep,Karateev:2018oml,Kravchuk:2018htv}. In the notation of \cite{Kravchuk:2018htv}, we can schematically write
\be 
\<\cO_1\cdots \cO_n\>=\int d\cO \int d^dx 
\<\cO_1(x_1)\cO_2(x_2)\cO(x)\>^aP^a_\cO(x_3,\dots x_n;x)
\label{eq: old syle partial wave expansion}
\ee 
for a generic $n-$point correlator. This corresponds to the following diagram
\be 
\includegraphics[scale=1.2]{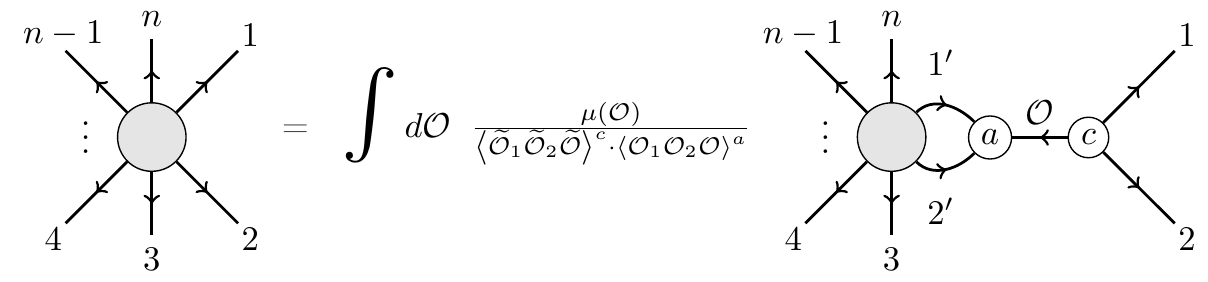},
\label{eq: old style partial wave expansion}
\ee
where we identify
\be 
\includegraphics[scale=1.23]{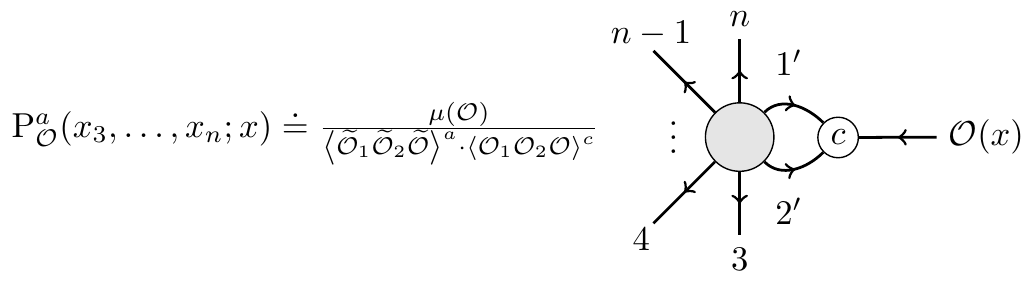}.
\ee
Here the integration measure is $d\cO=2\pi\Delta_{JJ'}\delta\left(s-s'\right)$ and it is defined over the principal series such that $\Delta=\frac{d}{2}+is$. We are glossing over the details in this quick review and refer the reader to \cite{Karateev:2018oml,Kravchuk:2018htv} for more details.

Let us consider this general expression in the case of $\<\cO_1\cO_2\cO_3\cO_4\>$. For four-point functions, we can decompose $P^a_\cO(x_3,x_4;x)$ in terms of three-point structures:
\be 
P^a_\cO(x_3,x_4;x)=\rho^{(s)}_{ab}(\cO)\<\cO_3(x_3)\cO_4(x_4)\tl\cO(x)\>^b.
\label{eq: definition of rho}
\ee 
Here $\rho^{(s)}_{ab}(\cO)$ are partial wave expansion coefficients and are related to OPE coefficients via \equref{eqn:OPEfunc_To_Coefs_schannel}.

\begin{figure}
	\centering
	\includegraphics[scale=1.4]{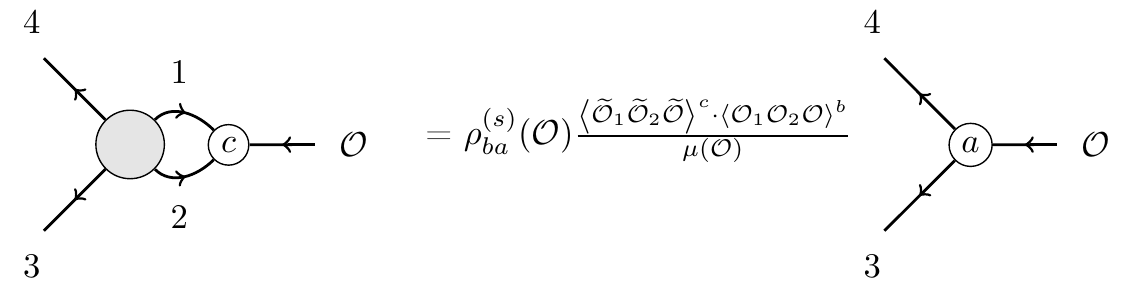}
	\caption{\label{fig:ope definition} We can take the definition of $\rho$ to be the coefficient of the three-point function $\<\cO_1\cO_2\cO\>^a$, which we obtain by pairing a four-point correlator $\<\cO_1\cO_2\cO_3\cO_4\>$ with a three-point structure $\<\cO_3\cO_4\cO\>^c$. Note that the overall coefficient also depends on the bubble coefficient $\cB$ which is a calculable kinematic term. By pairing both sides with $\<\cO_1\cO_2\tl\cO\>^e$, we can reduce this relation to the more standard definition generally used in the literature, such as (2.33) of \cite{Liu:2018jhs}, (2.40) of \cite{Karateev:2018oml}, or (1.6) of \cite{Simmons-Duffin:2017nub}. Note that these references use different conventions so the formulas are not entirely the same.}
\end{figure}

With \equref{eq: partial wave convention}, we can use the equation above to obtain the partial wave expansion\footnote{In some papers $P^a_\cO(x_3,\dots,x_n;x)$ is referred to as conformal partial wave as well. We will not be using these objects in this paper and will reserve this term for $\Psi^{ab}_\cO$ defined in \equref{eq: partial wave convention}.} of four-point function:
\be 
\<\cO_1\cO_2\cO_3\cO_4\>=
\<\cO_1\cO_2\>\<\cO_3\cO_4\>+\int_{\cC}d\cO \rho^{(s)}_{ab}(\cO)\Psi^{(s)ab}_{\cO}(x_i).
\label{eq: partial wave expansion}
\ee 

Note that we are explicitly writing the identity contribution as the identity block is actually orthogonal to the partial waves, hence it cannot be expanded in terms of them \cite{Caron-Huot:2017vep}. It is further argued in \cite{Simmons-Duffin:2017nub} that there may be other non-normalizable contributions to the four-point function that need to be written out explicitly. In particular, any scalar operator with $\Delta<\frac{d}{2}$ gives such a contribution. We will assume that either there is no scalar with $\Delta<\frac{d}{2}$ in the spectrum of the theory or that their contributions can be obtained by analytic continuation from the principal series.

In \equref{eq: partial wave expansion}, we also specified that the integration is over the contour $C$, where recall that we defined
\be 
\label{eq: contours}
\int_{\cC}d\cO\equiv \sum\limits_{J_\cO=0}^\infty\int\limits_{\frac{d}{2}}^{\frac{d}{2}+i\infty}\frac{d\Delta_{\cO}}{2\pi i}\;,\quad \int_{\cC'}d\cO\equiv \sum\limits_{J_\cO=0}^\infty\int\limits_{\frac{d}{2}-i\infty}^{\frac{d}{2}+i\infty}\frac{d\Delta_{\cO}}{2\pi i}
\ee 
for convenience.  Also, note that we give the expansion in terms of s-channel partial waves. This is indicated by the explicit $(s)$ superscript on $\rho$ and $\Psi$. Additionally, we leave the dependence of $\rho$ and $\psi$ on external operators implicit.

\begin{figure}
	\centering
	\includegraphics[scale=.24]{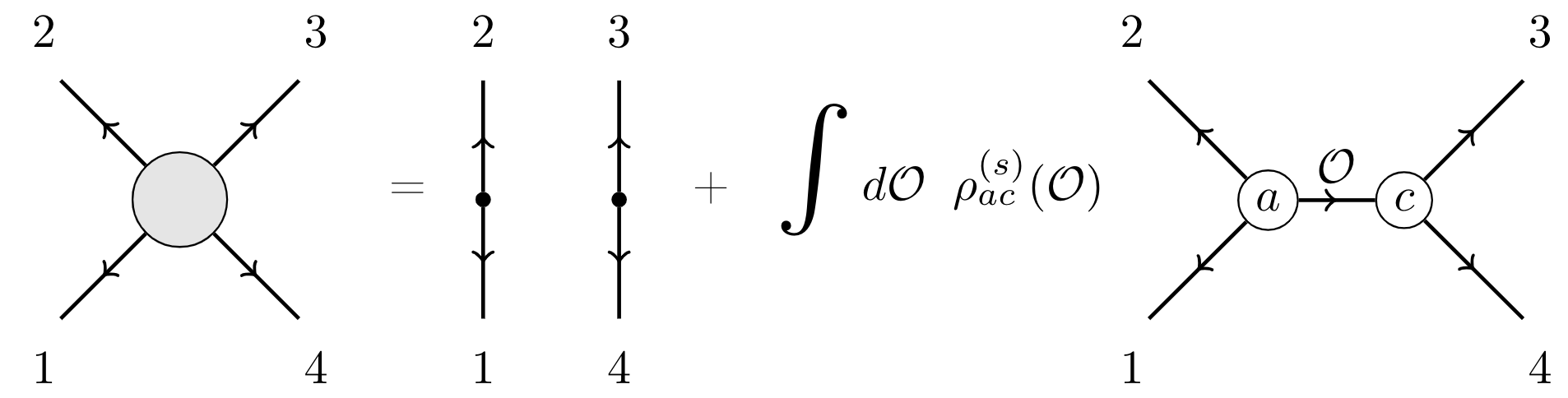}
	\caption{\label{fig: partial wave expansion} Diagrammatic illustration of the s-channel partial wave expansion of the four-point function, assuming that the identity contribution is the only non-normalizable contribution. Instead of separating it, we can deform it onto the principal series and deform back after the analytic continuation from principal series to physical poles.
}
\end{figure}

The definition of $\rho$ in \equref{eq: definition of rho} is diagrammatically shown in figure \ref{fig:ope definition}. We can pair both sides with a three-point function and obtain the Euclidean inversion formula:
\be 
\rho_{ab}^{(s)}(\cO_5)
=
\frac{\int d^{d}x_{1}...d^{d}x_5\mu(\cO_5)\<\tl\cO_1\tl\cO_2\tl\cO_5\>^c\.\<\cO_1\cO_2\cO_3\cO_4\>\.\<\tl\cO_3\tl\cO_4\cO_5\>^d}{\left(\<\tl\cO_1\tl\cO_2\tl\cO_5\>^a,\<\cO_1\cO_2\cO_5\>^c\right)\left(\<\cO_3\cO_4\tl\cO_5\>^d,\<\tl\cO_3\tl\cO_4\cO_5\>^b\right)}.
\label{eq: Euclidean inversion formula}
\ee 
Note that we can rewrite this as 
\be 
\label{eq: Euclidean inversion formula 2}
\rho_{ab}^{(s)}(\cO)
=
\frac{\int d^{d}x_{1}...d^{d}x_5\<\tl\cO_1\tl\cO_2\tl\cO_5\>^c\.\<\tl\cO_3\tl\cO_4\cO_5\>^d\.\<\cO_1\cO_2\cO_3\cO_4\>}{\frac{(-1)^{2J_{\cO_5}}}{\mu(\cO_5)}\left(\<\tl\cO_1\tl\cO_2\tl\cO_5\>^a,\<\cO_1\cO_2\cO_5\>^c\right)\left(\<\tl\cO_3\tl\cO_4\cO\>^b,\<\cO_3\cO_4\tl\cO_5\>^d\right)},
\ee 
which we recognize to be
\be 
\rho_{ab}^{(s)}(\cO)
= \eta_{(ac)(bd)}^\cO
\left(\Psi^{\tl{(s)}cd}_{\tl\cO}(x_i), \<\cO_1\cO_2\cO_3\cO_4\>\right).
\label{eq: Euclidean inversion formula}
\ee 

We could have derived this result by starting from the partial wave expansion of figure \ref{fig: partial wave expansion}, pairing it with $\Psi^{\tl{(s)}cd}_{\tl\cO}$, and utilizing the orthogonality of the partial waves.\footnote{\label{footnote: tadpole}When we pair the partial wave expansion \equref{eq: partial wave expansion} with a partial wave, there is actually another term coming from the pairing of identity exchange with the partial wave. However the pairing of identity exchange with the partial wave of the same channel is proportional to a tadpole diagram:
	\be 
	\label{fig:tadpole}
	\includegraphics[scale=1]{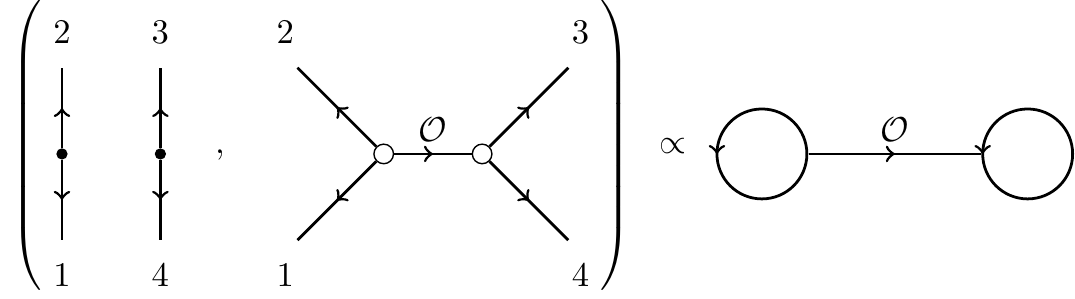}
	\ee
	Such diagrams are zero by the irreducibility of the representations unless $\cO=\mathbb{1}$, which is never the case for the partial waves on the principal series.
}

We would like to  remind the reader that the diagrams, albeit useful, are to be considered as schematic expressions only. In particular, they are agnostic to possible signs associated to the orderings of fermions. As an example, consider \equref{eq: old style partial wave expansion}: the equation it stands for is
\begin{subequations}
	\begin{equation}
\hspace{-.2in}	\<\cO_1\cO_2\cO_3\cdots\cO_n\>=\int d\cO \mu(\cO) \int d^dx d^dx_{1}'d^dx_{2}' \frac{\<\cO_1\cO_2\cO(x)\>^a\.\<\tl\cO(x)\tl\cO'_{2}\tl\cO'_{1}\>^c\.\<\cO'_{1}\cO'_{2}\cO_3\cdots\cO_n\>}{\left(\<\tl\cO\tl\cO_{2}\tl\cO_{1}\>^a,\<\cO_{1}\cO_{2}\cO\>^c\right)}\;,
	\label{eq: correct equation}
	\end{equation}
but not
\begin{equation}
\hspace{-.2in}	\<\cO_1\cO_2\cO_3\cdots\cO_n\>\neq\int d\cO \mu(\cO) \int d^dx d^dx_{1}'d^dx_{2}' \frac{\<\cO_1\cO_2\cO(x)\>^a\.\<\cO'_{1}\cO'_{2}\cO_3\cdots\cO_n\>\.\<\tl\cO(x)\tl\cO'_{2}\tl\cO'_{1}\>^c}{\left(\<\tl\cO\tl\cO_{2}\tl\cO_{1}\>^a,\<\cO_{1}\cO_{2}\cO\>^c\right)}\;.
\end{equation}
However, one cannot deduce this from the diagram alone.
\end{subequations}

\section{Symmetries of $6j$ Symbols}
\label{sec: symmetries of 6j symbols}

\begin{figure}
	\centering
	\includegraphics[scale=1.4]{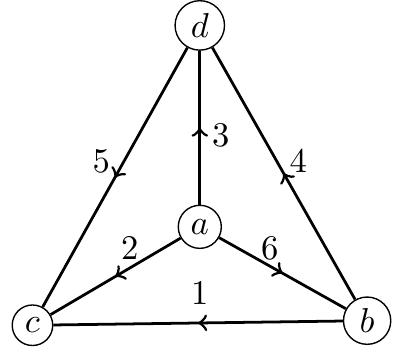}
	\caption{\label{fig: tetrahedron} Diagrammatic form of $6j$ symbol $\left(\Psi^{\tl{(s)}cd}_{\tl\cO_5},\Psi^{(t)ab}_{\cO_6}\right)$ as a tetrahedron.}
\end{figure}

By representing the $6j$ symbol as a tetrahedron as in figure \ref{fig: tetrahedron}, we can reveal its symmetries, as was done in \cite{Liu:2018jhs}. Explicitly, we can consider the three transformations:
\begin{enumerate}
	\begin{subequations}
		\label{eq: 6j symmetries}
		\item[\textbf{S1}] Rotation around the axis that passes through the vertex $a$ and the center of the triangle $\De bcd$, generated by the permutation $(1\tl 5\tl 4)(\tl 154)(236)(\tl 2\tl 3\tl 6)(bcd)$:
		\be 
		\label{eq: 6j symmetry 1}
		\left\{\begin{aligned}
			\cO_1  \cO_2  \cO_6 \\
			\cO_3  \cO_4  \cO_5
		\end{aligned}\right\}^{cdab} =(-1)^{2l_{\cO_6}} 
		\left\{\begin{aligned}
			\tl\cO_5  \cO_3  \cO_2 \\
			\cO_6  \tl\cO_1  \cO_4
		\end{aligned}\right\}^{dbac}.
		\ee 	
		
		\item[\textbf{S2}] Rotation around the axis that passes through the vertex $c$ and the center of the triangle $\De abd$, generated by the permutation $(125)(\tl 1\tl 2\tl 5)(\tl3 46)(3\tl 4\tl 6)(bad)$:
		\be 
		\label{eq: 6j symmetry 2}
		\left\{\begin{aligned}
			\cO_1  \cO_2  \cO_6 \\
			\cO_3  \cO_4  \cO_5
		\end{aligned}\right\}^{cdab} =(-1)^{2l_{\cO_1}}  \left\{\begin{aligned}
			\cO_2  \cO_5  \tl \cO_3 \\
			\tl\cO_4  \cO_6  \cO_1
		\end{aligned}\right\}^{cbda}.
		\ee 
		
		\item[\textbf{S3}] Reflection with respect to the plane that passes through the points $c$, $d$, and the midpoint of the line segment $\overline{ab}$, generated by the permutation $(12)(\tl1\tl2)(34)(\tl3\tl4)(6\tl6)(ba)$:
		\be 
		\label{eq: 6j symmetry 3}
		\left\{\begin{aligned}
			\cO_1  \cO_2  \cO_6 \\
			\cO_3  \cO_4  \cO_5
		\end{aligned}\right\}^{cdab} =(-1)^{2l_{\cO_6}}  
		\left\{\begin{aligned}
			\cO_2  \cO_1  \tl\cO_6 \\
			\cO_4  \cO_3  \cO_5
		\end{aligned}\right\}^{cdba}.
		\ee 
	\end{subequations}
\end{enumerate}
The overall phases in the front follow from the fermionic nature of the correlators and can be checked explicitly.

The validity of \equref{eq: 6j symmetries} depends on the choice of three point basis. For example, the first two equalities require us to work in a basis which respects the cyclic permutations; i.e., we should have $\<\cO_1\cO_2\cO_3\>^a=\<\cO_2\cO_3\cO_1\>^a=\<\cO_2\cO_3\cO_1\>^a$. Generically, we can always find a basis which respects this property.

The equality \equref{eq: 6j symmetry 3} on the other hand requires the basis to respect inversions, i.e. we should have $\<\cO_1\cO_2\cO_3\>^a=\<\cO_2\cO_1\cO_3\>^a$. We can always choose a basis to respect this \emph{unless} we have $l_{\cO_1}=l_{\cO_2}$. In that case, we can no longer choose two independent bases $\<\cO_1\cO_2\cO_3\>^a$ and $\<\cO_2\cO_1\cO_3\>^a$ to satisfy the required equality; we need the same basis to satisfy this condition. However, if we work in a parity definite basis, all \emph{nonzero} $6j$ symbols will have an even number of parity odd three-point structures, therefore the equality holds. Assuming we are in such a basis, we can use following relations to derive all permutations:
\be 
\textbf{S1}^3=\textbf{S2}^3=\textbf{S3}^2=\textbf{E} \text{ with } (\textbf{S2}*\textbf{S1})^2=(\textbf{S3}*\textbf{S1})^4=(\textbf{S3}*\textbf{S2})^2=\textbf{E}.
\ee 

In summary, we can derive \equref{eq: 6j symmetries} and similar identities by considering the inversions and rotations of the tetrahedron and are valid in a parity definite basis with a cyclic property. These conditions are trivially satisfied for external scalars as there is only one three-point structure. 

An interesting set of transformations is the one that does not move the edges $5,6$. There are only three such permutations:
\be 
(12)(\tl 1\tl 2)(3 4)(\tl 3\tl 4)(6 \tl 6)(a b),\\
(5\tl 5)(2 3)(\tl 2\tl 3)(1 4)(\tl 1 \tl 4)(c d),\\
(13)(\tl 1\tl 3)(24)(\tl 2\tl 4)(5\tl 5)(6\tl 6)(ab)(cd),
\ee 
which yields
\small 
\be 
\left\{\begin{aligned}
	\cO_1  \cO_2  \cO_6 \\
	\cO_3  \cO_4  \cO_5
\end{aligned}\right\}^{cdab} =(-1)^{2l_{\cO_6}}  
\left\{\begin{aligned}
	\cO_2  \cO_1  \tl\cO_6 \\
	\cO_4  \cO_3  \cO_5
\end{aligned}\right\}^{cdba}
=(-1)^{2l_{\cO_5}} 
\left\{\begin{aligned}
	\cO_4  \cO_3  \cO_6 \\
	\cO_2  \cO_1  \tl \cO_5
\end{aligned}\right\}^{dcab}
=(-1)^{2l_{\cO_5}+2l_{\cO_6}} 
\left\{\begin{aligned}
	\cO_3  \cO_4  \tl\cO_6 \\
	\cO_1  \cO_2  \tl\cO_5
\end{aligned}\right\}^{dcba} .
\ee 
\normalsize
But we also know how to relate $\left\{\begin{aligned}
\cO_1  \cO_2  \cO_6 \\
\cO_3  \cO_4  \tl\cO_5
\end{aligned}\right\}$ to $\left\{\begin{aligned}
\cO_1  \cO_2  \cO_6 \\
\cO_3  \cO_4  \cO_5
\end{aligned}\right\}$, and likewise for $\cO_6$, due to shadow symmetry of the partial waves.

We may also be interested in interchanging $\cO_{5,6}$ in the $6j$ symbol, and this can be achieved with the transformation $\mathfrak{T}=(1\tl 1)(2\tl 4)(3\tl 3)(56)(ad)(bc)$:
\be 
\label{eq: 6j symmetry regarding exchange of 5 and 6}
\left\{\begin{aligned}
	\cO_1  \cO_2  \cO_6 \\
	\cO_3  \cO_4  \cO_5
\end{aligned}\right\}^{cdab} =(-1)^{2\left(l_{\cO_5}+l_{\cO_6}\right)}  
\left\{\begin{aligned}
	\tl\cO_1  \tl\cO_4  \cO_5 \\
	\tl\cO_3  \tl\cO_2  \cO_6
\end{aligned}\right\}^{badc}.
\ee 
Likewise, with the transformation $\mathfrak{T}=(1\tl 3)(2\tl 2)(4\tl 4)(5 \tl6)(ac)(bd)$, we get
\be 
\left\{\begin{aligned}
	\cO_1  \cO_2  \cO_6 \\
	\cO_3  \cO_4  \cO_5
\end{aligned}\right\}^{cdab} =  
\left\{\begin{aligned}
	\tl\cO_3  \tl\cO_2  \tl\cO_5 \\
	\tl\cO_1  \tl\cO_4  \tl\cO_6
\end{aligned}\right\}^{abcd}.
\ee

\section{OPE Coefficients}
\subsection{MFT Coefficients for Higher Twist Towers}
\label{app:MFT_higher_twist}

In section~\ref{sec:MFTOPE} we presented the MFT coefficients for the leading twist towers; in this appendix, we present the results for higher twist towers.

\paragraph{For $\<\f\psi\psi\f\>:$}	
	\scriptsize
	\bea 
	P_{11}^{(s)}\left([\phi\psi]_{n,J}\right)=&\,
	\frac{(2 J+1) 2^{2 \Delta _{\phi }-4} \Gamma (J+1) \cos \left(\pi  \left(\Delta _{\psi }+\Delta
		_{\phi }\right)\right) \Gamma \left(n+\Delta _{\psi }-1\right) \Gamma \left(n+\Delta _{\phi }-\frac{1}{2}\right) \Gamma
		\left(J+n+\Delta _{\psi }\right)}{\pi ^{3/2} n! \Gamma \left(J+\frac{3}{2}\right) \Gamma
		\left(\Delta _{\psi }-1\right) \Gamma \left(\Delta _{\psi }+\frac{1}{2}\right) \Gamma \left(2 \Delta _{\phi }-1\right)
		\Gamma (J+n+1)}
	\nn\\
	&\x \frac{\left(2 \Delta _{\psi }+2 \Delta _{\phi }+4 J+4 n-3\right) \Gamma \left(J+n+\Delta
		_{\phi }-\frac{1}{2}\right) \Gamma \left(-2 n-\Delta _{\phi }-\Delta _{\psi }+\frac{7}{2}\right) }{\Gamma \left(2 J+2 n+\Delta _{\phi
		}+\Delta _{\psi }-\frac{1}{2}\right)}
	\nn\\
	&\x\frac{\Gamma \left(n+\Delta
		_{\phi }+\Delta _{\psi }-\frac{5}{2}\right) \Gamma \left(J+n+\Delta _{\phi }+\Delta _{\psi }-\frac{3}{2}\right) \Gamma
		\left(J+2 n+\Delta _{\phi }+\Delta _{\psi }-1\right)}{\Gamma \left(J+2 n+\Delta _{\phi }+\Delta _{\psi }-\frac{3}{2}\right)},
	\\
	P_{22}^{(s)}\left([\phi\partial\psi]_{n,J}\right)=&\,
	P_{11}^{(s)}\left([\phi\psi]_{n,J}\right) \frac{\left(\Delta _{\psi }+n-1\right) \left(\Delta _{\phi }+J+n-\frac{1}{2}\right) \left(\Delta _{\psi }+\Delta _{\phi
		}+n-\frac{5}{2}\right) \left(\frac{1}{2} \left(\Delta _{\psi }+\Delta _{\phi }+J-1\right)+n\right)}{2 (J+n+1)
		\left(\frac{1}{4} \left(2 \Delta _{\psi }+2 \Delta _{\phi }-5\right)+n\right) \left(\Delta _{\psi }+\Delta _{\phi }+J+2
		n-\frac{3}{2}\right) \left(J+\frac{1}{4} \left(2 \Delta _{\psi }+2 \Delta _{\phi }+4 n-3\right)\right)}.
	\eea 
	\normalsize
	
\paragraph{For $\<\psi_1\psi_2\psi_2\psi_1\>:$}
	\begin{multline}
	P^{(s)}_{[\psi_{1\a}\psi_{2\b}]_{n,J}}= -\begin{pmatrix}
	\frac{n \left(\Delta _1+\Delta _2+2 J+2 n-2\right) }{4 J}c_1 & c_2\\ c_2& \left(\Delta _1+\Delta _2+J+2 n-2\right) \left(\Delta _1+\Delta _2+2 J+2 n-2\right) c_3
	\end{pmatrix}
	\\\x \frac{\Gamma \left(J+n+\Delta
		_2-\frac{1}{2}\right) \Gamma \left(J+n+\Delta _1+\Delta _2-\frac{5}{2}\right) \Gamma \left(J+2 n+\Delta _1+\Delta
		_2-3\right)}{\Gamma \left(2
		n+\Delta _1+\Delta _2-3\right) \Gamma \left(J+2 n+\Delta _1+\Delta _2-\frac{5}{2}\right) \Gamma \left(2 J+2 n+\Delta
		_1+\Delta _2-1\right)}
	\\\x  \frac{\Gamma \left(J+\frac{3}{2}\right) \Gamma \left(n+\Delta _1-1\right) \Gamma \left(n+\Delta _2-1\right) \Gamma
		\left(n+\Delta _1+\Delta _2-3\right) \Gamma \left(J+n+\Delta _1-\frac{1}{2}\right) }{(J+1) n! \Gamma \left(\Delta _1-1\right) \Gamma \left(\Delta _1+\frac{1}{2}\right) \Gamma \left(\Delta
		_2-1\right) \Gamma \left(\Delta _2+\frac{1}{2}\right) \Gamma (J) \Gamma \left(J+n+\frac{3}{2}\right) },
	\end{multline} 
	and
	\begin{multline}
	P^{(s)}_{[\psi_{1\a}\psi_2^\a]_{n,J}}=-2^{2 \Delta _1+2 \Delta _2-6}\begin{pmatrix}
	\frac{1+J}{2}d_1&d_2\\-d_2&-\frac{J}{2}d_3
	\end{pmatrix}
	\\\x
	\frac{\Gamma \left(J+\frac{3}{2}\right) \left(\Delta _1+\Delta _2+2 J+2 n-1\right) \Gamma
		\left(n+\Delta _1-1\right) \Gamma \left(n+\Delta _2-1\right) \Gamma \left(n+\Delta _1+\Delta _2-2\right) }{\pi  \left(2 \Delta _1-1\right) \left(2 \Delta
		_2-1\right) (J+1) n! \Gamma \left(2 \Delta _1-2\right) \Gamma \left(2 \Delta _2-2\right) \Gamma (J+1) \Gamma
		\left(J+n+\frac{3}{2}\right)}
	\\\x \frac{\Gamma
		\left(J+n+\Delta _1-\frac{1}{2}\right) \Gamma \left(J+n+\Delta _2-\frac{1}{2}\right) \Gamma \left(J+n+\Delta _1+\Delta
		_2-\frac{3}{2}\right) \Gamma \left(J+2 n+\Delta _1+\Delta _2-2\right)}{\Gamma \left(2 n+\Delta _1+\Delta _2-2\right)\Gamma \left(J+2 n+\Delta _1+\Delta
		_2-\frac{3}{2}\right) \Gamma \left(2 J+2 n+\Delta _1+\Delta _2\right)},
	\end{multline}
	where
	\be 
	c_1=&
	4 J^2 \left(\Delta _1+\Delta _2+n-3\right)+4 J \left(\Delta _1+\Delta _2+n-3\right) \left(\Delta _1+\Delta _2+2
	n-2\right)\\&+\left(\Delta _1+\Delta _2+2 n-3\right) \left(2 \Delta _1+2 \Delta _2+2 n-5\right),
	\\
	c_2=&4 n \left(\Delta _1+\Delta _2+n-\frac{7}{2}\right) \left(\frac{1}{2} \left(\Delta _1+\Delta _2+J-2\right)+n\right)
	\left(J+\frac{1}{2} \left(\Delta _1+\Delta _2+2 n-2\right)\right),
	\\
	c_3=&  J^2+J \left(\Delta _1+\Delta
	_2+2 n-2\right)+(2 n+1) \left(\Delta _1+\Delta _2+n-3\right),
	\\
	d_1=& 2 J^2 \left(\Delta _1+\Delta _2+2 n-2\right)+\left(2 \Delta _1+2 n-1\right) \left(\Delta
	_1+\Delta _2+2 n-2\right) \left(2 \Delta _2+2 n-1\right)
	\\&+J \left(-9 \Delta _1-9 \Delta _2+2 \left(\Delta _1^2+\Delta _2^2+4 \Delta _1
	\Delta _2+6 n^2+\left(6 \Delta _1+6 \Delta _2-9\right) n\right)+6\right),
	\\
	d_2=& \left(\Delta _1-\Delta _2\right) J (J+1) \left(\Delta _1+\Delta _2+J+2 n-\frac{3}{2}\right),
	\\
	d_3=&J \left(-11 \Delta _1-11 \Delta _2+2 \left(\Delta _1^2+\Delta _2^2+4 \Delta
	_1 \Delta _2+6 n^2+\left(6 \Delta _1+6 \Delta _2-11\right) n\right)+10\right)
	\\&+4 \left(\Delta _1+n-1\right) \left(\Delta
	_2+n-1\right) \left(\Delta _1+\Delta _2+2 n-1\right)
	+ 2 J^2 \left(\Delta _1+\Delta _2+2 n-2\right).
	\ee 
	
\paragraph{For $\<\psi\psi\psi\psi\>:$}
\begin{multline}
P^{(s)}_{[\psi_\a\psi_\b]_{n,J}}
=
n\frac{(-1)^{J+1} \left((-1)^J+1\right) 2^{-2 J-4 n+1} \Gamma \left(J+\frac{3}{2}\right) \Gamma \left(n+\Delta _{\psi
	}-1\right) \Gamma \left(n+2 \Delta _{\psi }-3\right) }{\left(1-2 \Delta _{\psi
	}\right)^2 \Gamma (J+2) \Gamma (n+1) \Gamma \left(2 \left(\Delta _{\psi }-1\right)\right){}^2 \Gamma
	\left(J+n+\frac{3}{2}\right)}
\\\x
\frac{\Gamma \left(J+n+\Delta _{\psi }-\frac{1}{2}\right) \Gamma
	\left(J+n+2 \Delta _{\psi }-\frac{5}{2}\right) \Gamma \left(J+2 n+2 \Delta _{\psi }-3\right)}{\Gamma \left(n+\Delta _{\psi }-\frac{3}{2}\right) \Gamma \left(J+n+\Delta _{\psi
	}-1\right) \Gamma \left(J+2 n+2 \Delta _{\psi }-\frac{5}{2}\right)}
\begin{pmatrix}
c_1&  c_2\\  c_2&\frac{1}{n}c_3
\end{pmatrix},
\end{multline}
and
\begin{multline}
P^{(s)}_{[\psi_\a\psi^\a]_{n,J}}
=
\frac{(-1)^J 2^{-2 J-4 n} \Gamma \left(J+\frac{3}{2}\right) \Gamma \left(n+\Delta _{\psi }\right) \Gamma \left(n+2
	\Delta _{\psi }-2\right)\Gamma \left(J+2 n+2 \Delta _{\psi }\right)}{\left(1-2 \Delta _{\psi }\right){}^2 \Gamma (J+2)
	\Gamma (n+1) \Gamma \left(2 \Delta _{\psi }-2\right){}^2 \Gamma \left(J+n+\frac{3}{2}\right)\Gamma \left(J+2 n+2 \Delta _{\psi
	}-\frac{3}{2}\right)}
\\\x \frac{\Gamma \left(J+n+\Delta _{\psi }+\frac{1}{2}\right) \Gamma \left(J+n+2 \Delta _{\psi
	}-\frac{3}{2}\right)}{\Gamma \left(n+\Delta
	_{\psi }-\frac{1}{2}\right) \Gamma \left(J+n+\Delta _{\psi }\right)}\left(
\begin{array}{cc}
-\frac{\left(1+(-1)^J\right) (J+1)}{J+2 \left(n+\Delta _{\psi }-1\right)} & 0 \\
0 & \frac{\left(-1+(-1)^J\right) J}{J+2 n+2 \Delta _{\psi }-1} \\
\end{array}
\right),
\end{multline}
where
\be 
c_1=&4 J^2 \left(2 \Delta _{\psi }+n-3\right)+8 J \left(\Delta _{\psi }+n-1\right) \left(2 \Delta _{\psi
}+n-3\right)+\left(2 \Delta _{\psi }+2 n-3\right) \left(4 \Delta _{\psi }+2 n-5\right),\\
c_2=&2J \left(4 \Delta _{\psi }+2 n-7\right) \left(J+2 \left(\Delta _{\psi }+n-1\right)\right),\\
c_3=&4J \left(J+2 \left(\Delta _{\psi }+n-1\right)\right) \left(J^2+2 J \left(\Delta _{\psi }+n-1\right)+(2 n+1) \left(2
\Delta _{\psi }+n-3\right)\right).
\ee 

\subsection{Corrections to OPE Coefficients for $[\phi\psi]^+_0$}
\label{app:OPE}
In \equref{eq: OPE function in terms of scalar 6j symbol} we related the OPE function of spinning operators to the scalar $6j$ symbols, which we reproduce for reader's convenience:
\begin{align}
\rho_{ac}^{(s)}(\cO)S^c_b(\cO_3\cO_4[\tl\cO])\evaluated_{G^{(t),fg}_{\cO_6}}=\lambda_{326,f}\lambda_{146,g}\sum\limits_{\f_i,\cO',\cO'_{6}}
&\opeFuncDecomp{fg}{ab}{\cO_1}{\cO_2}{\cO_3}{\cO_4}{\cO}{\cO_6}{\f_1}{\f_2}{\f_3}{\f_4}{\cO'}{\cO'_{6}}
\nonumber \\ & \qquad\frac{S(\phi_3\phi_4[\widetilde{\cO'}])}{\eta^{(s)}_{\cO'}}\sixjBlock{\f_1}{\f_2}{\f_3}{\f_4}{\cO'}{\cO'_{6}}.
\end{align}
By taking the double poles in $\De$ on both sides, we can extract $\delta hP$ for spinning operators in terms of scalar data, which we detailed and illustrated in section~\ref{sec:applications}. In this appendix, we will use this equation to extract correction to OPE coefficients for double twist operators $[\phi\psi]^+_0$ due to an exchange of a scalar in the crossed channel.

We see in \equref{eq: parity even tower of two fermions and two scalars} that the $\delta h P$ for $[\f\psi]^+_{0,J}$ reads as
\be
(\delta h P)_{11}([\f\psi]^+_{0,J})\evaluated_{G^{(t)}_{\phi_6}}=&
-i \lambda_{\f\f\f_6}\lambda_{\psi\psi\f_6}^1\mathfrak{dp}_1^{J-\half,0}(\psi^{\half},\f,\f_6),
\\
(\delta h P)_{12}([\f\psi]^+_{0,J})\evaluated_{G^{(t)}_{\phi_6}}=&(\delta h P)_{21}([\f\psi]^+_{0,J})\evaluated_{G^{(t)}_{\phi_6}}=(\delta h P)_{22}([\f\psi]^+_{0,J})\evaluated_{G^{(t)}_{\phi_6}}=0,
\ee
as only the first term in \equref{eq: K coefficients for 2f2s with an exchange of a scalar t channel} contributes. For $(\delta P)$ on the other hand, we do not need double poles (single poles are sufficient) and there are also cross terms, hence we have
\be
(\delta P)_{11}([\f\psi]^+_{0,J})\evaluated_{G^{(t)}_{\phi_6}}=&	-i \lambda_{\f\f\f_6}\lambda_{\psi\psi\f_6}^1\mathfrak{p}_{1,+}^{J-\half,0},
\\
(\delta P)_{12}([\f\psi]^+_{0,J})\evaluated_{G^{(t)}_{\phi_6}}=&-\lambda_{\f\f\f_6}\lambda_{\psi\psi\f_6}^3\bigg(
\frac{i \left(2 l_5+1\right)}{\Delta _6-1}\mathfrak{p}_{1,-}^{J+\half,0}
\\&\quad+
\frac{8 i \left(\Delta _{\psi }+l_5-1\right) \left(\Delta _{\phi }+l_5-1\right) \left(\Delta _{\psi }+\Delta
	_{\phi }+l_5-2\right)}{\left(\Delta _6-1\right) \left(2 \Delta _{\psi }+2 \Delta _{\phi }+4 l_5-5\right)
	\left(2 \Delta _{\psi }+2 \Delta _{\phi }+4 l_5-3\right)}\mathfrak{p}_{1,-}^{J-\half,0}
\bigg),
\\
(\delta P)_{12}([\f\psi]^+_{0,J})\evaluated_{G^{(t)}_{\phi_6}}=&i\lambda_{\f\f\f_6}\lambda_{\psi\psi\f_6}^3\frac{1}{\Delta _6-1}\mathfrak{p}_{2,-}^{J-\half,0},
\\
(\delta P)_{22}([\f\psi]^+_{0,J})\evaluated_{G^{(t)}_{\phi_6}}=&0,
\ee
where we define the shorthand notation
\be 
\mathfrak{p}_{i,\pm}^{J,n}\equiv \mathfrak{p}_i^{J,n}(\psi^{\pm},\f,\f,\psi^{\half},\f_6)
\ee 
for
\small
\bea 
\mathfrak{p}_1^{J,n}(\f_1,\f_2,\f_3,\f_4,\cO_6)\equiv &
\lim\limits_{\De\rightarrow\De_1+\De_2+J+2n} \left(\De -\De_1-\De_2-J-2n\right) \frac{S(\f_3 \f_4 [\widetilde{\cO}_{\De,J}])}{\eta^{(s)}_{\cO_{\De,J}}} \sixjBlock{\f_1}{\f_2}{\f_3}{\f_4}{\cO_{\De,J}}{\cO_{6}},
\\
\mathfrak{p}_2^{J,n}(\f_1,\f_2,\f_3,\f_4,\cO_6)\equiv &
\lim\limits_{\De\rightarrow\De_3+\De_4+J+2n} \left(\De -\De_3-\De_4-J-2n\right) \frac{S(\f_3 \f_4 [\widetilde{\cO}_{\De,J}])}{\eta^{(s)}_{\cO_{\De,J}}} \sixjBlock{\f_1}{\f_2}{\f_3}{\f_4}{\cO_{\De,J}}{\cO_{6}}.
\eea
\normalsize

We can compute $\mathfrak{p}$ similar to $\mathfrak{dp}$ and include both perturbative and nonperturbative corrections to OPE coefficients. For brevity, we only reproduce the leading piece of the perturbative correction at large $l$:
\begin{multline}
(\delta P)([\f\psi]^+_{0,\ell_5})\evaluated_{G^{(t)}_{\phi_6}}= \frac{i \sqrt{\pi } (-1)^{l_5-\frac{1}{2}} 2^{-\Delta _{\psi }-\Delta _{\phi }+\Delta _6-2 l_5+\frac{5}{2}}
	l_5^{\Delta _{\psi }+\Delta _{\phi }-\Delta _6-1}}{\Gamma \left(-\frac{\Delta _6}{2}+\Delta _{\psi
	}+\frac{1}{2}\right) \Gamma \left(\Delta _{\phi }-\frac{\Delta _6}{2}\right)}
\\\x 
\begin{pmatrix}
-\lambda_{\f\f\f_6}\lambda_{\psi\psi\f_6}^1\frac{\Gamma \left(\frac{\Delta _6+1}{2}\right) H_{\frac{\Delta
			_6-2}{2}}}{\sqrt{\pi } \Gamma \left(\frac{\Delta _6}{2}\right)}
		& \lambda_{\f\f\f_6}\lambda_{\psi\psi\f_6}^3\frac{\Gamma \left(-\frac{\Delta _6}{2}+\Delta _{\psi }+\frac{1}{2}\right)}{2 \sqrt{l_5} \Gamma \left(\Delta
			_{\psi }-\frac{\Delta _6}{2}\right)}
		\\
		\lambda_{\f\f\f_6}\lambda_{\psi\psi\f_6}^3\frac{\Gamma \left(-\frac{\Delta _6}{2}+\Delta _{\psi }+\frac{1}{2}\right)}{2 \sqrt{l_5} \Gamma \left(\Delta
			_{\psi }-\frac{\Delta _6}{2}\right)} 
		& 0
\end{pmatrix},
\end{multline}
where $H_a$ is the Harmonic number. As a consistency check, we see that setting
\be 
\De_6\rightarrow 0\;,\quad \lambda_{\f\f\f_6}\rightarrow 1\;,\quad \lambda_{\psi\psi\f_6}^1\rightarrow i\;,\quad \lambda_{\psi\psi\f_6}^3 \rightarrow 0
\ee 
reduces the result to the MFT coefficient \equref{eq: OPE coefficient squared for fermion scalar parity even double twist family}.

\section{$\cK$ Coefficients}
\label{sec: K coefficients}

In this appendix, we present the explicit expression for the $\cK$ coefficients defined in \equref{eq: K coefficients} for $\<\psi\f\f\psi\>$. As there are a different number of three-point tensor structures depending on whether $l_{5,6}=0$, the minimal complete set of nonzero $\cK$ coefficients differs for each case. We already presented the results for $\<\psi\f\f\psi\>$ with $l_6=0$ in \equref{eq: K coefficients for 2f2s with an exchange of a scalar t channel}, so we will detail the $\<\psi\f\f\psi\>$ with $l_6\ne 0$ below. For $\<\psi_1\psi_2\psi_2\psi_1\>$ and $\<\psi\psi\psi\psi\>$, the coefficients become quite lengthy so we do not reproduce them here; please see the attached \texttt{Mathematica} file for their explicit expressions.

For the correlator $\<\psi\phi\phi\psi\>$, the list below constitutes a sufficient set of nonzero $\cK$ coefficients  if the exchanged operator in t-channel is not a scalar.\footnote{For scalar exchange in the t-channel, see \equref{eq: K coefficients for 2f2s with an exchange of a scalar t channel}.} For convenience, we use the same shorthand notation as earlier:
\be 
\De_{abc}=\De_{a}+\De_{b}+\De_{c}\;,\quad \Delta_{ab}^{c}=\De_{a}+\De_{b}-\De_{c}.
\ee 

The coefficients are:
\begin{subequations}
\begin{equation}
\begin{array}{cc}
\opeFuncDecomp{\uniq 2}{11}{\psi_1}{\f_2}{\f_2}{\psi_1}{\cO_5}{\cO_6}{\f_1^{-\half}}{\f_2}{\f_2}{\f_1^{-\half}}{\cO_{5}^{-\half,-\half}}{\cO_6} 
& \frac{i
	\left(\Delta _5-\frac{3}{2}\right) \left(\Delta_{125}+l_5-\frac{9}{2}\right){}^2
	\left(\Delta_{15}^{2}+l_5-\frac{3}{2}\right){}^2}{16 \left(\Delta _5-2\right)
	\left(\Delta _6-1\right) l_6 \left(\Delta _5+l_5-2\right) \left(\Delta _5+l_5-1\right)}
\\
\opeFuncDecomp{\uniq 2}{22}{\psi_1}{\f_2}{\f_2}{\psi_1}{\cO_5}{\cO_6}{\f_1^{-\half}}{\f_2}{\f_2}{\f_1^{-\half}}{\cO_{5}^{-\half,\half}}{\cO_6} 
& -\frac{i
	\left(\Delta _5-\frac{3}{2}\right) \left(l_5+\frac{1}{2}\right) \left(-\Delta
	_{125}+l_5+\frac{11}{2}\right){}^2 \left(-\Delta_{15}^{2}+l_5+\frac{5}{2}\right){}^2}{16 \left(\Delta _5-2\right) \left(\Delta _6-1\right)
	\left(l_5+1\right) l_6 \left(-\Delta _5+l_5+2\right) \left(-\Delta _5+l_5+3\right)} 
\\
\opeFuncDecomp{\uniq 2}{22}{\psi_1}{\f_2}{\f_2}{\psi_1}{\cO_5}{\cO_6}{\f_1^{-\half}}{\f_2}{\f_2}{\f_1^{-\half}}{\cO_{5}^{\half,-\half}}{\cO_6} 
& -\frac{i
	\left(\Delta_{12}^{5}+l_5-\frac{3}{2}\right){}^2}{4 \left(\Delta _6-1\right) l_6} 
\\
\opeFuncDecomp{\uniq 2}{11}{\psi_1}{\f_2}{\f_2}{\psi_1}{\cO_5}{\cO_6}{\f_1^{-\half}}{\f_2}{\f_2}{\f_1^{-\half}}{\cO_{5}^{\half,\half}}{\cO_6} 
& \frac{i
	\left(l_5+\frac{1}{2}\right) \left(-\Delta_{12}^{5}+l_5+\frac{5}{2}\right){}^2}{4
	\left(\Delta _6-1\right) \left(l_5+1\right) l_6} 
\\
\opeFuncDecomp{\uniq 3}{12}{\psi_1}{\f_2}{\f_2}{\psi_1}{\cO_5}{\cO_6}{\f_1^{-\half}}{\f_2}{\f_2}{\f_1^{\half}}{\cO_{5}^{-\half,-\half}}{\cO_6} 
& \frac{i
	\left(\Delta _5-\frac{3}{2}\right) \left(\Delta_{125}+l_5-\frac{9}{2}\right)
	\left(\Delta_{15}^{2}+l_5-\frac{3}{2}\right) \left(\Delta_{25}^{1}+l_5-\frac{3}{2}\right)}{8 \left(\Delta _5-2\right) \left(\Delta _6-1\right)
	\left(\Delta _5+l_5-2\right) \left(\Delta _5+l_5-1\right)} 
\\
\opeFuncDecomp{\uniq 4}{12}{\psi_1}{\f_2}{\f_2}{\psi_1}{\cO_5}{\cO_6}{\f_1^{-\half}}{\f_2}{\f_2}{\f_1^{\half}}{\cO_{5}^{-\half,-\half}}{\cO_6} 
& \frac{i
	\left(\Delta _5-\frac{3}{2}\right) \left(\Delta_{125}+l_5-\frac{9}{2}\right)
	\left(\Delta_{15}^{2}+l_5-\frac{3}{2}\right) \left(\Delta_{25}^{1}+l_5-\frac{3}{2}\right)}{8 \left(\Delta _5-2\right) l_6 \left(\Delta
	_5+l_5-2\right) \left(\Delta _5+l_5-1\right)} 
\\
\opeFuncDecomp{\uniq 3}{21}{\psi_1}{\f_2}{\f_2}{\psi_1}{\cO_5}{\cO_6}{\f_1^{-\half}}{\f_2}{\f_2}{\f_1^{\half}}{\cO_{5}^{-\half,\half}}{\cO_6} 
& \frac{i
	\left(\Delta _5-\frac{3}{2}\right) \left(l_5+\frac{1}{2}\right) \left(-\Delta
	_{125}+l_5+\frac{11}{2}\right) \left(-\Delta_{15}^{2}+l_5+\frac{5}{2}\right)
	\left(-\Delta_{25}^{1}+l_5+\frac{5}{2}\right)}{8 \left(\Delta _5-2\right) \left(\Delta
	_6-1\right) \left(l_5+1\right) \left(-\Delta _5+l_5+2\right) \left(-\Delta
	_5+l_5+3\right)} 
\\
\opeFuncDecomp{\uniq 4}{21}{\psi_1}{\f_2}{\f_2}{\psi_1}{\cO_5}{\cO_6}{\f_1^{-\half}}{\f_2}{\f_2}{\f_1^{\half}}{\cO_{5}^{-\half,\half}}{\cO_6}
& \frac{i
	\left(\Delta _5-\frac{3}{2}\right) \left(l_5+\frac{1}{2}\right) \left(-\Delta
	_{125}+l_5+\frac{11}{2}\right) \left(-\Delta_{15}^{2}+l_5+\frac{5}{2}\right)
	\left(-\Delta_{25}^{1}+l_5+\frac{5}{2}\right)}{8 \left(\Delta _5-2\right)
	\left(l_5+1\right) l_6 \left(-\Delta _5+l_5+2\right) \left(-\Delta _5+l_5+3\right)} 
\\
\opeFuncDecomp{\uniq 3}{21}{\psi_1}{\f_2}{\f_2}{\psi_1}{\cO_5}{\cO_6}{\f_1^{-\half}}{\f_2}{\f_2}{\f_1^{\half}}{\cO_{5}^{\half,-\half}}{\cO_6}& \frac{i
	\left(\Delta_{12}^{5}+l_5-\frac{3}{2}\right)}{2 \left(\Delta _6-1\right)} 
\\
\opeFuncDecomp{\uniq 4}{21}{\psi_1}{\f_2}{\f_2}{\psi_1}{\cO_5}{\cO_6}{\f_1^{-\half}}{\f_2}{\f_2}{\f_1^{\half}}{\cO_{5}^{\half,-\half}}{\cO_6}
& \frac{i
	\left(\Delta_{12}^{5}+l_5-\frac{3}{2}\right)}{2 l_6} 
\\
\opeFuncDecomp{\uniq 3}{12}{\psi_1}{\f_2}{\f_2}{\psi_1}{\cO_5}{\cO_6}{\f_1^{-\half}}{\f_2}{\f_2}{\f_1^{\half}}{\cO_{5}^{\half,\half}}{\cO_6}
& \frac{i
	\left(l_5+\frac{1}{2}\right) \left(-\Delta_{12}^{5}+l_5+\frac{5}{2}\right)}{2
	\left(\Delta _6-1\right) \left(l_5+1\right)} 
\\
\opeFuncDecomp{\uniq 4}{12}{\psi_1}{\f_2}{\f_2}{\psi_1}{\cO_5}{\cO_6}{\f_1^{-\half}}{\f_2}{\f_2}{\f_1^{\half}}{\cO_{5}^{\half,\half}}{\cO_6}
& \frac{i
	\left(l_5+\frac{1}{2}\right) \left(-\Delta_{12}^{5}+l_5+\frac{5}{2}\right)}{2
	\left(l_5+1\right) l_6} 
\\
\opeFuncDecomp{\uniq 3}{21}{\psi_1}{\f_2}{\f_2}{\psi_1}{\cO_5}{\cO_6}{\f_1^{\half}}{\f_2}{\f_2}{\f_1^{-\half}}{\cO_{5}^{-\half,-\half}}{\cO_6}
& \frac{i
	\left(\Delta _5-\frac{3}{2}\right) \left(\Delta_{125}+l_5-\frac{9}{2}\right)
	\left(\Delta_{15}^{2}+l_5-\frac{3}{2}\right) \left(\Delta_{25}^{1}+l_5-\frac{3}{2}\right)}{8 \left(\Delta _5-2\right) \left(\Delta _6-1\right)
	\left(\Delta _5+l_5-2\right) \left(\Delta _5+l_5-1\right)} \\
\opeFuncDecomp{\uniq 4}{21}{\psi_1}{\f_2}{\f_2}{\psi_1}{\cO_5}{\cO_6}{\f_1^{\half}}{\f_2}{\f_2}{\f_1^{-\half}}{\cO_{5}^{-\half,-\half}}{\cO_6}
& \frac{i
	\left(\Delta _5-\frac{3}{2}\right) \left(1-\Delta _6\right) \left(\Delta
	_{125}+l_5-\frac{9}{2}\right) \left(\Delta_{15}^{2}+l_5-\frac{3}{2}\right) \left(\Delta_{25}^{1}+l_5-\frac{3}{2}\right)}{8 \left(\Delta _5-2\right) \left(\Delta _6-1\right) l_6
	\left(\Delta _5+l_5-2\right) \left(\Delta _5+l_5-1\right)} \\
\opeFuncDecomp{\uniq 3}{12}{\psi_1}{\f_2}{\f_2}{\psi_1}{\cO_5}{\cO_6}{\f_1^{\half}}{\f_2}{\f_2}{\f_1^{-\half}}{\cO_{5}^{-\half,\half}}{\cO_6}
& \frac{i
	\left(\Delta _5-\frac{3}{2}\right) \left(l_5+\frac{1}{2}\right) \left(-\Delta
	_{125}+l_5+\frac{11}{2}\right) \left(-\Delta_{15}^{2}+l_5+\frac{5}{2}\right)
	\left(-\Delta_{25}^{1}+l_5+\frac{5}{2}\right)}{8 \left(\Delta _5-2\right) \left(\Delta
	_6-1\right) \left(l_5+1\right) \left(-\Delta _5+l_5+2\right) \left(-\Delta
	_5+l_5+3\right)} \\
\opeFuncDecomp{\uniq 4}{12}{\psi_1}{\f_2}{\f_2}{\psi_1}{\cO_5}{\cO_6}{\f_1^{\half}}{\f_2}{\f_2}{\f_1^{-\half}}{\cO_{5}^{-\half,\half}}{\cO_6}
& \frac{i
	\left(\Delta _5-\frac{3}{2}\right) \left(1-\Delta _6\right) \left(l_5+\frac{1}{2}\right)
	\left(-\Delta_{125}+l_5+\frac{11}{2}\right) \left(-\Delta_{15}^{2}+l_5+\frac{5}{2}\right) \left(-\Delta_{25}^{1}+l_5+\frac{5}{2}\right)}{8
	\left(\Delta _5-2\right) \left(\Delta _6-1\right) \left(l_5+1\right) l_6 \left(-\Delta
	_5+l_5+2\right) \left(-\Delta _5+l_5+3\right)} \\
\opeFuncDecomp{\uniq 3}{12}{\psi_1}{\f_2}{\f_2}{\psi_1}{\cO_5}{\cO_6}{\f_1^{\half}}{\f_2}{\f_2}{\f_1^{-\half}}{\cO_{5}^{\half,-\half}}{\cO_6}
& \frac{i
	\left(\Delta_{12}^{5}+l_5-\frac{3}{2}\right)}{2 \left(\Delta _6-1\right)} \\
\opeFuncDecomp{\uniq 4}{12}{\psi_1}{\f_2}{\f_2}{\psi_1}{\cO_5}{\cO_6}{\f_1^{\half}}{\f_2}{\f_2}{\f_1^{-\half}}{\cO_{5}^{\half,-\half}}{\cO_6}
& \frac{i
	\left(1-\Delta _6\right) \left(\Delta_{12}^{5}+l_5-\frac{3}{2}\right)}{2 \left(\Delta
	_6-1\right) l_6} \\
\opeFuncDecomp{\uniq 3}{21}{\psi_1}{\f_2}{\f_2}{\psi_1}{\cO_5}{\cO_6}{\f_1^{\half}}{\f_2}{\f_2}{\f_1^{-\half}}{\cO_{5}^{\half,\half}}{\cO_6}
& \frac{i
	\left(l_5+\frac{1}{2}\right) \left(-\Delta_{12}^{5}+l_5+\frac{5}{2}\right)}{2
	\left(\Delta _6-1\right) \left(l_5+1\right)} 
\\
\opeFuncDecomp{\uniq 4}{21}{\psi_1}{\f_2}{\f_2}{\psi_1}{\cO_5}{\cO_6}{\f_1^{\half}}{\f_2}{\f_2}{\f_1^{-\half}}{\cO_{5}^{\half,\half}}{\cO_6}
& \frac{i
	\left(1-\Delta _6\right) \left(l_5+\frac{1}{2}\right) \left(-
	\Delta_{12}^{5}+l_5+\frac{5}{2}\right)}{2 \left(\Delta _6-1\right) \left(l_5+1\right) l_6} \\
\opeFuncDecomp{\uniq 1}{22}{\psi_1}{\f_2}{\f_2}{\psi_1}{\cO_5}{\cO_6}{\f_1^{\half}}{\f_2}{\f_2}{\f_1^{\half}}{\cO_{5}^{-\half,-\half}}{\cO_6}
& -\frac{i
	\left(\Delta _5-\frac{3}{2}\right) \left(\Delta_{25}^{1}+l_5-\frac{3}{2}\right){}^2}{4
	\left(\Delta _5-2\right) \left(\Delta _5+l_5-2\right) \left(\Delta _5+l_5-1\right)} 
\end{array}
\end{equation}

\begin{equation}
\begin{array}{cc}
\opeFuncDecomp{\uniq 2}{22}{\psi_1}{\f_2}{\f_2}{\psi_1}{\cO_5}{\cO_6}{\f_1^{\half}}{\f_2}{\f_2}{\f_1^{\half}}{\cO_{5}^{-\half,-\half}}{\cO_6}
& \frac{i
	\left(\Delta _5-\frac{3}{2}\right) \left(\Delta _1+\frac{1}{2} \left(-\Delta
	_6+l_6-1\right)\right) \left(\Delta _1+\frac{1}{2} \left(\Delta _6-l_6-4\right)\right)
	\left(\Delta_{25}^{1}+l_5-\frac{3}{2}\right){}^2}{4 \left(\Delta _5-2\right) \left(\Delta
	_6-1\right) l_6 \left(\Delta _5+l_5-2\right) \left(\Delta _5+l_5-1\right)} \\
\opeFuncDecomp{\uniq 1}{11}{\psi_1}{\f_2}{\f_2}{\psi_1}{\cO_5}{\cO_6}{\f_1^{\half}}{\f_2}{\f_2}{\f_1^{\half}}{\cO_{5}^{-\half,\half}}{\cO_6}
& \frac{i
	\left(\Delta _5-\frac{3}{2}\right) \left(l_5+\frac{1}{2}\right) \left(-\Delta_{25}^{1}+l_5+\frac{5}{2}\right){}^2}{4 \left(\Delta _5-2\right) \left(l_5+1\right)
	\left(-\Delta _5+l_5+2\right) \left(-\Delta _5+l_5+3\right)} \\
\opeFuncDecomp{\uniq 2}{11}{\psi_1}{\f_2}{\f_2}{\psi_1}{\cO_5}{\cO_6}{\f_1^{\half}}{\f_2}{\f_2}{\f_1^{\half}}{\cO_{5}^{-\half,\half}}{\cO_6}
& -\frac{i
	\left(\Delta _5-\frac{3}{2}\right) \left(l_5+\frac{1}{2}\right) \left(\Delta
	_1+\frac{1}{2} \left(-\Delta _6+l_6-1\right)\right) \left(\Delta _1+\frac{1}{2}
	\left(\Delta _6-l_6-4\right)\right) \left(-\Delta_{25}^{1}+l_5+\frac{5}{2}\right){}^2}{4
	\left(\Delta _5-2\right) \left(\Delta _6-1\right) \left(l_5+1\right) l_6 \left(-\Delta
	_5+l_5+2\right) \left(-\Delta _5+l_5+3\right)} \\
\opeFuncDecomp{\uniq 1}{11}{\psi_1}{\f_2}{\f_2}{\psi_1}{\cO_5}{\cO_6}{\f_1^{\half}}{\f_2}{\f_2}{\f_1^{\half}}{\cO_{5}^{\half,-\half}}{\cO_6}
& i \\
\opeFuncDecomp{\uniq 2}{11}{\psi_1}{\f_2}{\f_2}{\psi_1}{\cO_5}{\cO_6}{\f_1^{\half}}{\f_2}{\f_2}{\f_1^{\half}}{\cO_{5}^{\half,-\half}}{\cO_6}
& -\frac{i
	\left(\Delta _1+\frac{1}{2} \left(-\Delta _6+l_6-1\right)\right) \left(\Delta
	_1+\frac{1}{2} \left(\Delta _6-l_6-4\right)\right)}{\left(\Delta _6-1\right) l_6} \\
\opeFuncDecomp{\uniq 1}{22}{\psi_1}{\f_2}{\f_2}{\psi_1}{\cO_5}{\cO_6}{\f_1^{\half}}{\f_2}{\f_2}{\f_1^{\half}}{\cO_{5}^{\half,\half}}{\cO_6}
& -\frac{i
	\left(l_5+\frac{1}{2}\right)}{l_5+1} \\
\opeFuncDecomp{\uniq 2}{22}{\psi_1}{\f_2}{\f_2}{\psi_1}{\cO_5}{\cO_6}{\f_1^{\half}}{\f_2}{\f_2}{\f_1^{\half}}{\cO_{5}^{\half,\half}}{\cO_6}
& \frac{i
	\left(l_5+\frac{1}{2}\right) \left(\Delta _1+\frac{1}{2} \left(-\Delta
	_6+l_6-1\right)\right) \left(\Delta _1+\frac{1}{2} \left(\Delta
	_6-l_6-4\right)\right)}{\left(\Delta _6-1\right) \left(l_5+1\right) l_6}
\end{array}
\end{equation}
\end{subequations}

% END
\bibliography{collectiveReferenceLibrary}{}
\bibliographystyle{utphys}
\end{document}